\documentstyle[twocolumn,seceq,epsbox]{jpsj}

\def\runtitle{Pseudogap Phenomena and Superconducting Fluctuations 
in Hubbard Model for High-$T_{{\rm c}}$ Cuprates} 
\def\runauthor{Youichi {\sc Yanase}, Kosaku {\sc Yamada} }

\title{ 
Pseudogap Phenomena and Superconducting Fluctuations \\ 
in Hubbard Model for High-$T_{{\rm c}}$ Cuprates 
} 

\author{Youichi {\sc Yanase}\footnote{E-mail: yanase@hosi.phys.s.u-tokyo.ac.jp},\footnote{Present address: Department of Physics, University of Tokyo, Tokyo 
113-0033}
and Kosaku {\sc Yamada}}

\inst{Department of Physics, Kyoto University, Kyoto 606-8502}

\recdate{October 2, 2000}

\abst
{
 The pseudogap phenomena in High-$T_{{\rm c}}$ cuprates are investigated 
on the basis of the Hubbard model which includes only the on-site repulsive 
interaction $U$. 
 We consider the pairing scenario for the pseudogap. The pseudogap arises from 
the resonance scattering due to the strong superconducting fluctuations.  
 First, the electronic state and the anti-ferromagnetic spin fluctuations 
are calculated by using the fluctuation exchange (FLEX) approximation. 
 The T-matrix which is the propagator of the superconducting fluctuations 
is calculated by extending the $\acute{{\rm E}}$liashberg equation. 
 The characteristics of the superconducting fluctuations 
due to the pairing interaction 
arising from the spin fluctuations are represented in the T-matrix. 
 The self-energy due to the superconducting fluctuations 
is calculated by the T-matrix approximation. 
 The pseudogap is shown in the single particle properties and the magnetic 
properties. 
 Moreover, a comprehensive explanation of the doping dependence 
of the pseudogap is obtained. 
 Furthermore, we apply the theory to the electron-doped cuprates and 
obtain the consistent results with the recent experiments. 
 Finally, the self-consistent calculation for the 
spin fluctuations, superconducting fluctuations and the single particle 
properties are carried out within the FLEX and the self-consistent T-matrix 
approximations. 
 The relation between the superconducting fluctuations and the spin 
fluctuations are clarified. 
 The calculated superconducting critical temperature $T_{{\rm c}}$ is  
remarkably reduced from the results of the mean field (FLEX) calculation. 
 In particular, it is shown that 
the critical temperature decreases with decreasing doping 
in the under-doped region with large $U$. 
}

\kword
{High-$T_{{\rm c}}$ cuprates; Hubbard model;  Pseudogap;
Strong coupling superconductivity; \\
Superconducting fluctuation; spin fluctuation }

\begin{document}
\sloppy
\maketitle

\section{Introduction}

  The pseudogap phenomena in under-doped High-$T_{{\rm c}}$ cuprates 
have been interested for many years. 
  The pseudogap is considered to be a key issue for the comprehensive 
understanding of the High-$T_{{\rm c}}$ superconductivity. 

 The pseudogap phenomena mean the suppression of the spectral weight 
above $T_{{\rm c}}$ without any long range order. 
 They are universal phenomena observed in various compounds of 
High-$T_{{\rm c}}$ cuprates in the under-doped region. 
 First, the pseudogap was found in the magnetic excitation 
channel by the nuclear magnetic resonance (NMR) experiment.~\cite{rf:yasuoka} 
 At present, the pseudogap phenomena have been observed in various quantities
which include NMR,~\cite{rf:yasuoka,rf:NMR,rf:takigawa,rf:itoh,rf:ishida,rf:tokunaga,rf:goto}  
neutron scattering,~\cite{rf:neutron} 
transport,~\cite{rf:transport,rf:odatransport} 
optical spectrum,~\cite{rf:homes} 
electronic specific heat,~\cite{rf:momono}
density of states~\cite{rf:renner}  
and the single particle spectral weight.~\cite{rf:ARPES}  
 The experimental results are reviewed in ref. 15.

 There are many theoretical scenarios proposed for the pseudogap phenomena. 
 An important one is the resonating valence bond (RVB) theory. 
The RVB theory is an approach from the non-Fermi liquid state and  
attributes the pseudogap to the singlet pairing of the 
spinons.~\cite{rf:tanamoto} 
 As an approach from the Fermi liquid state, the magnetic scenario has been 
investigated.~\cite{rf:pines,rf:SDW}  In this scenario, the pseudogap 
is an anti-ferromagnetic gap formation 
or its precursor.

 In this paper, we derive the pseudogap phenomena which arise from 
the strong superconducting fluctuations. 
 The strong superconducting fluctuations originates from the strong coupling 
superconductivity and the quasi-two dimensionality.~\cite{rf:yanasePG} 
 This theory belongs to the pairing scenario in which 
the pseudogap is a precursor of the superconductivity. 
 Some aspects of the observed pseudogap have supported the pairing scenario. 
 In particular, the measurements of the single particle properties  
have given the important features of the pseudogap. 
 The same energy scale and the same momentum dependence 
between the pseudogap and the superconducting gap have been shown 
by the angle-resolved photo-emission spectroscopy (ARPES).~\cite{rf:ARPES}   
 The tunneling spectroscopy have also shown the smooth change from the 
pseudogap state to the superconducting state.~\cite{rf:renner}
 The close relation between the pseudogap state and 
the superconducting state clearly indicates that the pseudogap is a 
precursor of the superconductivity.

 The paring scenarios have been proposed~\cite{rf:randeriareview,rf:emery} 
and actively investigated in recent years.~\cite{rf:yanasePG,rf:phase,rf:phase2,rf:haussman,rf:stintzing,rf:koikegamiNSR,rf:kobayashiNSR,rf:geshkenbein,rf:janko,rf:levin,rf:jujo,rf:jujoyanase,rf:yanaseSC,rf:yanaseMG,rf:micnas,rf:dagotto,rf:kobayasi,rf:onoda,rf:koikegami,rf:dahmpseudogap,rf:perali,rf:metzner}   
 They are classified into some kinds. 
 The scenario based on the phase fluctuations has been proposed by 
Emery and Kivelson~\cite{rf:emery} and calculated by other 
authors.~\cite{rf:phase,rf:phase2} 
 The strong phase fluctuations in the ordered state are expected 
in the under-doped region since the London penetration depth $\lambda$ is 
long.~\cite{rf:uemura} 
 The small London constant $\Lambda =1 /4 \pi \lambda^{2}$ means the small 
phase stiffness.

 The importance of the strong coupling superconductivity has been 
pointed out by Randeria {\it et al}~\cite{rf:randeriareview} 
on the basis of the Nozi$\grave{{\rm e}}$res and Schmitt-Rink 
(NSR) theory.~\cite{rf:Nozieres,rf:tokumitu} 
 The NSR theory describes the crossover from the conventional BCS 
superconductivity to the Bose condensation of tightly-bound pre-formed pairs. 
 The concept of the NSR theory is that the crossover is described by only 
shifting the chemical potential $\mu$. 
 In case of the strong attractive interaction, the fermions form the 
pre-formed pairs without phase coherence. In this case, the gap opens in the 
spectrum of the fermions. The low energy excitations correspond to  
the pre-formed bosons. 
 The NSR theory has been investigated by many authors 
because the gap in the fermionic excitations is considered to be 
the pseudogap.~\cite{rf:randeriareview,rf:haussman,rf:stintzing,rf:koikegamiNSR,rf:kobayashiNSR} 
 The existence of the pre-formed pairs has been proposed phenomenologically 
by Geshkenbein {\it et al.}~\cite{rf:geshkenbein} 
with reference to the sign problem of the fluctuational 
Hall effect.~\cite{rf:aronov}

 The strength of the superconducting coupling is indicated by the ratio 
$T_{{\rm c}}^{{\rm MF}}/\varepsilon_{{\rm F}}$. Here, $\varepsilon_{{\rm F}}$ 
is the effective Fermi energy, and $T_{{\rm c}}^{{\rm MF}}$ is the 
superconducting critical temperature obtained by the mean field theory. 
 Because the assumption of the weak coupling 
$T_{{\rm c}}^{{\rm MF}}/\varepsilon_{{\rm F}} \ll 1 $ is justified 
in usual superconductors, the BCS mean field theory correctly describes the 
superconducting phase transition. 
 The strong coupling superconductivity means that the assumption is violated 
and therefore the superconducting fluctuations play an important role.  
 The effects of the fluctuations are furthermore strong 
in the quasi-two dimensional systems.  
 The small $\varepsilon_{{\rm F}}$, high-$T_{{\rm c}}$ and the quasi-two 
dimensionality are the characteristics of the cuprates. 
 Therefore, it is natural to consider the strong superconducting fluctuations 
in High-$T_{{\rm c}}$ cuprates.

 The NSR theory takes into account the correction of the superconducting 
fluctuations on the chemical potential $\mu$. 
 However, we have asserted that the NSR theory is justified in the low density 
limit and not in High-$T_{{\rm c}}$ cuprates which are high density 
systems.~\cite{rf:yanasePG} 
 Therefore, the NSR scenario should be ruled out as a candidate 
of the pseudogap phenomena. 

 In the high density systems, the effect of the strong superconducting 
fluctuations manifests in a different way. 
 It has been shown that the pseudogap phenomena are derived by 
the self-energy correction due to the strong superconducting 
fluctuations.~\cite{rf:yanasePG,rf:janko} 
 The resonance scattering by the low energy superconducting fluctuations 
gives rise to the anomalous features of the self-energy and leads to the 
pseudogap. 
 These effects have been neglected in the NSR theory. 
 The large weight of the superconducting fluctuations at the low energy 
necessarily exists when the strong coupling superconductivity 
occurs in the quasi-two dimensional systems.~\cite{rf:yanasePG,rf:jujo} 
 We adopt the resonance scattering scenario also in this paper. 

 We furthermore emphasize that our scenario is clearly differentiated 
from the NSR scenario. 
 The phase transition in the NSR theory is regarded as the Bose 
condensation of the pre-formed pairs. 
 However, the phase transition is the superconductivity with strong 
fluctuations in our scenario.~\cite{rf:jujoyanase,rf:yanaseSC} 
 The later is the realistic situation in High-$T_{{\rm c}}$ cuprates.

 The effects of the resonance scattering are especially weak in the 
usual weak coupling superconductors. 
 In these cases, the self-energy correction has not been emphasized in the 
theory of the superconducting fluctuations, 
since they are less singular compared with the correction on 
the two-body correlation function, such as the AL term and 
the MT term.~\cite{rf:AL,rf:MT} 
 However, the superconducting fluctuations have strong effects on the 
electronic state in the strong coupling case.

 The resonance scattering scenario properly explains the magnetic field 
dependences of the pseudogap phenomena measured by the high field 
NMR experiments.~\cite{rf:zheng,rf:gorny,rf:eschrig,rf:zheng2} 
 The calculation gives a comprehensive explanation of the experimental 
results including their doping dependence.~\cite{rf:yanaseMG}  
 Moreover, We have explained the pseudogap phenomena in  
various experiments such as ARPES, tunneling spectroscopy, NMR, 
in-plane and {\it c}-axis transport and the optical 
conductivity.~\cite{rf:yanaseSC}

 Thus, the pseudogap phenomena are well explained on the basis of the 
resonance scattering scenario. 
 However, most of the calculations are based on the model with an attractive 
interaction. Our previous calculations are also based on the effective model 
in which the effectively strong $d$-wave pairing interaction affects 
the renormalized quasi-particles.~\cite{rf:yanasePG,rf:jujo,rf:jujoyanase,rf:yanaseSC,rf:yanaseMG}  
 Actually, the $d$-wave pairing interaction arises from the repulsive 
interaction between the electrons.

 Here, it is one of the most important issues to describe the microscopic 
theory which starts from the repulsive interaction and derives the 
superconducting fluctuations and the pseudogap phenomena. 
 It should be confirmed that the superconducting coupling is strong enough 
to lead to the pseudogap phenomena in the under-doped region. 
 Such microscopic calculation will naturally reproduce  
the doping dependence of the pseudogap phenomena.  
 As is mentioned before,~\cite{rf:yanasePG}  there are two important factors 
in order to realize the strong coupling superconductivity. 
 One is that the effective Fermi energy $\varepsilon_{{\rm F}}$ is 
renormalized by the electron-electron correlation. 
 The other is that the high critical temperature $T_{{\rm c}}$ is obtained 
by the pairing interaction mediated by the anti-ferromagnetic spin 
fluctuations.~\cite{rf:moriya,rf:monthoux} 
 The spin fluctuations also result from the 
strong electron-electron correlation. 
 Thus, the both factors are derived from the electron-electron correlation and 
should be described in a unified way. 
 The main purpose of this paper is describing the microscopic theory 
by which the superconducting fluctuations and the the pseudogap phenomena 
are derived from the electron-electron correlation.

 Some authors have calculated the pseudogap phenomena by starting from  
the model with an repulsive 
interaction.~\cite{rf:kobayasi,rf:koikegami,rf:dahmpseudogap} 
 However, the sufficient result describing the pseudogap phenomena 
has not been obtained without any phenomenology.

 In this paper, we start from the Hubbard model which is an typical model 
for the strongly correlated electron systems. 
 First, we describe the quasi-particles and the anti-ferromagnetic 
spin fluctuations by using the FLEX approximation. 
 The characteristic momentum dependence arising from the 
anti-ferromagnetic spin fluctuations~\cite{rf:stojkovic,rf:yanaseTR} 
are described by the FLEX approximation. 
 The superconducting fluctuations are derived from the pairing interaction 
mediated by the spin fluctuations. 
 The derived superconducting fluctuations include the characteristic 
momentum and frequency dependence of the spin fluctuations. 
 The self-energy due to the superconducting fluctuations is obtained by the 
one-loop approximation (T-matrix approximation). 
 As a result, the comprehensive understanding is obtained 
for the pseudogap phenomena on the basis of the resonance scattering scenario. 
 The doping dependence which includes the electron-doped cuprates is also 
explained consistently.

 This paper is constructed as follows. 
 In \S2, we explain the Hubbard Hamiltonian and the FLEX approximation. 
 In \S3.1, we explain the formalism which describes the superconducting 
fluctuations arising from the spin fluctuations and the self-energy due to the 
superconducting fluctuations. The calculated results for the single particle 
properties are shown in \S3.2. In \S3.3, we show the characteristics 
of the superconducting fluctuations including their doping dependence.  
 The magnetic properties in the pseudogap state are explained in \S3.4. 
 The consistent results with the NMR and the neutron scattering experiments 
are obtained. 
 In \S3.5, the theory of the superconductivity and the pseudogap phenomena 
is applied to the electron-doped cuprates. The relevant results including 
the particle-hole asymmetry are obtained. 
 In \S4.1, we clarify the relation between the superconducting fluctuations 
and the spin fluctuations in details. 
 In \S4.2, we show the results of the self-consistent calculation including 
the spin fluctuations, the superconducting fluctuations and the single 
particle properties. The critical temperature reduced by the fluctuations are 
calculated. As a result, the appropriate phase diagram is obtained. 
 In \S5, we summarize the obtained results and gives some discussions.

\section{Hubbard Model and FLEX Approximation}

 First, we explain the Hubbard model and the FLEX approximation. 
 The Hubbard model has been used for a long time as one of the typical models 
describing the strongly correlated electron systems. 
 The Hamiltonian is described as, 
\begin{eqnarray}
  \label{eq:model}
  H =  \sum_{\mbox{\boldmath$k$},s}  
        \varepsilon_{\mbox{{\scriptsize \boldmath$k$}}} 
c_{\mbox{{\scriptsize \boldmath$k$}},s}^{\dag} 
c_{\mbox{{\scriptsize \boldmath$k$}},s}  
 + U \sum_{\mbox{\boldmath$k$},\mbox{\boldmath$k'$},\mbox{\boldmath$q$}} 
c_{\mbox{{\scriptsize \boldmath$q$}}-\mbox{{\scriptsize \boldmath$k'$}},
\downarrow}^{\dag} 
c_{\mbox{{\scriptsize \boldmath$k'$}},\uparrow}^{\dag} 
c_{\mbox{{\scriptsize \boldmath$k$}},\uparrow} 
c_{\mbox{{\scriptsize \boldmath$q$}}-\mbox{{\scriptsize \boldmath$k$}},
\downarrow}.  
\nonumber \\
\end{eqnarray}
 The first term is the kinetic term. 
In this paper, we adopt the two-dimensional dispersion relation 
$\varepsilon_{\mbox{{\scriptsize \boldmath$k$}}}$ given by the 
tight-binding model for a square lattice including the nearest- and 
next-nearest-neighbor hopping $t$, $t'$, respectively, 
\begin{eqnarray}
 \label{eq:dispersion}
    \varepsilon_{\mbox{{\scriptsize \boldmath$k$}}} = 
                  -2 t (\cos k_{x} +\cos k_{y}) + 
                  4 t' \cos k_{x} \cos k_{y} - \mu. 
\nonumber \\
\end{eqnarray}
 We fix $t=0.5 $, $t'=0.25 t$ and the lattice constant $ a = 1$. 
 In this case, the band width $W=4$. 
 The chemical potential $\mu$ is determined according as the filling $n$. 
 These parameters well reproduce 
the typical Fermi surface of High-$T_{{\rm c}}$ cuprates (see Fig. 13). 
 We define the hole-doping concentration $\delta=1-n$. 
 The main part of this paper is concerned with the hole-doped cuprates  
$\delta > 0$. A brief application to the electron-doped cuprates $\delta < 0$ 
will be  carried out in \S3.5. The nearly half-filling systems, which are 
interested here, should be regarded as high-density systems. Therefore, 
the shift of the chemical potential, which is considered in the NSR scenario, 
is not important in this paper. 
 The second term in eq. (2.1) expresses the on-site repulsive interaction 
between the electrons. 
 All of the phenomena described in this paper result from the 
electron correlation effects.

 The FLEX approximation has been used in order to describe the 
strongly correlated electron systems near the anti-ferromagnetic 
instability.~\cite{rf:FLEX} 
 The FLEX approximation is a conserving approximation~\cite{rf:baym}  and 
describes the anti-ferromagnetic spin fluctuations 
including the mode coupling effects partly.~\cite{rf:SCR,rf:moriyaAD} 
 It has been shown that the superconducting critical temperature with a 
relevant order $T_{{\rm c}} \sim 100{\rm K}$ is obtained 
by the FLEX approximation 
for High-$T_{{\rm c}}$ cuprates.~\cite{rf:monthouxFLEX,rf:paoFLEX,rf:dahmFLEX,
rf:langerFLEX,rf:deiszFLEX,rf:koikegamiFLEX,rf:takimotoFLEX} 
 The relevant results are also obtained 
for the organic superconductor $\kappa$-(BEDT-TTF) 
compounds~\cite{rf:kinoFLEX,rf:kondoFLEX,rf:schmalianFLEX} 
and for the ladder type compound 
${\rm Sr}_{14-x}{\rm Ca}_{x}{\rm Cu}_{24}{\rm O}_{41}$.~\cite{rf:kontaniFLEX} 
 Moreover, the FLEX approximation is used to explain the anomalous properties 
resulting from the anti-ferromagnetic spin fluctuations,  
such as the Hall coefficient~\cite{rf:kontani,rf:kanki}, 
the neutron resonance peak and the dip-hump structure 
in ARPES.~\cite{rf:takimoto}

 The self-energy given by the FLEX approximation 
$ {\mit{\it \Sigma}}_{{\rm F}} (\mbox{\boldmath$k$}, {\rm i} \omega_{n}) $
is expressed by the one-loop diagram exchanging the normal vertex 
$ V_{\rm n} (\mbox{\boldmath$q$}, {\rm i} \Omega_{n}) $ (Fig. 1(a)). 
\begin{eqnarray}
 {\mit{\it \Sigma}}_{{\rm F}} (\mbox{\boldmath$k$}, {\rm i} \omega_{n}) & = & 
 T \sum_{\mbox{\boldmath$q$},{\rm i} \Omega_{n}} 
 V_{\rm n} (\mbox{\boldmath$q$}, {\rm i} \Omega_{n})
 G (\mbox{\boldmath$k-q$}, {\rm i} \omega_{n} - {\rm i} \Omega_{n}). 
\nonumber \\
\end{eqnarray}
 The normal vertex is expressed as follows, 
\begin{eqnarray}
 V_{\rm n} (\mbox{\boldmath$q$}, {\rm i} \Omega_{n}) & = & 
 U^{2} [\frac{3}{2} \chi_{{\rm s}} (\mbox{\boldmath$q$}, {\rm i} \Omega_{n}) 
      + \frac{1}{2} \chi_{{\rm c}} (\mbox{\boldmath$q$}, {\rm i} \Omega_{n}) 
      - \chi_{0} (\mbox{\boldmath$q$}, {\rm i} \Omega_{n})]. 
\nonumber \\
\end{eqnarray}
Here, $\chi_{{\rm s}} (\mbox{\boldmath$q$}, {\rm i} \Omega_{n})$ and 
$\chi_{{\rm c}} (\mbox{\boldmath$q$}, {\rm i} \Omega_{n})$ represent the 
spin and charge susceptibility, respectively, 
\begin{eqnarray}
\chi_{{\rm s}} (\mbox{\boldmath$q$}, {\rm i} \Omega_{n}) & = & 
 \frac{\chi_{0} (\mbox{\boldmath$q$}, {\rm i} \Omega_{n})}
      {1 - U \chi_{0} (\mbox{\boldmath$q$}, {\rm i} \Omega_{n})}, 
\nonumber \\
\chi_{{\rm c}} (\mbox{\boldmath$q$}, {\rm i} \Omega_{n}) & = &
 \frac{\chi_{0} (\mbox{\boldmath$q$}, {\rm i} \Omega_{n})}
      {1 + U \chi_{0} (\mbox{\boldmath$q$}, {\rm i} \Omega_{n})},    
\end{eqnarray}
where $\chi_{0} (\mbox{\boldmath$q$}, {\rm i} \Omega_{n})$ is the 
irreducible susceptibility, 
\begin{eqnarray}
  \chi_{0} (\mbox{\boldmath$q$}, {\rm i} \Omega_{n}) = 
  - T \sum_{\mbox{\boldmath$k$},{\rm i} \omega_{n}} 
  G (\mbox{\boldmath$k$}, {\rm i} \omega_{n}) 
  G (\mbox{\boldmath$k+q$}, {\rm i} \omega_{n} + {\rm i} \Omega_{n}). 
\nonumber \\
\end{eqnarray}

\begin{figure}[htbp]
  \begin{center}
   \epsfysize=2cm
    $$\epsffile{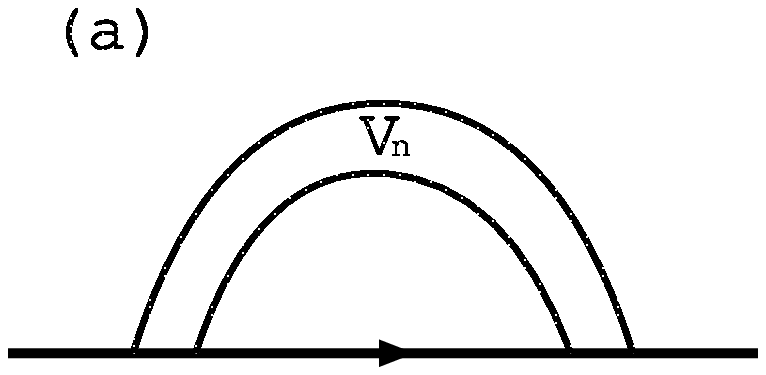}$$
\hspace{5mm}
   \epsfysize=2cm
    $$\epsffile{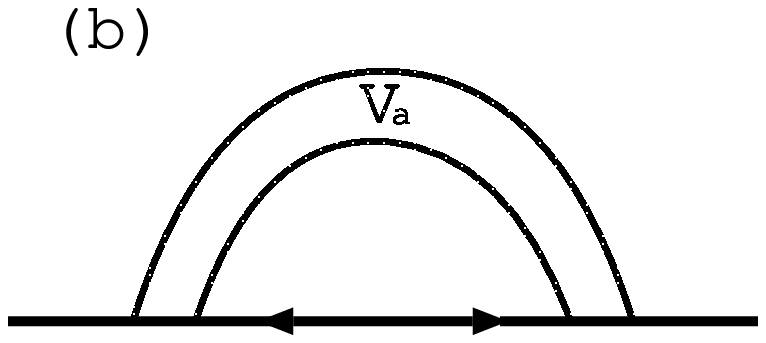}$$
    \caption{(a) The normal self-energy and (b) the anomalous self-energy 
              obtained by the FLEX approximation.}
  \end{center}
\end{figure}
 The first order term in  
$V_{\rm n} (\mbox{\boldmath$q$}, {\rm i} \Omega_{n})$ with respect to $U$ 
is eliminated because it gives only the trivial Hartree-Fock term. 
 In the FLEX approximation, the dressed Green 
function $G (\mbox{\boldmath$k$}, {\rm i} \omega_{n}) = 
( {\rm i} \omega_{n} - \varepsilon_{\mbox{{\scriptsize \boldmath$k$}}} - 
{\mit{\it \Sigma}}_{{\rm F}} (\mbox{\boldmath$k$}, {\rm i} \omega_{n}))^{-1} $
is used. 
 The self-energy and the spin susceptibility
are determined self-consistently. Equations. (2.3-6) are self-consistently 
solved by the numerical calculation. 
 In the main part of this paper, we divide the first Brillouin zone into  
$64 \times 64$ lattice points for the numerical calculation. 
 The spin susceptibility given by the FLEX approximation  
$ \chi_{{\rm s}} (\mbox{\boldmath$q$}, {\rm i} \Omega_{n})$ is enhanced 
near the anti-ferromagnetic wave vector $ \mbox{\boldmath$Q$} = (\pi,\pi)$. 
 The anti-ferromagnetic spin fluctuations described by 
$ \chi_{{\rm s}} (\mbox{\boldmath$q$}, {\rm i} \Omega_{n})$ play a dominant 
role in the FLEX approximation. The characteristic results of 
the nearly anti-ferromagnetic Fermi liquid theory~\cite{rf:pines,rf:moriya,
rf:monthoux,rf:stojkovic,rf:yanaseTR,rf:SCR,rf:moriyaAD,
rf:kontani,rf:kanki,rf:miyake}
are qualitatively reproduced within the FLEX approximation.

 The superconducting critical temperature $T_{{\rm c}}$ is determined 
as the temperature below which the linearized Dyson-Gor'kov equation 
has a non-trivial solution (Fig. 1(b)).  
 The criterion for $T_{{\rm c}}$ is described by 
the $\acute{{\rm E}}$liashberg equation which is the following 
eigenvalue equation, 
\begin{eqnarray}
  \lambda \phi (\mbox{\boldmath$k$}, {\rm i} \omega_{n}) = 
 - & T & \sum_{\mbox{\boldmath$p$},{\rm i} \omega_{m}} 
   V_{\rm a} (\mbox{\boldmath$k-p$}, {\rm i} \omega_{n} - {\rm i} \omega_{m}) 
\nonumber \\
 & \times &  |G (\mbox{\boldmath$p$}, {\rm i} \omega_{m})|^{2}
   \phi (\mbox{\boldmath$p$}, {\rm i} \omega_{m}). 
\end{eqnarray}
 Here, $V_{\rm a} (\mbox{\boldmath$q$}, {\rm i} \Omega_{n})$ is the anomalous 
vertex for the singlet channel. This is given by the FLEX approximation 
as follows, 
\begin{eqnarray}
   V_{\rm a} (\mbox{\boldmath$q$}, {\rm i} \Omega_{n}) & = & 
 U^{2} [\frac{3}{2} \chi_{{\rm s}} (\mbox{\boldmath$q$}, {\rm i} \Omega_{n}) 
      - \frac{1}{2} \chi_{{\rm c}} (\mbox{\boldmath$q$}, {\rm i} \Omega_{n})] 
      + U. 
\nonumber \\
\end{eqnarray}

 The critical temperature $T_{{\rm c}}$ is obtained as the temperature 
where the maximum eigenvalue $\lambda_{{\rm max}}$ becomes the unity 
$\lambda_{{\rm max}} = 1$. 
 The eigenfunction 
$\phi_{{\rm max}} (\mbox{\boldmath$p$}, {\rm i} \omega_{m})$ 
corresponding to the eigenvalue $\lambda_{{\rm max}}$ is the wave function 
of the Cooper pairs. 
In this paper, the symmetry of the superconductivity is always 
the $d_{x^{2}-y^{2}}$-wave.

 Here, we show the typical results of the FLEX approximation 
in Figs. 2-4. 
 The results for the analytically continuated self-energy  
${\mit{\it \Sigma}}_{{\rm F}}^{{\rm R}}(\mbox{\boldmath$k$}, \omega)$
are shown in Fig. 2. 
 In this paper, the analytic continuation is carried out by using 
the Pad$\acute{{\rm e}}$ approximation. 
 The real part 
${\rm Re} {\mit{\it \Sigma}}_{{\rm F}}^{{\rm R}}(\mbox{\boldmath$k$}, \omega)$ 
shows the negative slope and the imaginary part 
${\rm Im} {\mit{\it \Sigma}}_{{\rm F}}^{{\rm R}}(\mbox{\boldmath$k$}, \omega)$
has the minimum absolute value at the Fermi level. 
 They are the characteristic behaviors of the Fermi liquid theory which is 
violated in the pseudogap state. 

 Some notable features are present in the self-energy.  
 The imaginary part shows the $\omega$-linear dependence near the Fermi level
contrary to the $\omega$-square dependence in the conventional Fermi liquid
theory. 
 The behavior has the same origin as that of the $T$-linear 
resistivity.~\cite{rf:stojkovic,rf:yanaseTR}
 The imaginary part at $\omega=0$, that is the quasi-particle's damping, 
is large near $(\pi,0)$ ('hot spot') and is small near $(\pi/2,\pi/2)$ 
('cold spot'). The above momentum dependence plays a crucial role in the 
coherent in-plane transport and the incoherent {\it c}-axis 
transport.~\cite{rf:yanaseSC,rf:yanaseTR,rf:ioffe} 
  The renormalization factor 
$Z_{\mbox{{\scriptsize \boldmath$k$}}}^{-1} = 1 - \partial {\rm Re}
{\mit{\it \Sigma}}_{{\rm F}}^{{\rm R}}(\mbox{\boldmath$k$}, \omega)
/\partial \omega$ has the qualitatively same feature 
as the quasi-particle's damping.  
 A lot of the low energy states near $(\pi,0)$ arise from 
the Van Hove singularity and the large renormalization factor 
$Z_{\mbox{{\scriptsize \boldmath$k$}}}^{-1}$. 
 Because the Fermi surface is transformed by the electron correlation, 
the Van Hove singularity approaches the Fermi level much more. 
 The transformation of the Fermi surface is a characteristic feature of the 
nearly anti-ferromagnetic systems~\cite{rf:yanaseTR} and is confirmed 
by the FLEX approximation~\cite{rf:kontani}. 
 The effective Fermi energy for the superconducting fluctuations are 
renormalized by the above features.

\begin{figure}[htbp]
  \begin{center}
   \epsfysize=6.5cm
    $$\epsffile{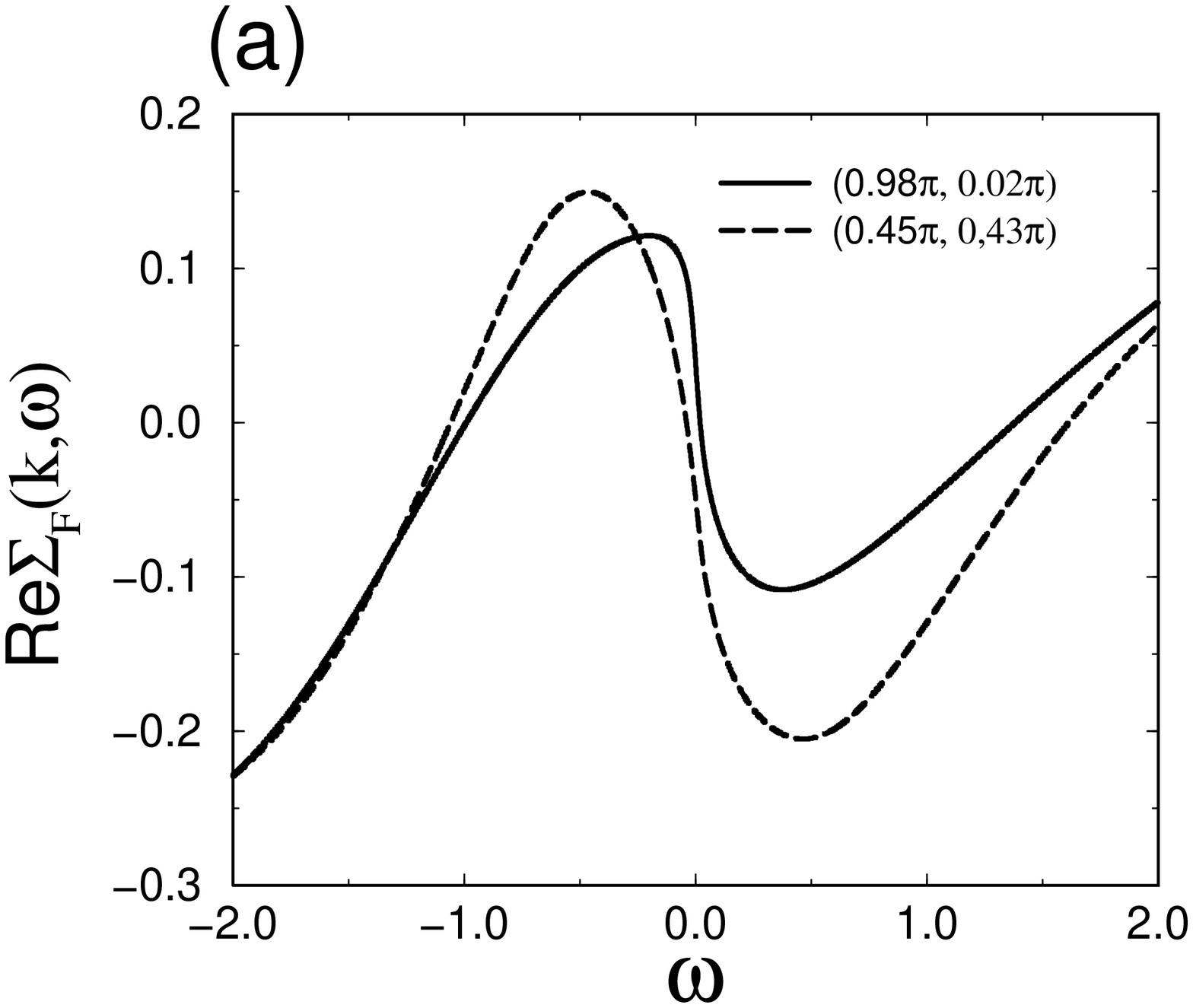}$$
\hspace{5mm}
   \epsfysize=6.5cm
    $$\epsffile{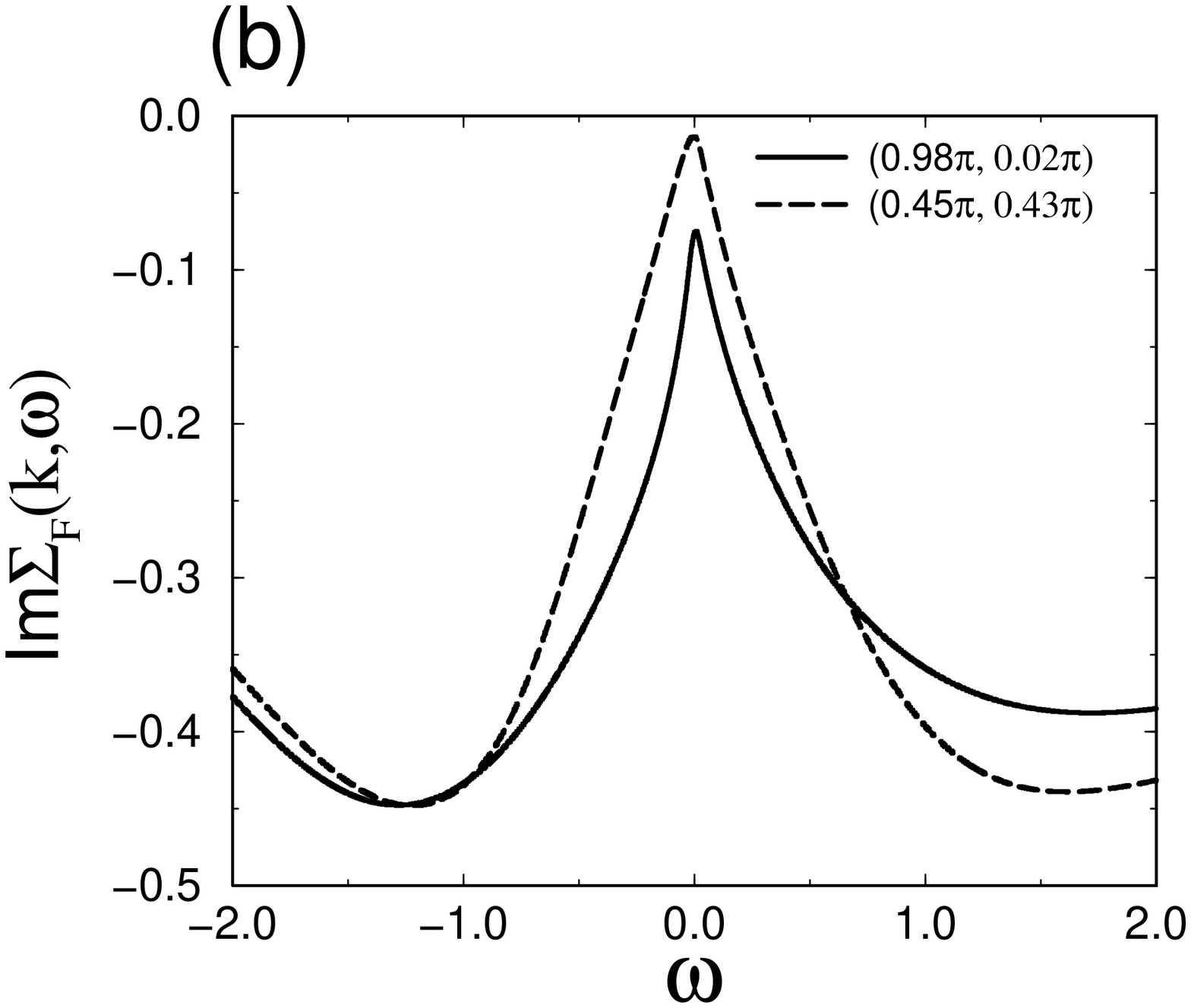}$$
    \caption{The self-energy obtained by the FLEX approximation. 
             (a) The real part. (b) The imaginary part.
             Here, $U=1.6$, $\delta=0.095$ and $T=0.010$. 
             The solid lines and the long-dashed lines correspond to 
             $(\frac{63}{64}\pi,\frac{1}{64}\pi)$ ('hot spot') and 
             $(\frac{29}{64}\pi,\frac{27}{64}\pi)$ ('cold spot'), 
             respectively. 
             The value ${\rm Re} {\mit{\it \Sigma}}_{{\rm F}}^{{\rm R}}
             (\mbox{\boldmath$k$}, 0)$ is positive at the 'hot spot' and 
             negative at the 'cold spot'. This means that the Fermi surface is 
             transformed to be more appropriate to the nesting. 
             }
  \end{center}
\end{figure}

 The results for the single particle spectral weight and the density of states
(DOS) are shown in Fig. 3. 
 In spite of the unconventional Fermi liquid behaviors, the typical 
single peak structure appears in the spectral weight. 
 Thus, the picture of the quasi-particles holds in the nearly 
anti-ferromagnetic Fermi liquid. In particular, the pseudogap is not seen. 
 The spectral weight is remarkably broad at the 'hot spot' and sharp 
at the 'cold spot' reflecting the momentum dependence of the quasi-particle's 
lifetime (see Fig. 2(b)). 
 The low energy states near $(\pi,0)$ give the large DOS near 
the Fermi level. 
 In this paper a small constant damping is added in the self-energy 
when calculating the DOS in order to exclude the finite size effect. 
 This treatment has no significant effect on the obtained results.

\begin{figure}[htbp]
  \begin{center}
   \epsfysize=6.5cm
    $$\epsffile{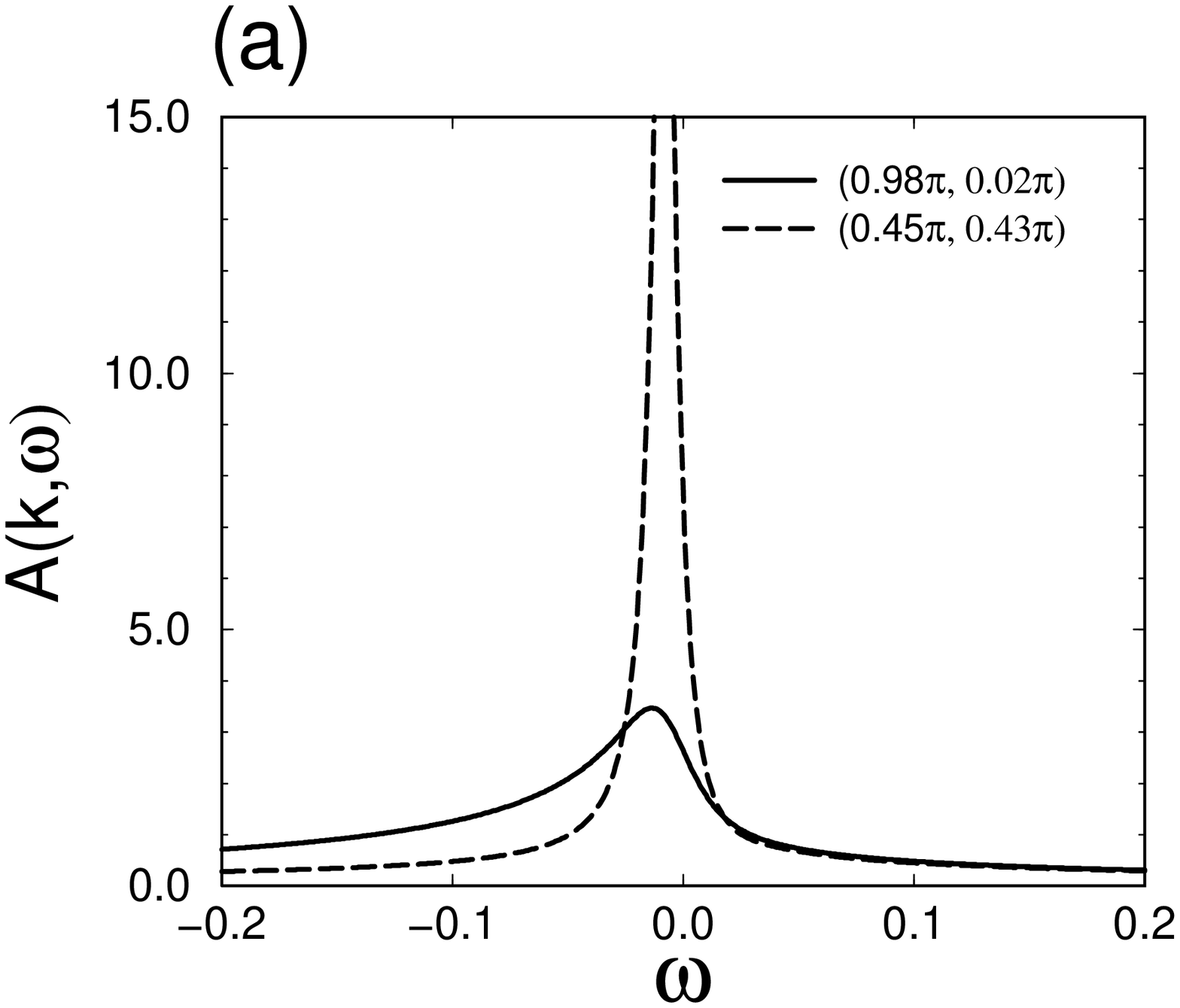}$$
   \epsfysize=6.5cm
    $$\epsffile{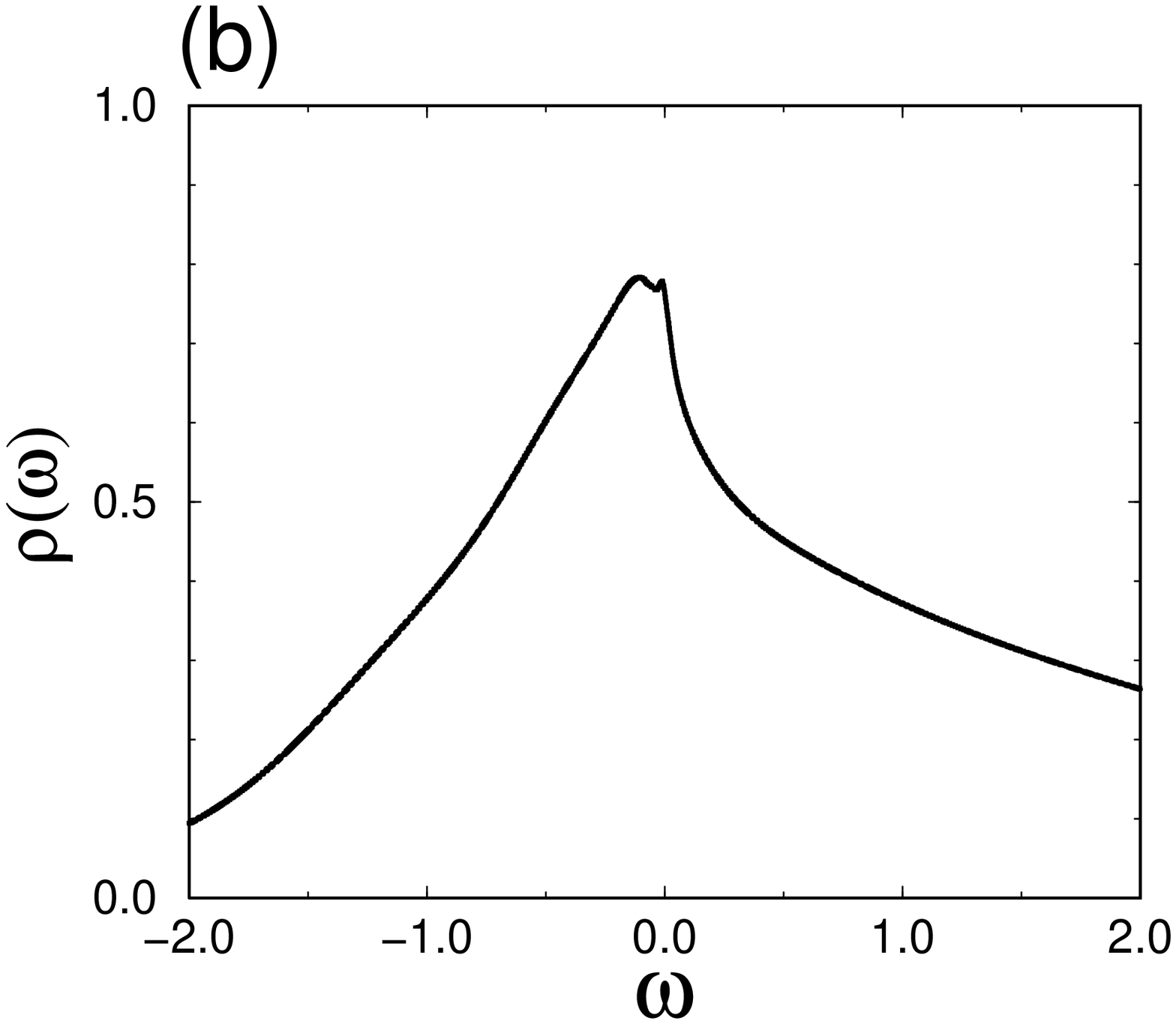}$$
    \caption{(a) The single particle spectral weight and 
             (b) the density of states obtained by the FLEX approximation. 
             The parameters are the same as those in the Fig. 2. 
             }
  \end{center}
\end{figure}

 The obtained results for the superconducting critical temperature 
$T_{{\rm c}}$ are shown in Fig. 4. 
 It is seen that the critical temperature increases with decreasing the 
doping concentration $\delta$, or increasing the repulsive interaction $U$. 
 In other words, the critical temperature increases with the development of 
the anti-ferromagnetic spin fluctuations. 
 The critical temperature tends to be saturated in the large $U$ and 
the small $\delta$ region. It is a common result of the FLEX calculation 
because the spin fluctuations have not only the pairing effects 
but also the de-pairing effects.

  \begin{figure}[htbp]
  \begin{center}
   \epsfysize=6cm
    $$\epsffile{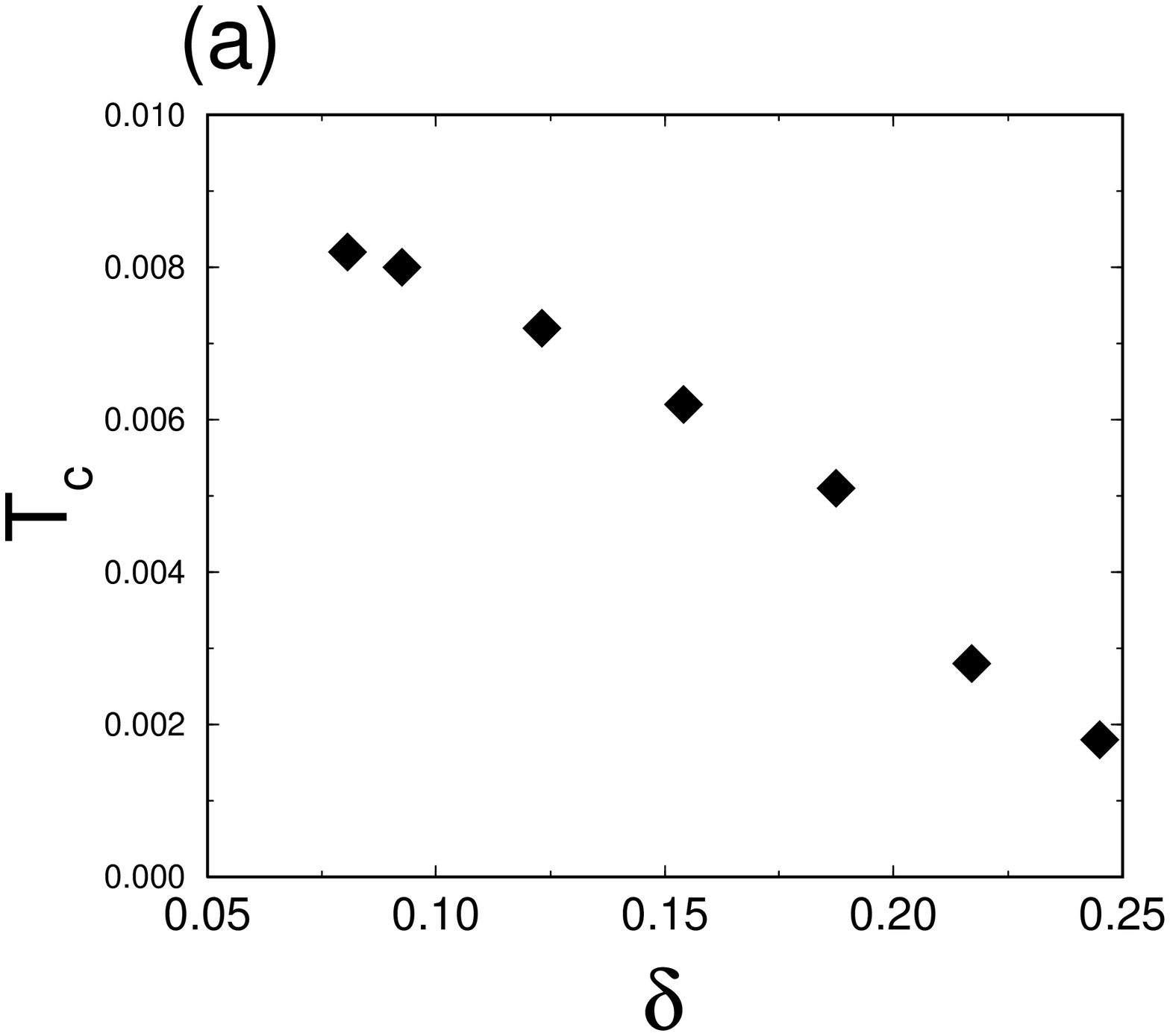}$$
   \epsfysize=6cm
    $$\epsffile{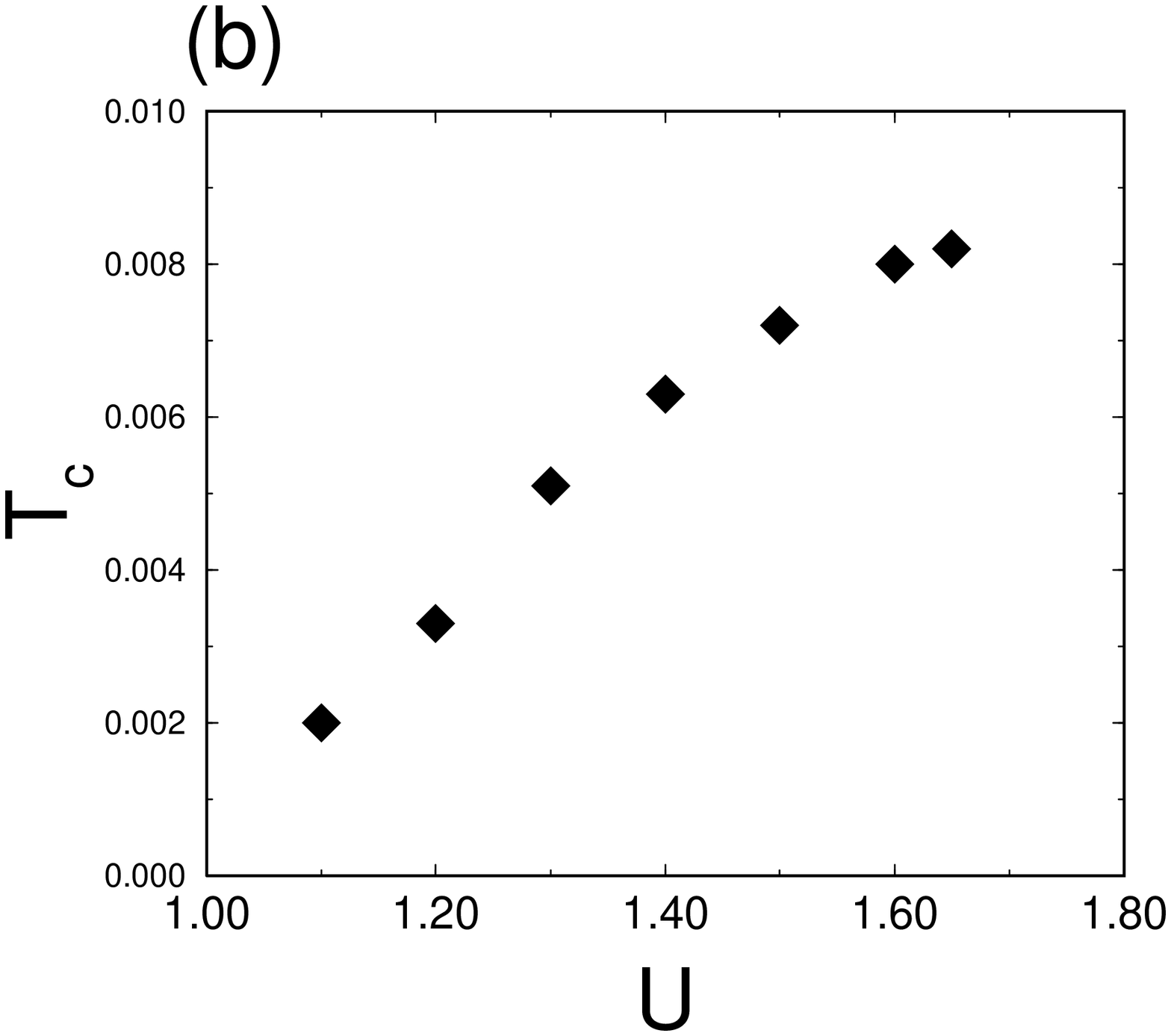}$$
    \caption{The superconducting critical temperature 
             obtained by the FLEX approximation. 
             (a) The doping concentration dependence for $U=1.6$. 
             (b) The repulsive interaction dependence for $\delta=0.09$.
             }
  \end{center}
\end{figure}

\section{FLEX+T-matrix Approximation}

 Hereafter, we consider the superconducting fluctuations and their effects on 
the single particle properties and the magnetic properties. 
 In our scenario, the superconducting fluctuations play a dominant role for 
the pseudogap phenomena. Therefore, we have to describe the superconducting 
fluctuations derived from the anomalous vertex 
$V_{\rm a} (\mbox{\boldmath$q$}, {\rm i} \Omega_{n})$ 
given by the FLEX approximation. 
 Although the calculation has been difficult,~\cite{rf:koikegami} 
we succeed in the description including the momentum and frequency 
dependence of the anti-ferromagnetic spin fluctuations. 
 We explain the method of the calculation bellow. 
 The obtained results well explain the pseudogap phenomena including 
their doping dependence and some characteristic properties. 
 In this section, the coupling constant $U$ is fixed to $U=1.6$ unless 
we specify.

\subsection{Formalism}

 First, we explain the formalism of the FLEX+T-matrix calculation 
which includes the effects of superconducting fluctuations within 
the lowest order. 
 The superconducting fluctuations are generally represented by the 
T-matrix which is the propagator of the superconducting fluctuations. 
The T-matrix is expressed by the ladder diagrams 
in the particle-particle channel (Fig. 5 (a)) and is derived 
from the following Bethe-Salpeter equation. 
\begin{eqnarray}
  T&(&\mbox{\boldmath$k_{1}$}, {\rm i}  \omega_{n}: 
\mbox{\boldmath$k_{2}$}, {\rm i} \omega_{m}: 
\mbox{\boldmath$q$},  {\rm i} \Omega_{n})  = 
  V_{{\rm a}}(\mbox{\boldmath$k_{1}$}-\mbox{\boldmath$k_{2}$}, 
 {\rm i} \omega_{n} - {\rm i} \omega_{m} )  
\nonumber \\
 & - & T \sum_{\mbox{\boldmath$k$}, \omega_{l}} 
  V_{{\rm a}}(\mbox{\boldmath$k_{1}$}-\mbox{\boldmath$k$}, 
 {\rm i} \omega_{n} - {\rm i} \omega_{l}) 
  G(\mbox{\boldmath$k$}, {\rm i} \omega_{l}) 
\nonumber \\
 & \times &
  G(\mbox{\boldmath$q$}-\mbox{\boldmath$k$}, 
    {\rm i} \Omega_{n} - {\rm i} \omega_{l}) 
  T(\mbox{\boldmath$k$}, {\rm i} \omega_{l}: 
\mbox{\boldmath$k_{2}$}, {\rm i} \omega_{m}: 
\mbox{\boldmath$q$},  {\rm i} \Omega_{n}) .
\end{eqnarray}
Here, the pairing interaction is given by the anomalous vertex 
obtained by the FLEX approximation. 
 Generally, it is difficult to solve the above integral equation except for 
the case where the separable pairing interaction 
is assumed.~\cite{rf:yanasePG,rf:jujo,rf:jujoyanase,rf:yanaseSC,rf:yanaseMG}
 Therefore, we carry out the following two approximations 
where the meaningful component as the $d_{x^{2}-y^{2}}$-wave 
superconducting fluctuations is properly taken out. 

\begin{figure}[htbp]
  \begin{center}
   \epsfysize=3cm
    $$\epsffile{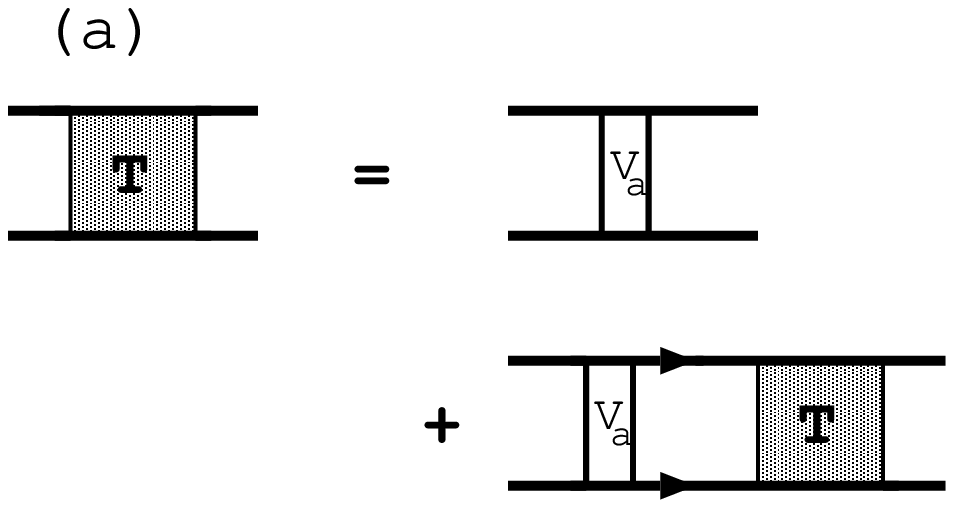}$$
   \epsfysize=2.5cm
    $$\epsffile{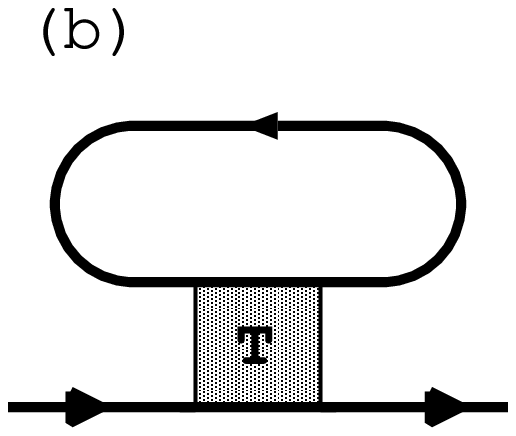}$$
    \caption{(a) The T-matrix. (b) The self-energy due to the superconducting 
             fluctuations.}    
  \end{center}
\end{figure}

 The T-matrix at $\mbox{\boldmath$q$} =  \Omega_{n} = 0$ is approximately 
decomposed into the eigenfunction with its respective eigenvalue 
of the $\acute{{\rm E}}$liashberg equation.~\cite{rf:dahmpseudogap} 
\begin{eqnarray}
  &T&(\mbox{\boldmath$k_{1}$}, {\rm i} \omega_{n}: 
\mbox{\boldmath$k_{2}$}, {\rm i} \omega_{m}: 
\mbox{\boldmath$q$} = {\rm i} \Omega_{n} =0) 
\nonumber \\
& = & \sum_{\alpha} \frac{g_{\alpha} 
                \phi_{\alpha}(\mbox{\boldmath$k_{1}$}, {\rm i} \omega_{n}) 
                \phi_{\alpha}^{*}(\mbox{\boldmath$k_{2}$}, {\rm i} \omega_{m})}
                       {1 - \lambda_{\alpha}} .
\end{eqnarray}

 The eigenvalue $\lambda_{\alpha}$ and the eigenfunction 
$ \phi_{\alpha}(\mbox{\boldmath$k$}, {\rm i} \omega_{n}) $ are derived from 
the $\acute{{\rm E}}$liashberg equation, eq. (2.7). 
 The index ${\alpha}$ denotes each mode included in the T-matrix. 
 Here, we take out the component with the maximum eigenvalue 
$\lambda_{{\rm max}}$ and the corresponding eigenfunction $\phi_{{\rm max}}$ 
which has the $d_{x^{2}-y^{2}}$-wave character. 
 By considering that the superconducting transition is 
determined by the condition $\lambda_{{\rm max}}=1$, 
we can understand that the mode described by $\lambda_{{\rm max}}$ and 
$\phi_{{\rm max}}$ represents 
the $d_{x^{2}-y^{2}}$-wave superconducting fluctuations. 
 The function $\phi_{{\rm max}}$ is the wave function of the fluctuating 
Cooper pairs in the fluctuating regime. 
 Hereafter, we neglect the other modes since they have no significant effect 
on the superconducting fluctuations and the electronic state 
at the low energy. 
 Hereafter, we neglect the index `max' for simplicity.  
 The approximation is justified when the superconducting fluctuations 
are strong. 
 By using the above approximation, the T-matrix is expressed 
by the following equations.~\cite{rf:koikegami} 
\begin{eqnarray}
  & & T(\mbox{\boldmath$k_{1}$}, {\rm i} \omega_{n}: 
\mbox{\boldmath$k_{2}$}, {\rm i} \omega_{m}: 
\mbox{\boldmath$q$}, {\rm i} \Omega_{n}) = 
\nonumber \\
 & & \frac{g(\mbox{\boldmath$q$}, {\rm i} \Omega_{n}) 
  \phi(\mbox{\boldmath$k_{1}$}, {\rm i} \omega_{n}: 
       \mbox{\boldmath$q$}, {\rm i} \Omega_{n}) 
  \phi^{*}(\mbox{\boldmath$k_{2}$}, {\rm i} \omega_{m}: 
       \mbox{\boldmath$q$}, {\rm i} \Omega_{n})} 
  {1 - \lambda(\mbox{\boldmath$q$}, {\rm i} \Omega_{n})}.  
\nonumber \\
\end{eqnarray}
 The function $\lambda(\mbox{\boldmath$q$}, {\rm i} \Omega_{n})$ is given by 
the maximum eigenvalue and 
$\phi(\mbox{\boldmath$k$}, {\rm i} \omega_{n}: \mbox{\boldmath$q$}, {\rm i} \Omega_{n})$ is the corresponding eigenfunction of the following equation, 
\begin{eqnarray}
  &\lambda&(\mbox{\boldmath$q$}, {\rm i} \Omega_{n}) 
  \phi(\mbox{\boldmath$k$}, {\rm i} \omega_{n}: 
        \mbox{\boldmath$q$}, {\rm i} \Omega_{n})  = 
  - T \sum_{\mbox{\boldmath$p$},{\rm i} \omega_{m}} 
   V_{\rm a} (\mbox{\boldmath$k-p$}, {\rm i} \omega_{n} - {\rm i} \omega_{m}) 
\nonumber \\
&\times & G (\mbox{\boldmath$p$}, {\rm i} \omega_{m}) 
   G (\mbox{\boldmath$q-p$}, {\rm i} \Omega_{n} - {\rm i} \omega_{m}) 
   \phi (\mbox{\boldmath$p$}, {\rm i} \omega_{m}: 
         \mbox{\boldmath$q$}, {\rm i} \Omega_{n}). 
\end{eqnarray}

 The function $1 - \lambda(\mbox{\boldmath$q$}, {\rm i} \Omega_{n}) $ 
describes the dispersion relation of the fluctuating Cooper pairs. 
 In the above expression, the wave function is normalized as, 
$\sum_{\mbox{\boldmath$k$},{\rm i} \omega_{n}} 
|\phi(\mbox{\boldmath$k$}, {\rm i} \omega_{n}: 
       \mbox{\boldmath$q$}, {\rm i} \Omega_{n})|^{2} = 1$, and 
the coupling constant is given as, 
$ g(\mbox{\boldmath$q$}, {\rm i} \Omega_{n}) = 
\sum_{\mbox{\boldmath$k_{1}$},{\rm i} \omega_{n}}
\sum_{\mbox{\boldmath$k_{2}$},{\rm i} \omega_{m}}
\phi^{*}(\mbox{\boldmath$k_{1}$}, {\rm i} \omega_{n}: 
       \mbox{\boldmath$q$}, {\rm i} \Omega_{n})
V_{\rm a} (\mbox{\boldmath$k_{1}-k_{2}$}, 
{\rm i} \omega_{n} - {\rm i} \omega_{m}) 
\phi(\mbox{\boldmath$k_{2}$}, {\rm i} \omega_{m}: 
       \mbox{\boldmath$q$}, {\rm i} \Omega_{n}) $.

 Here, we perform one more approximation, which is valid around 
$q=\Omega_{n}=0$. 
 We neglect the $q$- and $\Omega_{n}$ dependence of the eigenfunction 
$\phi (\mbox{\boldmath$k$}, {\rm i} \omega_{n}: \mbox{\boldmath$q$}, {\rm i} \Omega_{n})$. 
 The wave function is fixed to that at $q=\Omega_{n}=0$.  
 This approximation is well justified 
in the following way. 
 The main $q$- and $\Omega_{n}$ dependence of the eigenvalue 
$ \lambda(\mbox{\boldmath$q$}, {\rm i} \Omega_{n}) $ is derived from 
the differential of the Green function 
$G (\mbox{\boldmath$q-p$}, {\rm i} \Omega_{n} - {\rm i} \omega_{m}) $
in eq. (3.4). In other words, the dispersion of the fluctuating Cooper pairs, 
$1 - \lambda(\mbox{\boldmath$q$}, {\rm i} \Omega_{n}) $ is insensitive to 
the change of the eigenfunction. 
 It should be noticed that the wave function 
$\phi (\mbox{\boldmath$k$}, {\rm i} \omega_{n}: \mbox{\boldmath$q$}, {\rm i} \Omega_{n})$ is independent of $q$ and $\Omega_{n}$ when the separable paring 
interaction is considered.~\cite{rf:yanasePG}  

 We have explicitly solved the eigenvalue equation (eq. (3.4)).  
The results have confirmed that the $q$- and $\Omega_{n}$ dependence of 
the eigenfunction $\phi (\mbox{\boldmath$k$}, {\rm i} \omega_{n}: \mbox{\boldmath$q$}, {\rm i} \Omega_{n})$ is small in the region where the dominant 
contribution to the self-energy is yielded. 
 In particular, it is confirmed that the value 
$1 - \lambda(\mbox{\boldmath$q$}, {\rm i} \Omega_{n}) $ 
is found within the error of $10$-$20$ par cent by this approximation. 
 Thus, the second approximation only slightly underestimates the eigenvalue 
$ \lambda(\mbox{\boldmath$q$}, {\rm i} \Omega_{n}) $ around $q=\Omega_{n}=0$, 
since the precise eigenfunction is determined so as to optimize the 
eigenvalue. 
 Needless to say, the precise eigenvalue is obtained at $q=\Omega_{n}=0$. 
 In other wards, the TDGL parameter (see eq. (3.10)) $b$ is overestimated 
by the second approximation whereas the mass term $t_{0}$ is 
obtained precisely. 
 The approximation slightly underestimates the effects of the superconducting 
fluctuations. 
 Since the approximation is not precise for large $q$ and $\Omega_{n}$, 
the unphysical contribution to the self-energy is caused by the region. 
 Therefore, we eliminate the $q$- and $\Omega_{n}$ independent term 
$ g \phi(\mbox{\boldmath$k_{1}$}, {\rm i} \omega_{n}) 
\phi^{*}(\mbox{\boldmath$k_{2}$}, {\rm i} \omega_{m})$ in the T-matrix.

 By using the above two approximations, the T-matrix is expressed as follows, 
\begin{eqnarray}
  &T&(\mbox{\boldmath$k_{1}$}, {\rm i} \omega_{n}: 
\mbox{\boldmath$k_{2}$}, {\rm i} \omega_{m}: 
\mbox{\boldmath$q$}, {\rm i} \Omega_{n}) 
\nonumber \\
  & = & \frac{ g \lambda(\mbox{\boldmath$q$}, {\rm i} \Omega_{n})
  \phi(\mbox{\boldmath$k_{1}$}, {\rm i} \omega_{n}) 
  \phi^{*}(\mbox{\boldmath$k_{2}$}, {\rm i} \omega_{m})} 
  {1 - \lambda(\mbox{\boldmath$q$}, {\rm i} \Omega_{n})}, 
\end{eqnarray}
where 
\begin{eqnarray}
  \lambda(\mbox{\boldmath$q$}, {\rm i} \Omega_{n}) & = & 
  - T \sum_{\mbox{\boldmath$k$},{\rm i} \omega_{n}} 
      \sum_{\mbox{\boldmath$p$},{\rm i} \omega_{m}} 
  \phi^{*} (\mbox{\boldmath$k$}, {\rm i} \omega_{n})
   V_{\rm a} (\mbox{\boldmath$k-p$}, {\rm i} \omega_{n} - {\rm i} \omega_{m}) 
\nonumber \\
& & G (\mbox{\boldmath$p$}, {\rm i} \omega_{m}) 
   G (\mbox{\boldmath$q-p$}, {\rm i} \Omega_{n} - {\rm i} \omega_{m}) 
   \phi (\mbox{\boldmath$p$}, {\rm i} \omega_{m}). 
\end{eqnarray}

 Here, the coupling constant $g$ is defined as, 
\begin{eqnarray}
   g &=& 
\sum_{\mbox{\boldmath$k_{1}$},{\rm i} \omega_{n}}
\sum_{\mbox{\boldmath$k_{2}$},{\rm i} \omega_{m}}
\phi^{*}(\mbox{\boldmath$k_{1}$}, {\rm i} \omega_{n})
\nonumber \\
&\times& 
V_{\rm a} (\mbox{\boldmath$k_{1}-k_{2}$}, 
{\rm i} \omega_{n} - {\rm i} \omega_{m}) 
\phi(\mbox{\boldmath$k_{2}$}, {\rm i} \omega_{m}), 
\end{eqnarray}
and the wave function is normalized as, 
\begin{eqnarray}
  \sum_{\mbox{\boldmath$k$},{\rm i} \omega_{n}} 
|\phi(\mbox{\boldmath$k$}, {\rm i} \omega_{n})|^{2} = 1. 
\end{eqnarray}

 By taking out the freedom of the superconducting fluctuations, 
we define the pair susceptibility as 
$t(\mbox{\boldmath$q$}, {\rm i} \Omega_{n})  =  
g/(1 - \lambda(\mbox{\boldmath$q$}, {\rm i} \Omega_{n}))$ 
which we have termed the T-matrix in the previous 
papers.~\cite{rf:yanasePG,rf:yanaseMG,rf:yanaseSC,rf:jujoyanase,rf:jujo}

 The self-energy due to the superconducting fluctuations is  
given by the T-matrix approximation (Fig. 5(b)). 
\begin{eqnarray} 
 {\mit{\it \Sigma}}_{{\rm S}} (\mbox{\boldmath$k$}, {\rm i} \omega_{n}) =  
 &T& \sum_{\mbox{\boldmath$q$},{\rm i} \Omega_{n}} 
  T(\mbox{\boldmath$k$}, {\rm i} \omega_{n}: 
\mbox{\boldmath$k$}, {\rm i} \omega_{n}: 
\mbox{\boldmath$q$}, {\rm i} \Omega_{n}) 
\nonumber \\
&\times&
 G (\mbox{\boldmath$q-k$}, {\rm i} \Omega_{n} - {\rm i} \omega_{n}). 
\end{eqnarray} 

 In this section, we use the Green function determined by the FLEX 
approximation $ G_{{\rm F}} (\mbox{\boldmath$k$}, {\rm i} \omega_{n}) = 
( {\rm i} \omega_{n} - \varepsilon_{\mbox{{\scriptsize \boldmath$k$}}} - 
{\mit{\it \Sigma}}_{{\rm F}} (\mbox{\boldmath$k$}, {\rm i} \omega_{n}))^{-1}$ 
in calculating eqs. (3.5)-(3.9). 
 That is to say, we calculate the lowest order correction due to the 
superconducting fluctuations on the FLEX approximation. 
 We call the calculation FLEX+T-matrix approximation. 
 The self-energy is obtained by the summation, 
${\mit{\it \Sigma}} (\mbox{\boldmath$k$}, {\rm i} \omega_{n}) = 
{\mit{\it \Sigma}}_{{\rm F}} (\mbox{\boldmath$k$}, {\rm i} \omega_{n}) + 
{\mit{\it \Sigma}}_{{\rm S}} (\mbox{\boldmath$k$}, {\rm i} \omega_{n})$. 
 We show the results of the FLEX+T-matrix approximation in the following 
subsections. 
 Since the calculation is carried out with the fixed chemical potential $\mu$, 
the doping concentration $\delta$ decreases with decreasing the temperature 
$T$. However, the difference is smaller than $0.01$ and has no significant 
effect.

 In \S4, we carry out the self-consistent calculation in which the fully 
dressed Green function 
$ G (\mbox{\boldmath$k$}, {\rm i} \omega_{n}) = 
( {\rm i} \omega_{n} - \varepsilon_{\mbox{{\scriptsize \boldmath$k$}}} - 
{\mit{\it \Sigma}}_{{\rm F}} (\mbox{\boldmath$k$}, {\rm i} \omega_{n}) - 
{\mit{\it \Sigma}}_{{\rm S}} (\mbox{\boldmath$k$}, {\rm i} \omega_{n}))^{-1}$ 
is used everywhere. As a result of the self-consistency, the effect of the 
superconducting fluctuations are reduced. However, the qualitatively similar 
conclusions are obtained.

\subsection{Pseudogap in the single particle properties}

 In this subsection the pseudogap, which is the main subject of this paper, 
is derived. The calculated results well justify the pairing scenario 
based on the self-energy correction due to the superconducting 
fluctuations.~\cite{rf:yanasePG,rf:janko} 
 On the basis of the model with a $d$-wave attractive interaction, 
it has been shown that the anomalous properties of the self-energy 
give rise to the pseudogap.~\cite{rf:yanasePG,rf:janko}  
 The characteristics of the self-energy is that 
the real part has the positive slope and the imaginary part has 
the maximum absolute value at the Fermi level 
(see Fig. 8 in ref. 19). 
 They are anomalous compared with the typical behaviors in the 
Fermi liquid theory (see Fig. 2). The large imaginary part reduces 
the single particle spectral weight at the Fermi level and gives rise to
pseudogap. 
 The anomalous features seem to compete with the Fermi liquid behaviors 
obtained by the FLEX calculations. In particular, the negative slope of the 
real part which is related to the renormalization factor 
$Z_{\mbox{{\scriptsize \boldmath$k$}}}^{-1}$ is generally 
increased by the strong electron correlation. 
 However, the calculated results clarify the important mechanism of the 
pseudogap formation in the strongly correlated electron systems.

\begin{figure}[htbp]
\begin{center}
   \epsfysize=6.5cm
    $$\epsffile{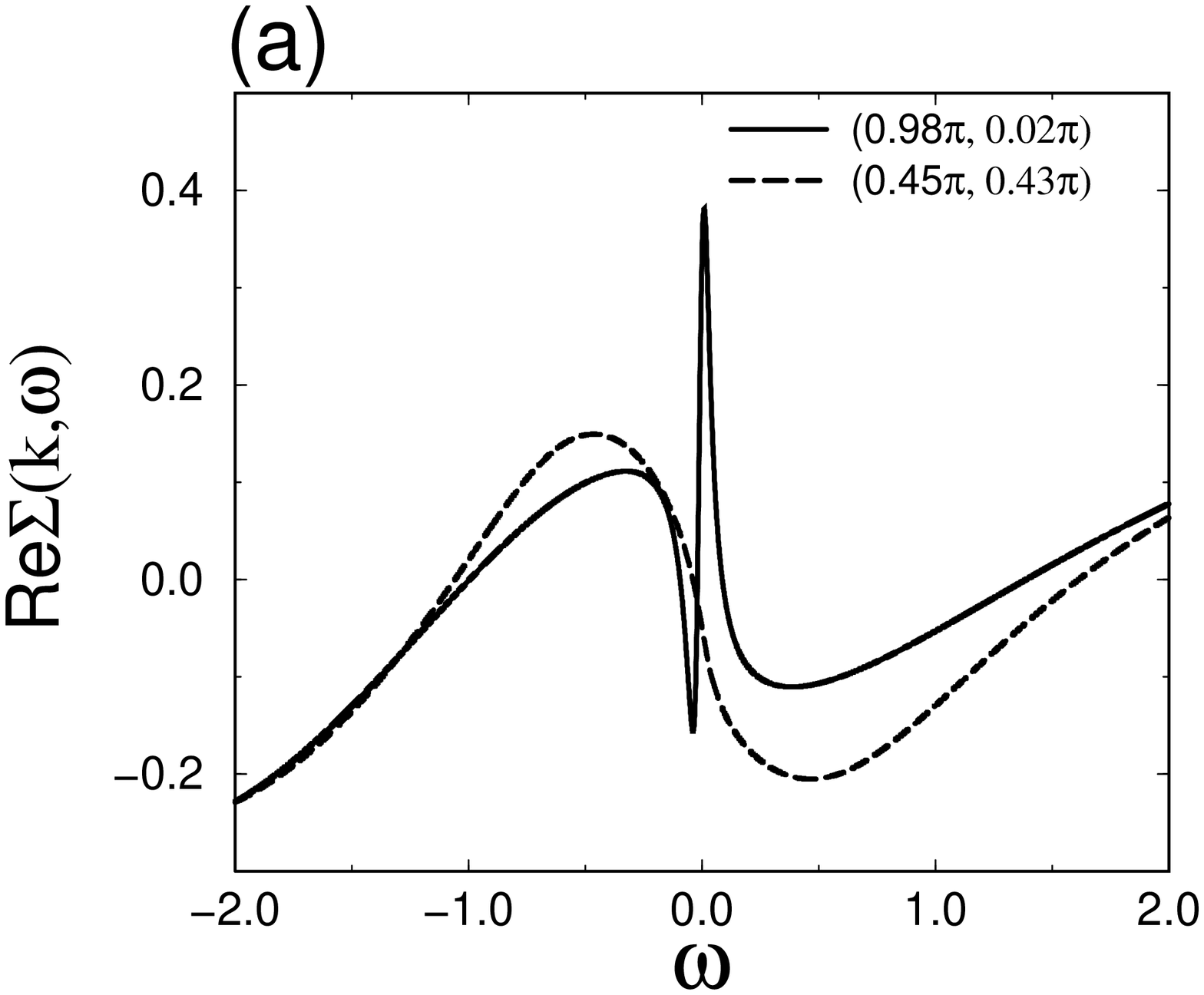}$$
   \epsfysize=6.5cm
    $$\epsffile{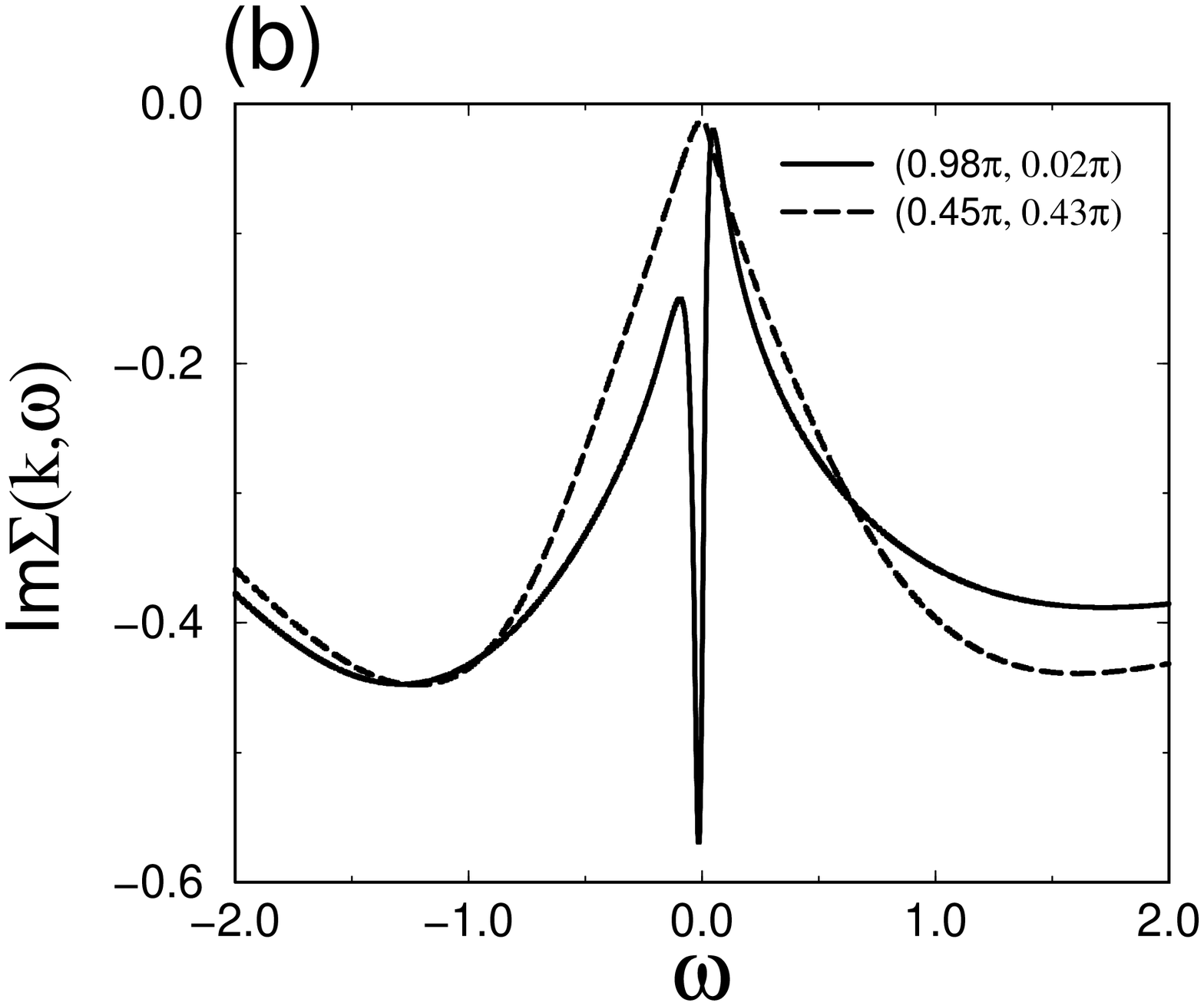}$$
    \caption{The self-energy obtained by the FLEX+T-matrix approximation. 
             (a)The real part. (b) The imaginary part. 
             Here, $U=1.6$, $\delta=0.095$ and $T=0.010$. 
             The solid line and the long-dashed line correspond to 
             $(\frac{63}{64}\pi,\frac{1}{64}\pi)$ (hot spot) and 
             $(\frac{29}{64}\pi,\frac{27}{64}\pi)$ (cold spot), respectively.
            }
  \end{center}
\end{figure}

 We show the obtained results for the analytically continuated self-energy 
${\mit{\it \Sigma}}^{{\rm R}}(\mbox{\boldmath$k$}, \omega)$ 
in Fig. 6. 
Here, the doping concentration corresponds to the under-doped cuprates. 
 It should be noticed that the Fermi liquid behaviors are seen when 
we look with the large energy scale $\omega \sim 0.5$. 
 However, the anomalous behaviors leading to the pseudogap are clearly seen 
in much smaller energy scale $\omega \sim 0.05$. 
 The anomalous behaviors vanish around $(\pi/2, \pi/2)$ because of the 
$d_{x^{2}-y^{2}}$ wave symmetry of the fluctuating Cooper pairs. 

 It is the important point that the superconductivity and the pseudogap 
take place in the renormalized quasi-particles near the Fermi surface. 
 In other words, the pseudogap occurs with much smaller energy scale 
than that of the electron systems. 
 The above results clearly show the smaller energy scale and justify the 
calculation based on the effective model in which a $d$-wave attractive 
interaction acts on the renormalized quasi-particles.

 The results for the spectral weight are shown in Fig. 7. 
The pseudogap clearly appears in the single particle spectral weight 
(Fig. 7(a)) and the DOS (Fig. 7(b)). 
 The $d_{x^{2}-y^{2}}$-wave form of the pseudogap is naturally obtained by the 
$d_{x^{2}-y^{2}}$-wave symmetry of the superconducting fluctuations 
(Fig. 7(a)).  
 The inset of Fig. 7(b) shows the much smaller energy scale of the pseudogap 
compared with the band width. 
 We wish to note again that the pseudogap is derived from the self-energy 
correction due to the superconducting fluctuations which are enhanced by the 
strong coupling superconductivity and 
the quasi-two dimensionality.~\cite{rf:yanasePG,rf:janko,rf:jujo} 
 The $\omega$-dependence of the wave function furthermore advances 
the pseudogap formation. 
 Thus, the resonance scattering scenario is justified 
by the microscopic calculation based on the Hubbard model. 
 It is microscopically confirmed that the superconducting fluctuations 
are sufficiently strong to give rise to the pseudogap.

 The energy scale of the obtained pseudogap is consistent with 
that of the superconducting gap.  
 The experimental results show that the energy scale of the pseudogap is 
nearly the same as that of the superconducting gap.~\cite{rf:renner,rf:ARPES}  
 The ratio $2 \Delta_{{\rm s}}/T_{{\rm c}} \sim 12 $ has been obtained 
by the FLEX approximation in the optimally-doped region.~\cite{rf:takimoto} 
 Here, $\Delta_{{\rm s}}$ is the maximum value of the gap function in the 
ordered state. 
 The larger ratio is expected in the under-doped region. 
 Our results for the under-doped cuprates show the ratio 
$2 \Delta_{{\rm pg}}/T_{{\rm c}} \sim 20 $ (Fig. 7(a)). 
 Here $\Delta_{{\rm pg}}$ is the energy scale of the pseudogap. 
 Thus, the pseudogap obtained by our calculation seems to have the relevant 
energy scale compared with the superconducting gap.

\begin{figure}[htbp]
\begin{center}
   \epsfysize=6.5cm
    $$\epsffile{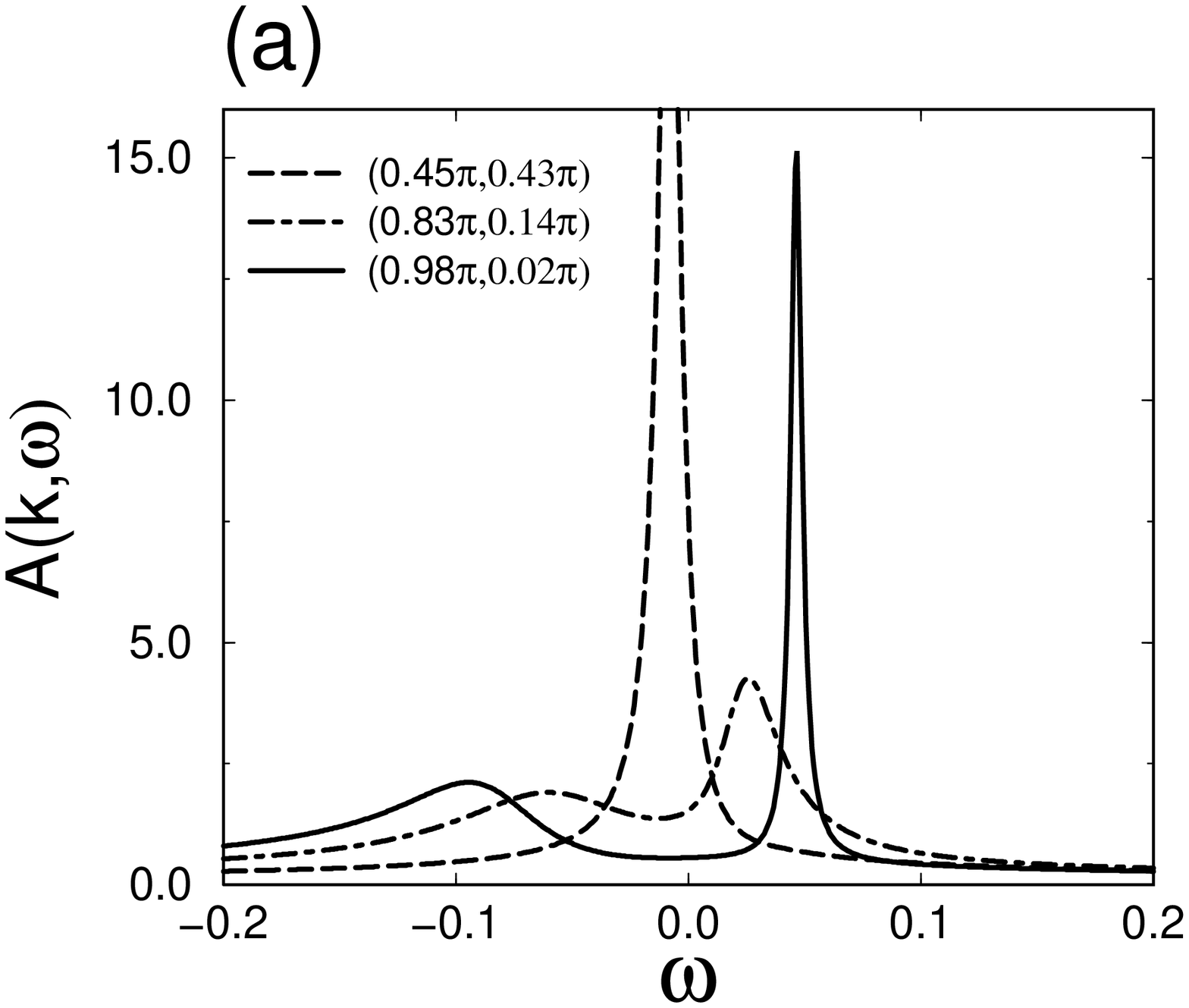}$$
   \epsfysize=6.5cm
    $$\epsffile{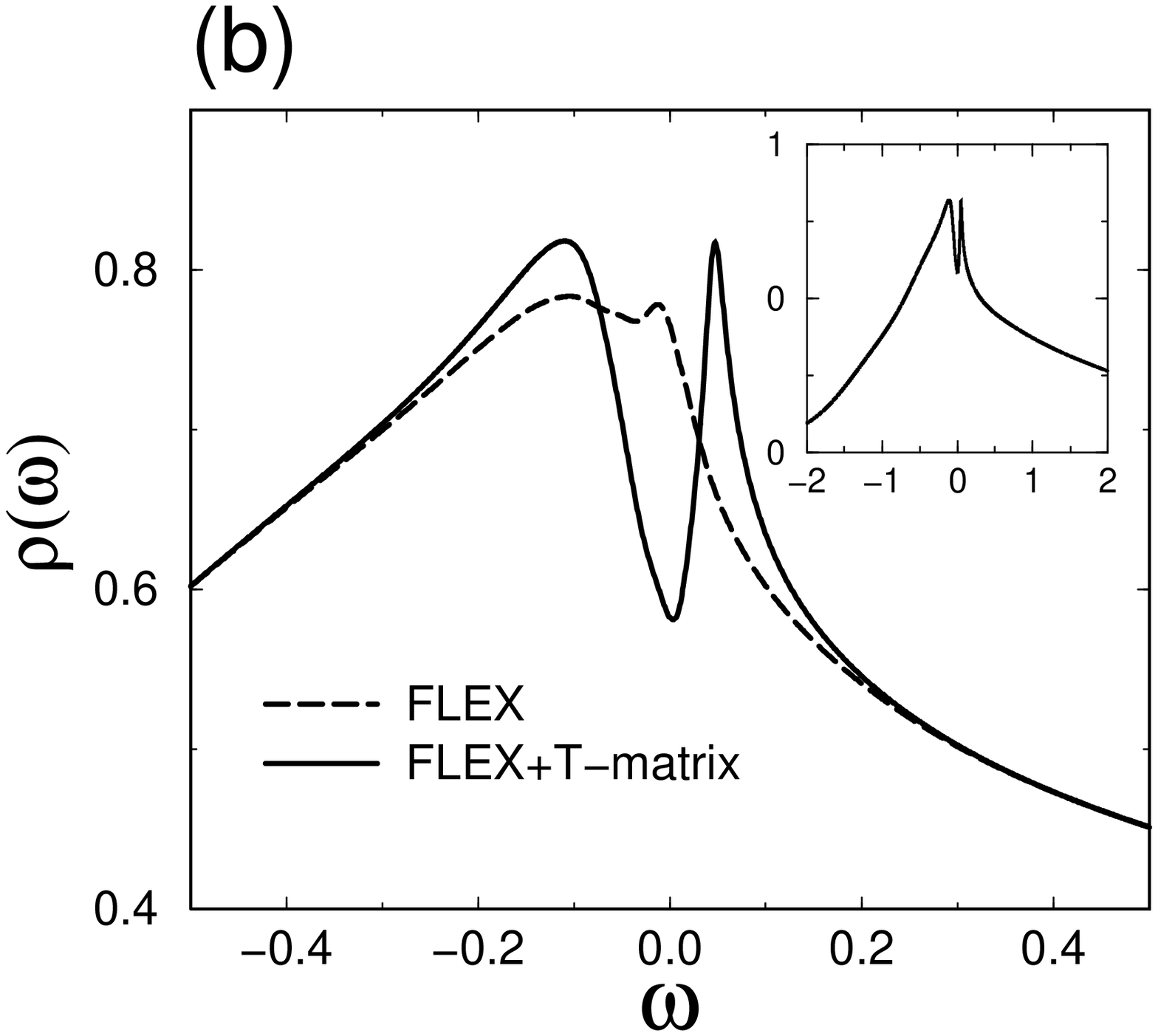}$$
    \caption{(a)The single particle spectral weight obtained 
             by the FLEX+T-matrix approximation. 
             The solid and long-dashed lines correspond to those in Fig. 6, 
             respectively. The dash-dotted line corresponds to 
             $(\frac{53}{64}\pi,\frac{9}{64}\pi)$. 
             (b) The DOS obtained 
             by the FLEX (long-dashed line) and the FLEX+T-matrix 
             (solid line) approximations. 
             The inset is the same result shown in the large energy scale. 
             The other parameters are the same as those in Fig. 6. 
             }
  \end{center}
\end{figure}

 We show the detailed results for the DOS in Fig. 8. 
 We choose the temperature $T = 1.25 T_{{\rm c}}$ in all figures of Fig. 8 
in order to fix the distance to the critical point. 
 The doping dependence is shown in Fig. 8(a). 
 It is shown that the pseudogap becomes weak with increasing the hole-doping. 
 The gap structure is filled up and the DOS near the Fermi level increases 
in the optimally-doped region.  
 The effects of the superconducting fluctuations almost disappear in the 
over-doped region $\delta > 0.2$. 
 This is because both the critical temperature and 
the renormalization of the quasi-particles are reduced by the hole-doping. 
 Since the ratio $T_{{\rm c}}/\varepsilon_{{\rm F}}$ decreases, 
the superconducting coupling also decreases. 
 Therefore, the effects of the superconducting fluctuations are reduced. 
 Thus, the above calculation well 
explains the doping dependence of the pseudogap phenomena.

\begin{figure}[htbp]
\begin{center}
   \epsfysize=6.5cm
    $$\epsffile{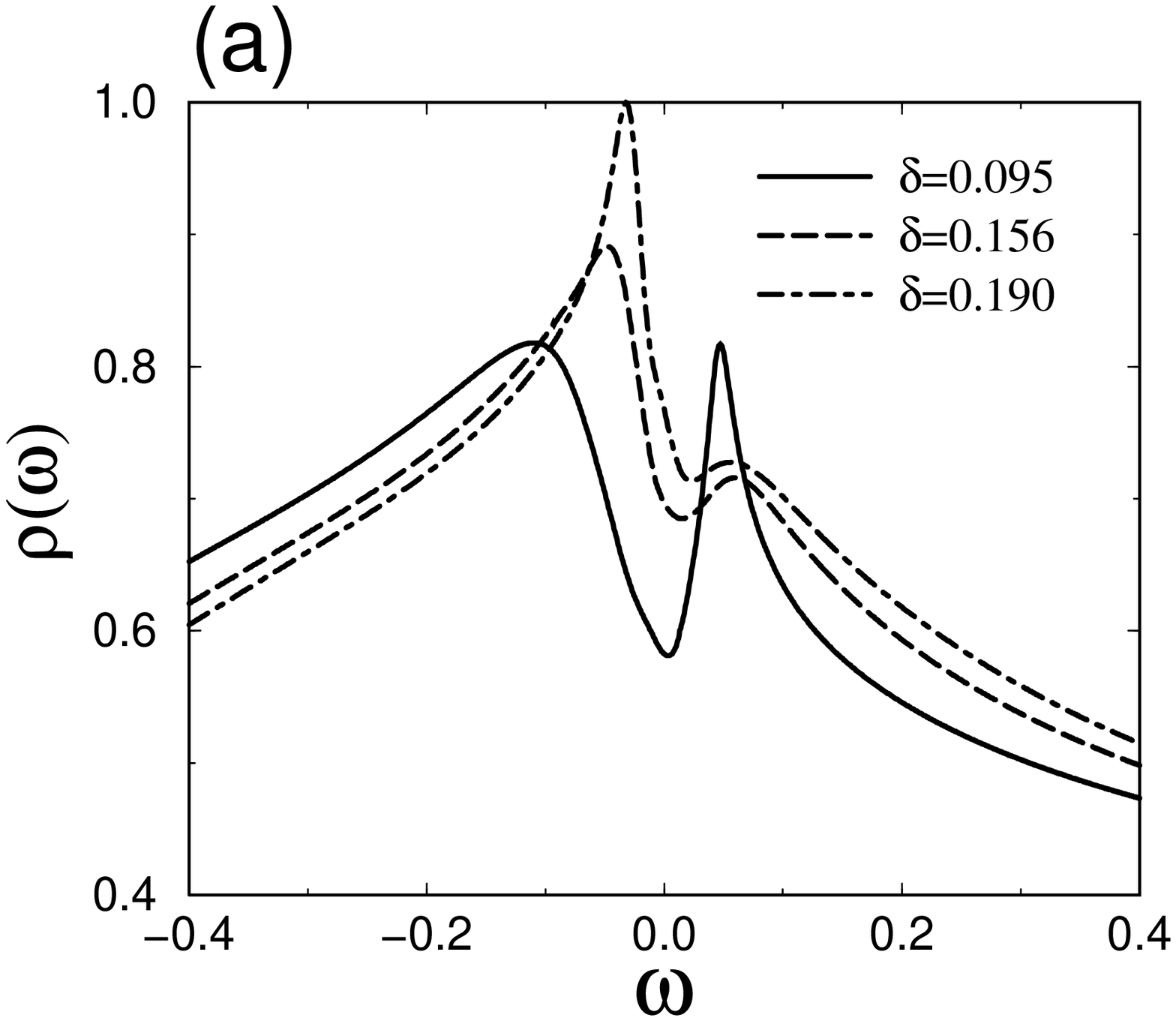}$$
   \epsfysize=6.5cm
    $$\epsffile{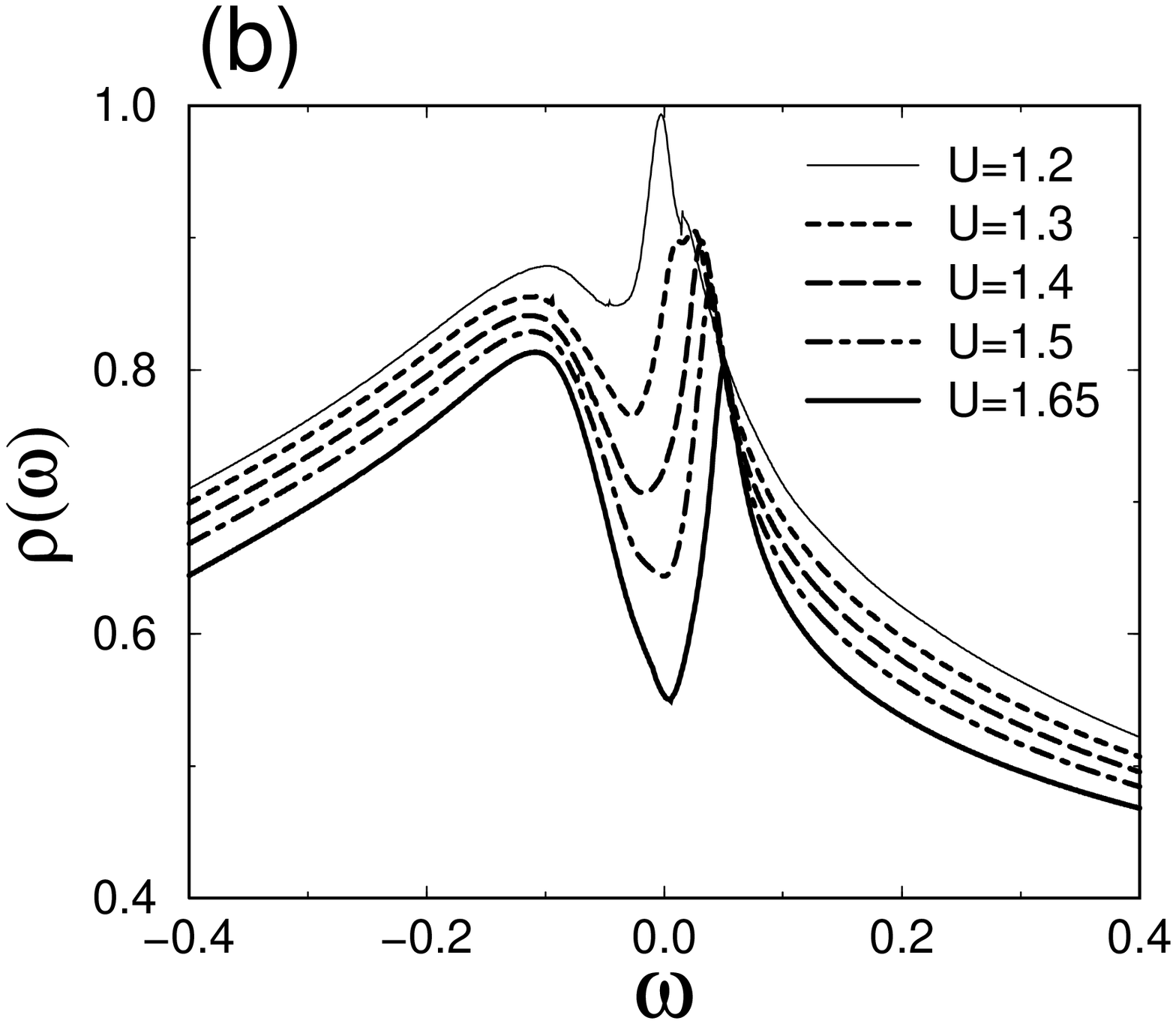}$$
    \caption{(a) The doping dependence of the DOS obtained by 
             the FLEX+T-matrix approximation. 
             The solid, long-dashed and dash-dotted lines correspond to 
             $\delta=0.095$, $0.156$ and $0.190$, respectively. 
             (b) The density of state for the various strength of $U$. 
             The thin solid, dashed, long-dashed, dash-dotted and thick solid 
             lines correspond to $U=1.2$, $1.3$, $1.4$, $1.5$ and 
             $1.65$, respectively. 
             In all figures, the temperature is chosen as 
             $T = 1.25 T_{{\rm c}}$. 
             }
  \end{center}
\end{figure}

 Here, the importance of the strong correlation is shown in Fig. 8(b). 
 It is shown that the pseudogap becomes remarkable by increasing $U$. 
 The reason is the same as that in the explanation of the doping 
dependence (Fig. 8(a)). 
 Thus, the strong superconducting fluctuations and the resultant pseudogap are 
the characteristics of the strongly correlated electron systems.

\subsection{Superconducting fluctuations}

 In this subsection, we discuss the character of the superconducting 
fluctuations 
in order to emphasize the importance of the strong coupling superconductivity.

 In our pervious papers,~\cite{rf:yanasePG,rf:yanaseMG} 
the TDGL expansion is used for the pair susceptibility in the following way. 
\begin{eqnarray}
  \label{eq:tdgl}
   t(\mbox{\boldmath$q$},\omega) = \frac{g}
   {t_{0} + b \mbox{\boldmath$q$}^{2} - (a_{1}+{\rm i}a_{2}) \Omega}, 
\end{eqnarray} 
where the factors arising from the wave function 
$\phi(\mbox{\boldmath$k$}, {\rm i} \omega_{n})$ are neglected.  
 In the notation of this paper, the TDGL expansion corresponds to 
the expansion for the eigenvalue function 
$\lambda(\mbox{\boldmath$q$}, \Omega)$ as  
$1 - \lambda(\mbox{\boldmath$q$}, \Omega) = 
t_{0} + b \mbox{\boldmath$q$}^{2} - (a_{1}+{\rm i}a_{2}) \Omega$.

 The TDGL expansion parameters generally describe the character of the 
superconducting fluctuations. 
 The detailed properties of the TDGL parameters are discussed in ref. 19. 
 Here, we discuss the TDGL parameters on the basis of 
the calculated results in this paper.

 The parameter 
$ t_{0} = 1 - \lambda(\mbox{\boldmath$0$},0) $ represents the distance to 
the phase transition, and is sufficiently small near $T_{{\rm c}}$. 
 The parameter $a_{2}$ expresses the time scale of the fluctuations. 
 The parameter $a_{1}$ is usually neglected within the weak coupling 
theory because it is higher order than $a_{2}$ with respect to 
the superconducting coupling $T_{{\rm c}}/\varepsilon_{{\rm F}}$.  
 However, the parameter $a_{1}$ should not be neglected  
in the strong coupling case.~\cite{rf:yanasePG} 
 The calculated results show that the sign of $a_{1}$ is 
negative in the under-doped region. 
 The sign of $a_{1}$ which is fixed by the particle-hole asymmetry 
determines the sign of the Hall conductivity arising from the 
superconducting fluctuations.~\cite{rf:aronov,rf:ebisawa} 
The Hall anomaly~\cite{rf:aronov} has been explained 
by assuming the existence of the electron-like pre-formed 
pairs.~\cite{rf:geshkenbein} 
 However, it has been pointed out that the explanation is not necessarily 
justified within the theory of the strong coupling 
superconductivity.~\cite{rf:yanasePG}

 The parameter $ b $ represents the dispersion relation of the fluctuations 
and is related to the superconducting coherence length $\xi_{0}$ as 
$b \propto \xi_{0}^{2}$. 
 The parameter $b$ is an important factor determining the strength of the 
fluctuations. 
 The small $ b $ generally means the strong superconducting fluctuations. 
 Therefore, we show the results of the TDGL parameter $b$ in Fig. 9.

\begin{figure}[htbp]
\begin{center}
   \epsfysize=6.5cm
    $$\epsffile{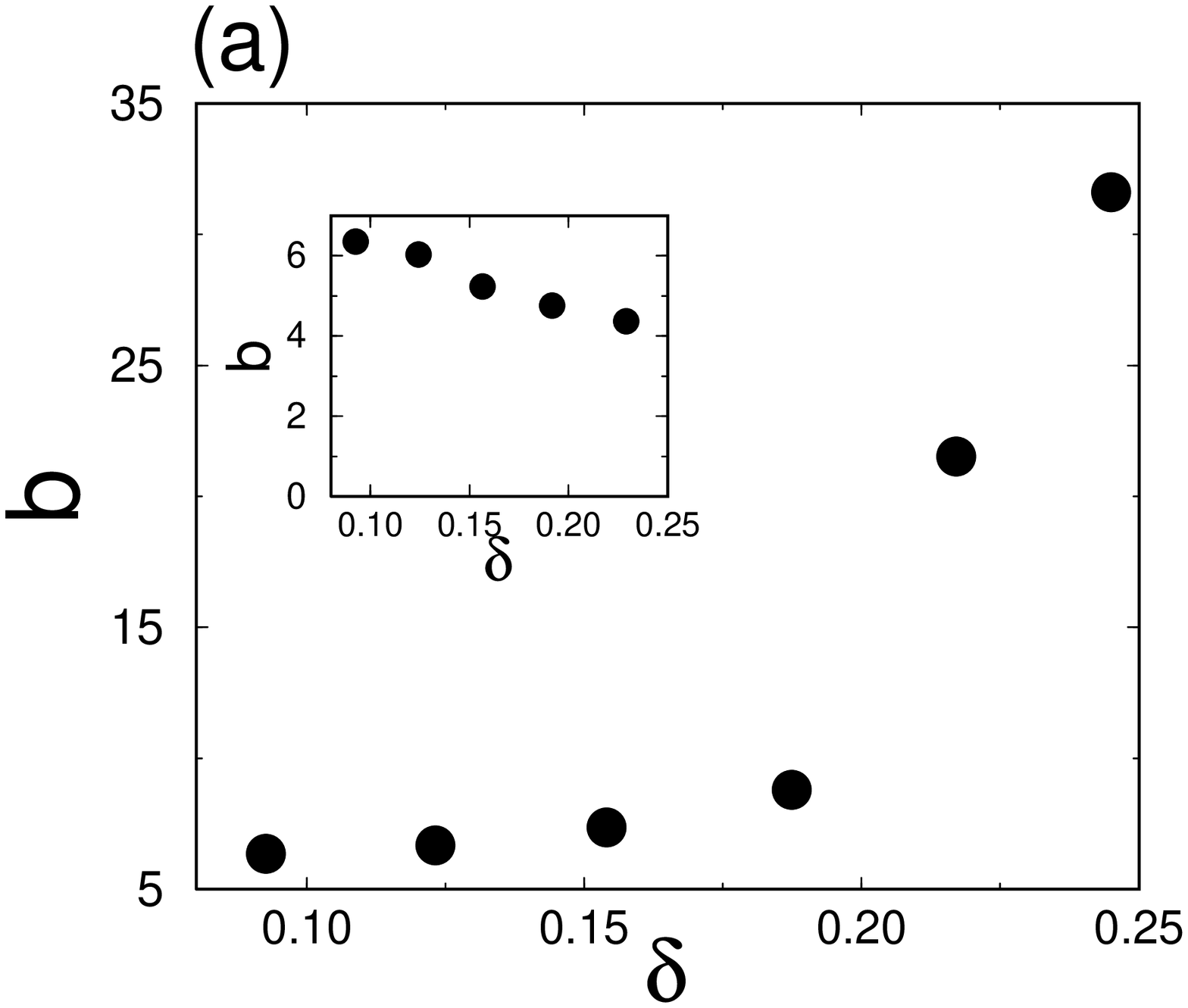}$$
   \epsfysize=6.5cm
    $$\epsffile{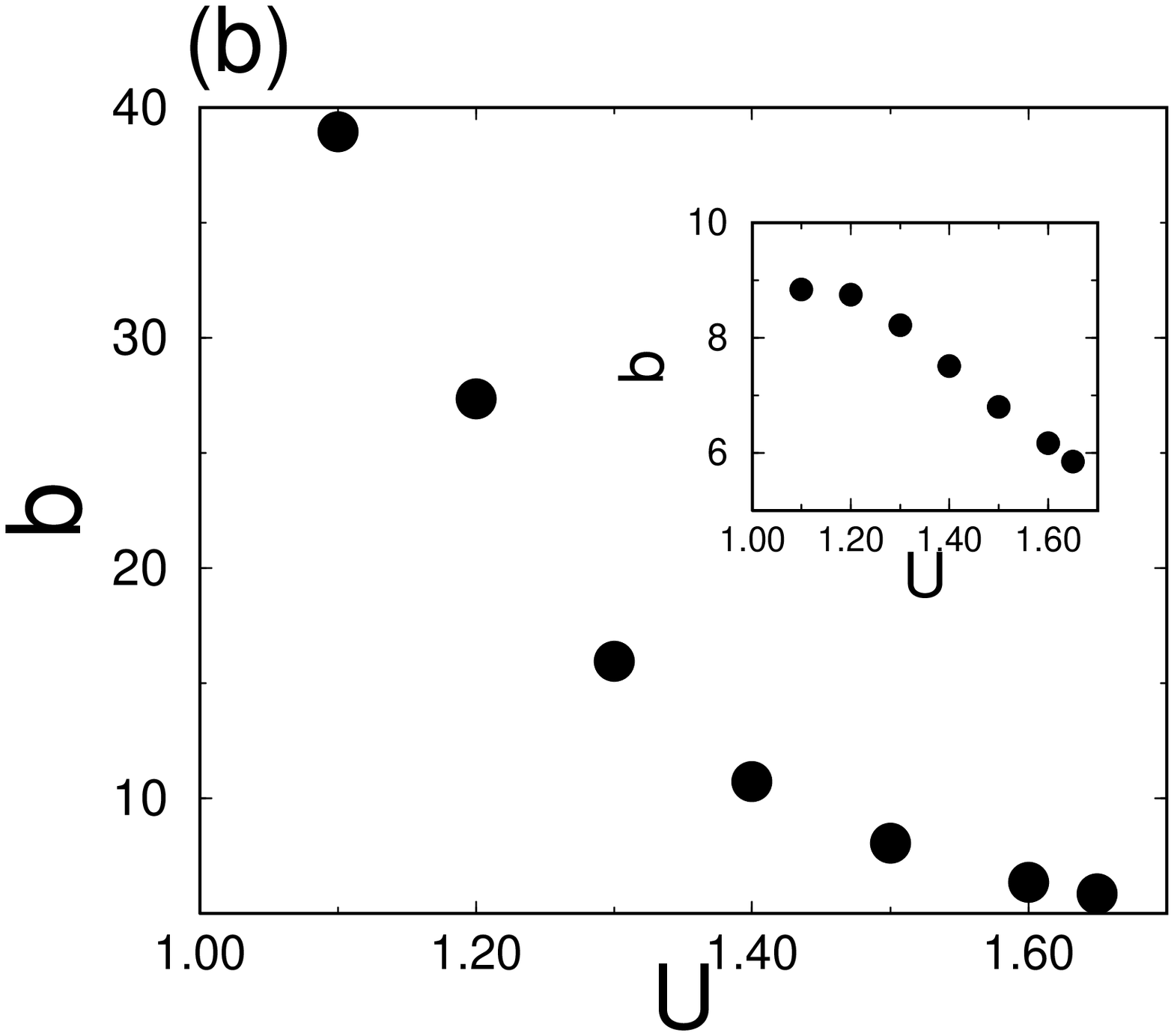}$$
    \caption{(a) The doping dependence of the TDGL parameter $b$ at the 
             critical point $T=T_{{\rm c}}$. 
             Here, the coupling constant is fixed to $U=1.6$. 
             The inset shows the same quantity at the fixed temperature 
             $T=0.0080$
             (b) The TDGL parameter $b$ for the various strength of $U$ at 
             $T=T_{{\rm c}}$. Here the doping concentration is fixed to 
             $\delta=0.09$. 
             The inset shows the same quantity at the fixed temperature 
             $T=0.0082$
             }
  \end{center}
\end{figure}

 The TDGL parameter $b$ is expressed by the Fermi liquid description within 
the weak coupling theory as $b = g_{{\rm d}} \tilde{\rho}_{{\rm d}}(0) 
\frac{7 \zeta(3)}{32 \pi^{2}}  \frac{\bar{v}_{{\rm F}}^{2}}{T^{2}}$. 
 Here, $g_{{\rm d}}$ is the $d$-wave component of the residual interaction 
$Z_{\mbox{{\scriptsize \boldmath$k$}}} 
V_{{\rm a}}(\mbox{\boldmath$k$} - \mbox{\boldmath$k'$}) 
Z_{\mbox{{\scriptsize \boldmath$k'$}}}$, namely 
$g_{{\rm d}} = \sum_{\mbox{\boldmath$k$}, \mbox{\boldmath$k'$}}
\phi_{{\rm d}}(\mbox{\boldmath$k$})
Z_{\mbox{{\scriptsize \boldmath$k$}}} 
V_{{\rm a}}(\mbox{\boldmath$k$} - \mbox{\boldmath$k'$}) 
Z_{\mbox{{\scriptsize \boldmath$k'$}}}
\phi_{{\rm d}}^{*}(\mbox{\boldmath$k'$})$. 
 The effective density of state of the quasi-particles 
$\tilde{\rho}_{{\rm d}}(\varepsilon)$ is defined as 
$\tilde{\rho}_{{\rm d}}(\varepsilon) = \sum_{\mbox{\boldmath$k$}} 
\delta(\varepsilon - E_{\mbox{{\scriptsize \boldmath$k$}}}) 
|\phi_{{\rm d}}(\mbox{\boldmath$k$})|^{2} $, which is enhanced by the 
renormalization. 
 Here, $E_{\mbox{{\scriptsize \boldmath$k$}}}$ is the quasi-particle's energy 
$E_{\mbox{{\scriptsize \boldmath$k$}}} =  
Z_{\mbox{{\scriptsize \boldmath$k$}}} 
(\varepsilon_{\mbox{{\scriptsize \boldmath$k$}}} + {\rm Re} 
{\mit{\it \Sigma}}^{{\rm R}}(\mbox{\boldmath$k$}, 0)) $. 
 The $\bar{v}_{{\rm F}}$ is the effective Fermi velocity 
for the $d$-wave symmetry. This is defined as 
$\bar{v}_{{\rm F}}^{2} = \int_{{\rm F}}  
\tilde{\rho}_{k_{{\rm F}}} \tilde{v}(k_{{\rm F}})^{2} {\rm d}k_{{\rm F}} /
\tilde{\rho}_{{\rm d}}(0) $, where the integration is carried out on the 
Fermi surface and $\tilde{\rho}_{k_{{\rm F}}} =  
|\phi_{{\rm d}}(k_{{\rm F}})|^{2} / \tilde{v}(k_{{\rm F}}) $. 
The velocity $\tilde{v}(k_{{\rm F}})$ is the absolute value of 
the quasi-particle's velocity 
$\tilde{\mbox{\boldmath$v$}}_{\mbox{{\scriptsize \boldmath$k$}}} = 
{\rm d}E_{\mbox{{\scriptsize \boldmath$k$}}}/{\rm d}\mbox{\boldmath$k$}$ 
on the Fermi surface. 
 It should be noticed that the quasi-particle's velocity is reduced by the 
renormalization, especially around $(\pi,0)$. The renormalization is 
caused by the mass renormalization $Z_{\mbox{{\scriptsize \boldmath$k$}}}$ and 
the transformation of the Fermi surface. The bare velocity 
$\mbox{\boldmath$v$}_{\mbox{{\scriptsize \boldmath$k$}}} = 
{\rm d}\varepsilon_{\mbox{{\scriptsize \boldmath$k$}}}/
{\rm d}\mbox{\boldmath$k$}$ is small at the `hot spot' on the transformed 
Fermi surface. 
 Since the effective Fermi 
velocity $\bar{v}_{{\rm F}}$ is mainly determined by the region around 
$(\pi,0)$, the velocity $\bar{v}_{{\rm F}}$ is remarkably reduced 
by the electron correlation.

 It should be noticed that the TDGL parameter $b$ is proportional to 
$\bar{v}_{{\rm F}}^{2}/T^{2}$, which is proportional to the inverse 
square of the superconducting coupling 
$T_{{\rm c}}/\varepsilon_{{\rm F}}$. In this sense, the effective Fermi energy
should be defined as $\varepsilon_{{\rm F}} \propto \bar{v}_{{\rm F}}$. 
 That is to say, 
the TDGL parameter $b$ decreases with increasing the superconducting coupling. 
 We can see in Fig. 9 that the TDGL parameter $b$ decreases 
with under-doping and/or with increasing $U$. 
 It is because the critical temperature $T_{{\rm c}}$ increases  
and the renormalization for the effective Fermi velocity 
$\bar{v}_{{\rm F}}$ becomes remarkable. 
 These effects reduce the TDGL parameter $b$ in spite of the increasing 
coupling constant $g_{{\rm d}} \tilde{\rho}_{{\rm d}}(0)$. 
 The effects of the quasi-particle's renormalization is confirmed by showing 
the results at the fixed temperature. (see the inset in Fig. 9(b)) 
 Although the coupling constant $g_{{\rm d}} \tilde{\rho}_{{\rm d}}(0)$ 
increases and the temperature $T$ is fixed, 
the parameter $b$ decreases with increasing $U$. 
 This is because the effective Fermi velocity $\bar{v}_{{\rm F}}$ is reduced. 

 Thus, the superconductivity becomes strong coupling one 
when the electron correlation is strong. 
 The above results microscopically justify our scenario about 
the doping dependence of the pseudogap phenomena. 
 The superconducting fluctuations become strong, and the pseudogap phenomena 
take place in the wide temperature region with decreasing the doping 
concentration $\delta$. 
 These doping dependence is consistent with the experimental results. 
 Moreover, the above doping dependence of the parameter $b$ is consistent with 
the magnetic field dependence of the pseudogap 
phenomena.~\cite{rf:yanaseMG,rf:zheng,rf:gorny,rf:eschrig,rf:zheng2}

\subsection{Magnetic properties}

 In this subsection, we show the results for the magnetic properties which 
are measured by NMR, neutron scattering and so on. 
 The anti-ferromagnetic spin fluctuations are one of the important 
characters of High-$T_{{\rm c}}$ cuprates. The development of the spin 
fluctuations are well described by the FLEX approximation. 
 One the other hand, the suppression of the low frequency spin fluctuations 
have been pointed out by NMR 
measurements.~\cite{rf:yasuoka,rf:NMR,rf:takigawa,rf:itoh,rf:ishida,rf:tokunaga,rf:goto} 
 The phenomena have been called 'spin gap' at the initial stage. 
 At present, it is known that not only the spin channel but also the single 
particle properties show the gap-like features.~\cite{rf:renner,rf:ARPES} 
 Therefore, the 'spin gap' is considered to be a result of the gap formation 
in the single particle properties, namely 'pseudogap'. 
 In this subsection, the pseudogap phenomena in the magnetic properties are 
explained by considering the effects of the superconducting fluctuations 
on the single particle properties.

 In the FLEX+T-matrix approximation, the spin susceptibility 
$\chi_{{\rm s}} (\mbox{\boldmath$q$}, \Omega)$ 
is obtained by extending the FLEX approximation, 
\begin{eqnarray}
    \chi_{{\rm s}}^{{\rm R}} (\mbox{\boldmath$q$}, \Omega) & = &
    \frac{\chi_{0}^{{\rm R}} (\mbox{\boldmath$q$}, \Omega)}
      {1 - U \chi_{0}^{{\rm R}} (\mbox{\boldmath$q$}, \Omega)}, 
\\
  \chi_{0} (\mbox{\boldmath$q$}, {\rm i} \Omega_{n}) & = &
  - T \sum_{\mbox{\boldmath$k$},{\rm i} \omega_{n}} 
  G (\mbox{\boldmath$k$}, {\rm i} \omega_{n}) 
  G (\mbox{\boldmath$k+q$}, {\rm i} \omega_{n} + {\rm i} \Omega_{n}), 
\nonumber \\
\end{eqnarray}
 where the dressed Green function 
$G (\mbox{\boldmath$k$}, {\rm i} \omega_{n}) = 
( {\rm i} \omega_{n} - \varepsilon_{\mbox{{\scriptsize \boldmath$k$}}} - 
{\mit{\it \Sigma}}_{{\rm F}} (\mbox{\boldmath$k$}, {\rm i} \omega_{n}) - 
{\mit{\it \Sigma}}_{{\rm S}} (\mbox{\boldmath$k$}, {\rm i} \omega_{n}))^{-1} $
is used. 
 The effects of the superconducting fluctuations are included in the 
self-energy,  
${\mit{\it \Sigma}}_{{\rm S}} (\mbox{\boldmath$k$}, {\rm i} \omega_{n})$.

 The NMR spin-lattice relaxation rate $1/T_{1}$ and spin-echo decay rate 
$1/T_{2{\rm G}}$ are obtained by the following formula. 
\begin{eqnarray}
  1/T_{1}T & = & \sum_{\mbox{\boldmath$q$}} F_{\perp}(\mbox{\boldmath$q$})  
             [\frac{1}{\omega} 
             {\rm Im} \chi_{{\rm s}}^{{\rm R}} (\mbox{\boldmath$q$}, \omega) 
             \mid_{\omega \to 0}],
\\
  1/T_{2G}^{2} & = & \sum_{\mbox{\boldmath$q$}} 
                     [F_{\parallel}(\mbox{\boldmath$q$}) 
        {\rm Re} \chi_{{\rm s}}^{{\rm R}} (\mbox{\boldmath$q$}, 0)]^{2}
                   - [\sum_{\mbox{\boldmath$q$}} 
                      F_{\parallel}(\mbox{\boldmath$q$}) 
        {\rm Re} \chi_{{\rm s}}^{{\rm R}} (\mbox{\boldmath$q$}, 0)]^{2}.
\nonumber \\
\end{eqnarray}
 Here, $F_{\perp}(\mbox{\boldmath$q$}) = \frac{1}{2} [\{A_{1} + 2 B 
(\cos q_{x}+\cos q_{y})\}^{2} + \{A_{2} + 2 B (\cos q_{x}+\cos q_{y})\}^{2}] $
 and $F_{\parallel}(\mbox{\boldmath$q$}) = 
\{A_{2} + 2 B (\cos q_{x}+\cos q_{y})\}^{2}$.
 The hyperfine coupling constants $A_{1}, A_{2}$ and $B$ are evaluated as 
$A_{1} = 0.84 B$ and $A_{2} = -4 B$.~\cite{rf:barzykin}

 The calculated results for the NMR $1/T_{1}T$, $1/T_{2{\rm G}}$ and 
the static spin susceptibility  are shown in Fig. 10. 
 The results show the pseudogap in the NMR $1/T_{1}T$ (Fig. 10(a)). 
 In the FLEX calculation, the NMR $1/T_{1}T$ increases with decreasing the 
temperature (see the inset in Fig. 10(a)). 
This corresponds to the Curie-Weiss low of the spin fluctuations. 
 In the FLEX+T-matrix calculation, the NMR $1/T_{1}T$ increases with 
decreasing the temperature from high temperature, shows its peak at $T^{*}$ 
and decreases with decreasing the temperature. 
 This decrease above the critical point $T=T_{{\rm c}}$ is the well-known 
pseudogap in NMR 
measurements.~\cite{rf:yasuoka,rf:NMR,rf:takigawa,rf:itoh,rf:ishida,rf:tokunaga,rf:goto} 
 This phenomenon is caused by the superconducting fluctuations. 
 Since the DOS is reduced by the superconducting fluctuations, 
the low frequency spin fluctuations are suppressed. 
 Because the NMR $1/T_{1}T$ measures the low frequency component of the spin 
fluctuations, $1/T_{1}T$ decreases with approaching the critical point. 
 Thus, the pseudogap observed in NMR $1/T_{1}T$ takes place through the 
pseudogap in the single particle properties.

\begin{figure}[htbp]
\begin{center}
   \epsfysize=6.5cm
    $$\epsffile{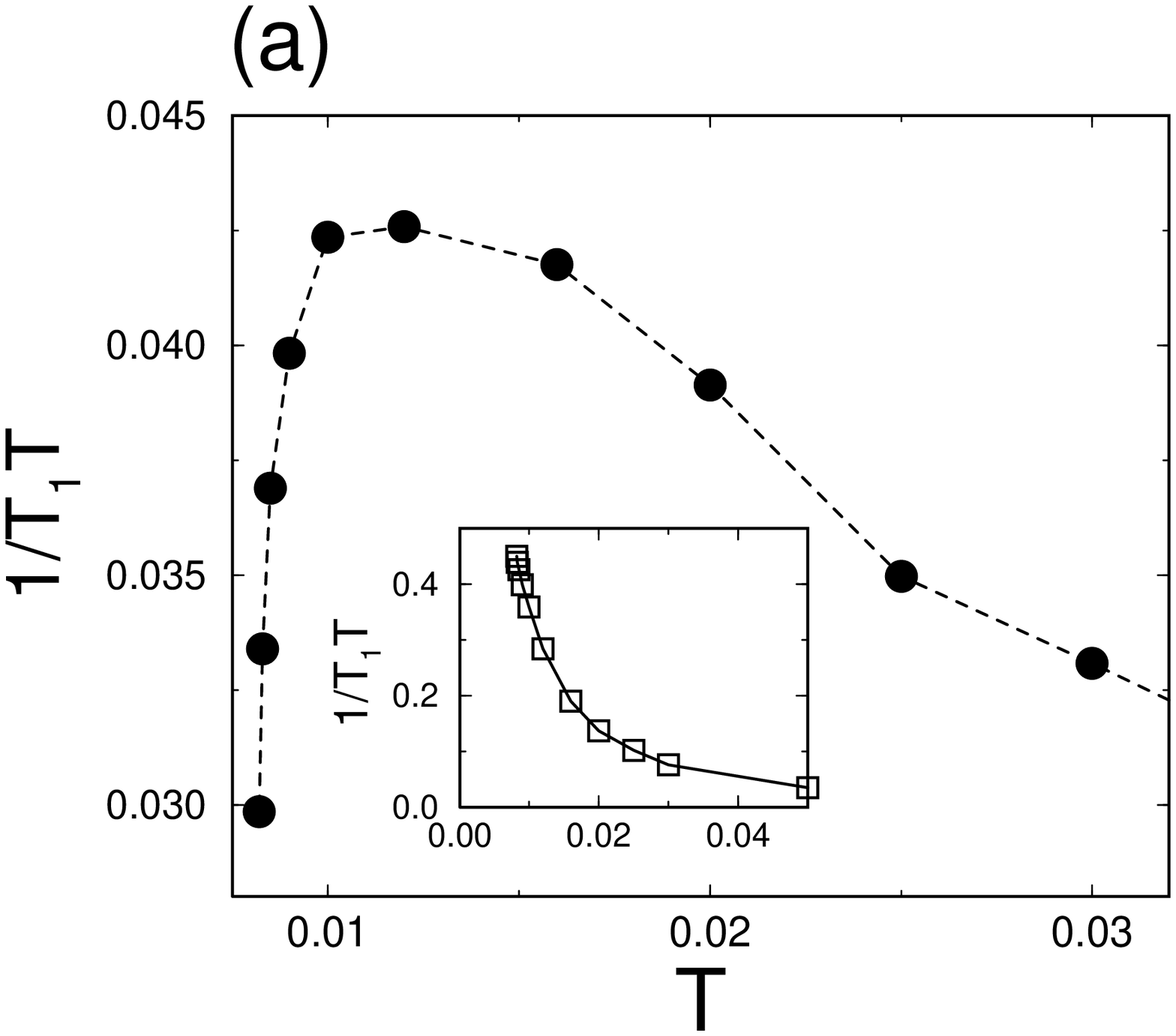}$$
   \epsfysize=6.5cm
    $$\epsffile{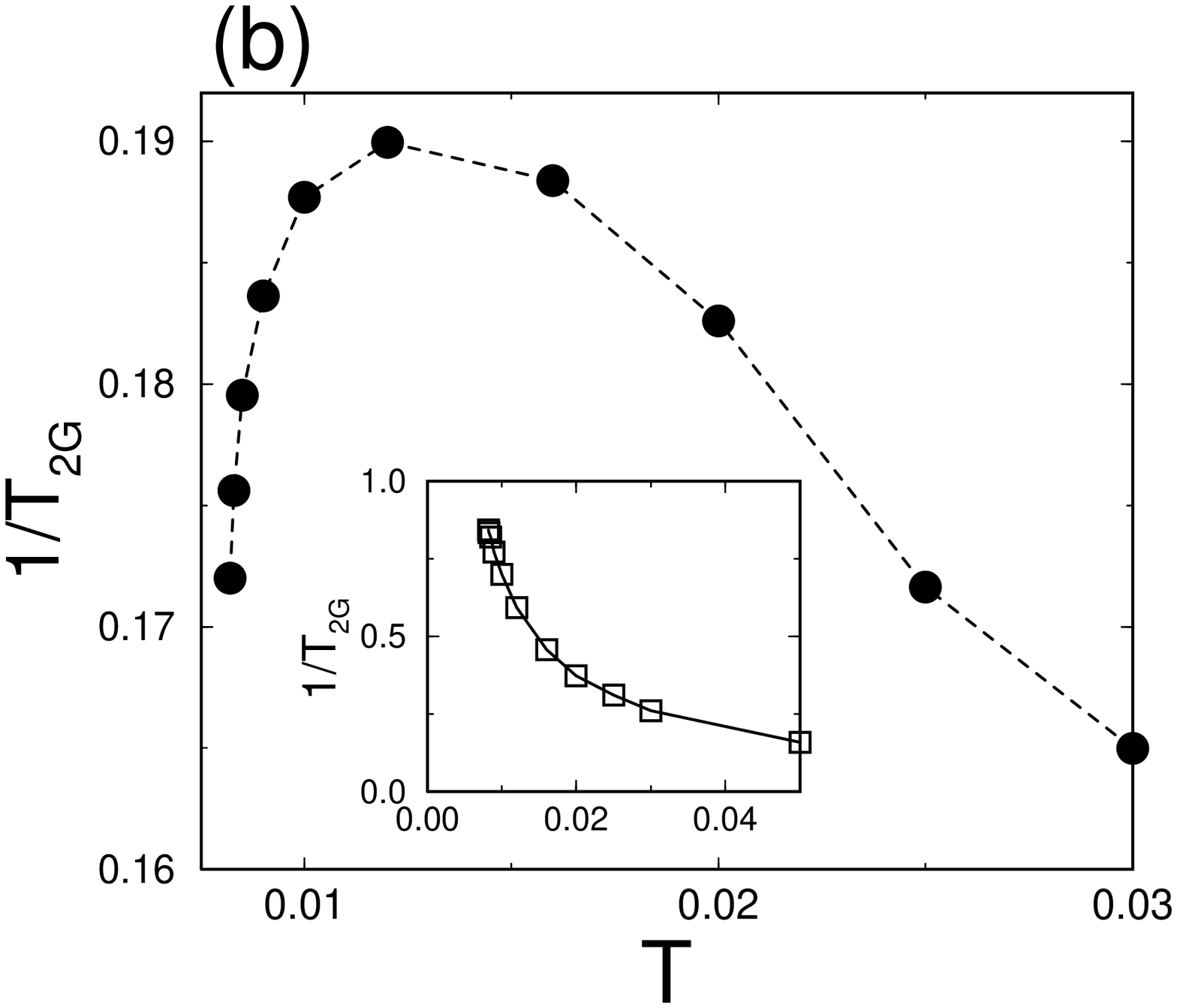}$$
   \epsfysize=6.5cm
    $$\epsffile{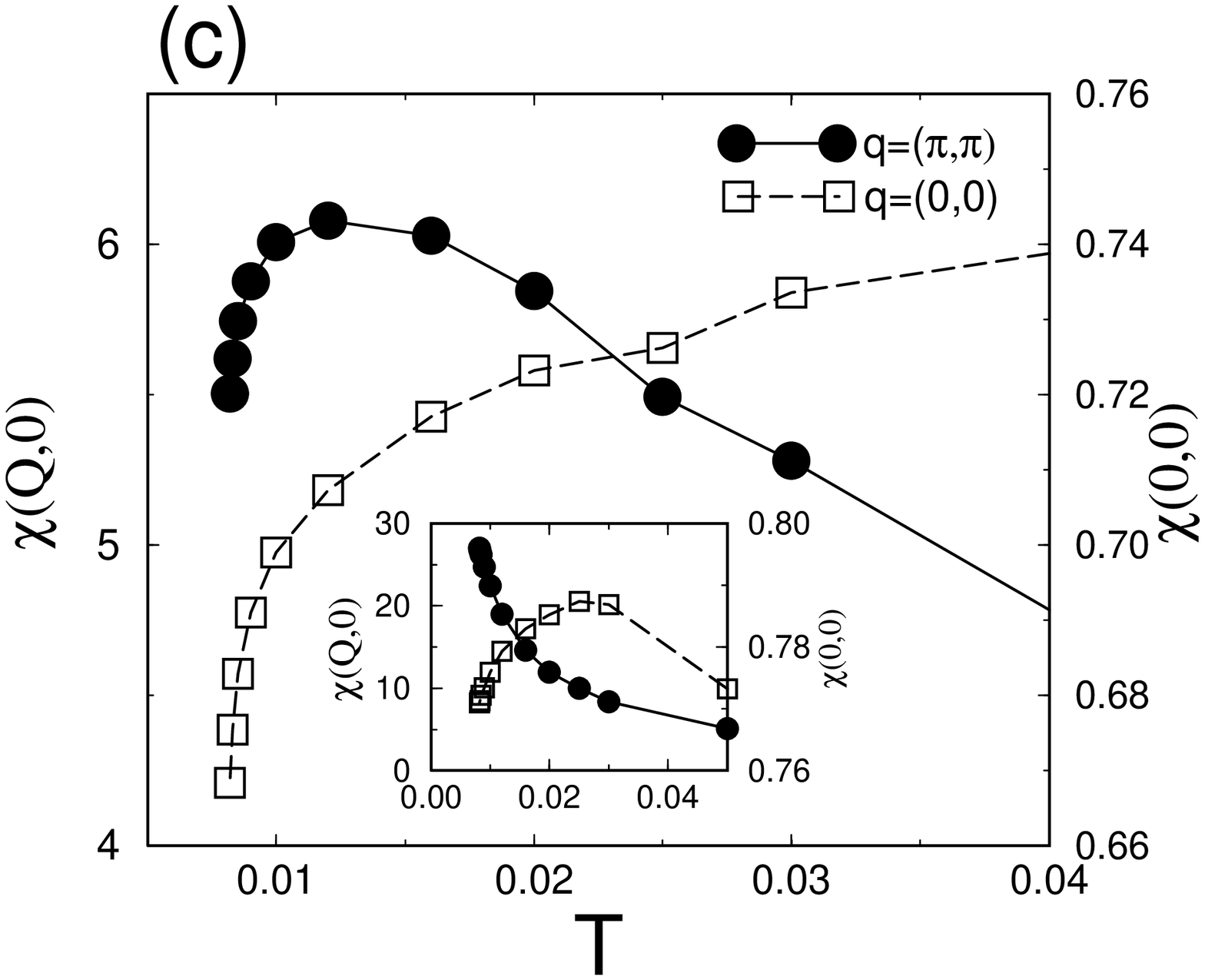}$$
    \caption{The temperature dependence of (a) the NMR $1/T_{1}T$ 
             and (b) the NMR $1/T_{2{\rm G}}$ calculated 
             by the FLEX+T-matrix approximation. 
             (c) The temperature dependence of static spin susceptibility 
             $\chi_{{\rm s}}^{{\rm R}} (\mbox{\boldmath$q$}, 0)$ at 
             $\mbox{\boldmath$q$} = (0,0)$ (Open squares) and at 
             $\mbox{\boldmath$q$} = (\pi,\pi)$ (Closed circles). 
             Here, the doping concentration is fixed to the under-doped region 
             $\delta = 0.093 \sim 0.103$. 
             The inset in (a), (b) and (c) shows the same quantities 
             calculated by the FLEX approximation. 
             }
  \end{center}
\end{figure}

 The NMR $1/T_{2{\rm G}}$ also shows the pseudogap (Fig. 10(b)) with the same 
onset temperature $T^{*}$ as that in $1/T_{1}T$.  
 The NMR $1/T_{2{\rm G}}$ decreases with approaching the critical point. 
 This is also an effect of the superconducting fluctuations. 
 However, the pseudogap in the NMR $1/T_{2{\rm G}}$ is weak compared with that 
in the NMR $1/T_{1}T$. This is because the NMR $1/T_{2{\rm G}}$ measures 
the static spin susceptibility which reflects the total weight of the spin 
fluctuations.~\cite{rf:yanaseSC,rf:yanaseMG} 
 It should be noticed that the pseudogap suppresses only the low frequency 
component of the spin fluctuations. This is a natural result because the 
superconductivity has a smaller energy scale than that of the spin 
fluctuations. 
 Thus, it is no wonder that the scaling relation of the spin 
fluctuations is violated in the pseudogap state.~\cite{rf:takigawa,rf:itoh,rf:tokunaga,rf:goto} 
 The above results for the NMR give the qualitatively consistent explanation 
for the experimental results.~\cite{rf:yasuoka,rf:NMR,rf:takigawa,rf:itoh,rf:ishida,rf:tokunaga,rf:goto} 
 Minutely speaking, the different behaviors of $1/T_{2G}$ have been 
reported for different High-$T_{{\rm c}}$ 
compounds.~\cite{rf:takigawa,rf:itoh,rf:tokunaga,rf:goto} 
 There is an idea that attributes the difference to the effects of 
the interlayer coupling.~\cite{rf:itoh,rf:goto} 
 The results of the single layer compounds show the decrease of $1/T_{2G}$ 
in the pseudogap state.~\cite{rf:itoh} 
 Anyway, the relatively weak effect of the pseudogap 
on $1/T_{2G}$ than on $1/T_{1}T$ is observed in common, which is consistent 
with our results.

 The similar features of the NMR $1/T_{1}T$ and $1/T_{2{\rm G}}$ have been 
observed in the superconducting state.~\cite{rf:itoh} 
 In particular, the $1/T_{2{\rm G}}$ remains even in the low temperature, 
although the $1/T_{1}T$ rapidly decreases. 
 These features are the characteristics of the $d$-wave 
superconductivity.~\cite{rf:yanaseSC,rf:bulut} 
 Therefore, the above results for the pseudogap state are natural 
because the pseudogap is a precursor of the $d$-wave superconductivity.

 While the many quantities 
show the pseudogap with the same onset temperature 
$T^{*}$,~\cite{rf:odatransport} 
the uniform spin susceptibility 
$\chi_{{\rm s}}^{{\rm R}} (\mbox{\boldmath$0$}, 0)$ 
decreases from the much higher temperature than $T^{*}$.~\cite{rf:oda} 
 However, the decrease of the uniform susceptibility becomes more rapid
near $T^{*}$.~\cite{rf:ishida} 
 We consider that the rapid decrease is caused by the superconducting 
fluctuations. The calculated results confirm the consideration. 
 We show the results for the uniform susceptibility 
$\chi_{{\rm s}}^{{\rm R}} (\mbox{\boldmath$0$}, 0)$ and 
the staggard susceptibility 
$\chi_{{\rm s}}^{{\rm R}} (\mbox{\boldmath$Q$}, 0)$ in Fig. 10(c). 
 The staggard susceptibility shows the pseudogap at the pseudogap onset 
temperature $T^{*}$ observed in $1/T_{1}T$.  
 On the other hand, the uniform susceptibility 
$\chi_{{\rm s}}^{{\rm R}} (\mbox{\boldmath$0$}, 0)$ decreases with decreasing 
the temperature from much higher temperature. 
 The decrease becomes rapid near $T^{*}$. 
 These results well explain results of the NMR 
measurements.~\cite{rf:ishida}

 The slight decrease of the uniform susceptibility is shown even in the FLEX 
approximations. (see the inset in Fig. 10(c).) 
 Therefore, the decrease of the uniform susceptibility is not necessarily 
attributed to the superconducting fluctuations. 
 However, the superconducting fluctuations significantly affect the uniform  
susceptibility and remarkably reduce the quantity near the critical point.

 The frequency dependence of the spin susceptibility well describes the 
character of the pseudogap in the magnetic properties. 
 The results for the dynamical spin susceptibility  
$\chi_{{\rm s}}^{{\rm R}} (\mbox{\boldmath$q$}, \Omega)$ 
at $\mbox{\boldmath$q$} = \mbox{\boldmath$Q$}$ is shown in Fig. 11. 
 The real part is suppressed at the low frequency in the pseudogap state 
($T=0.0082$ in Fig. 11(a)). 
 Thus, the magnetic order is suppressed by the superconducting fluctuations. 

 The imaginary part has been measured by the inelastic neutron scattering and 
shows the pseudogap.~\cite{rf:neutron} 
 The calculated results show that the imaginary part is remarkably suppressed 
at the low frequency in the pseudogap state (Fig. 11(b)). 
 This is the pseudogap phenomenon observed by the neutron scattering 
measurements. 
 On the other hand, the spin fluctuations develop in higher frequency region. 
In other words, the pseudogap transfers the spectral weight of the spin 
fluctuations from the low frequency region to the high frequency region. 
 Therefore, the total weight is not so reduced by the pseudogap. 
 These features are consistent with the experimental results~\cite{rf:neutron} 
and with the above explanation for the 
NMR $1/T_{1}T$ and $1/T_{2{\rm G}}$. 
 It is notable that the pairing interaction arising from the spin fluctuations 
originates in the relatively wide frequency region. 
 Therefore, the $d$-wave pairing interaction is not so reduced 
by the pseudogap. The interplay between the superconducting fluctuations and 
the spin fluctuations will be discussed in \S4.1.

\begin{figure}[htbp]
\begin{center}
   \epsfysize=6.5cm
    $$\epsffile{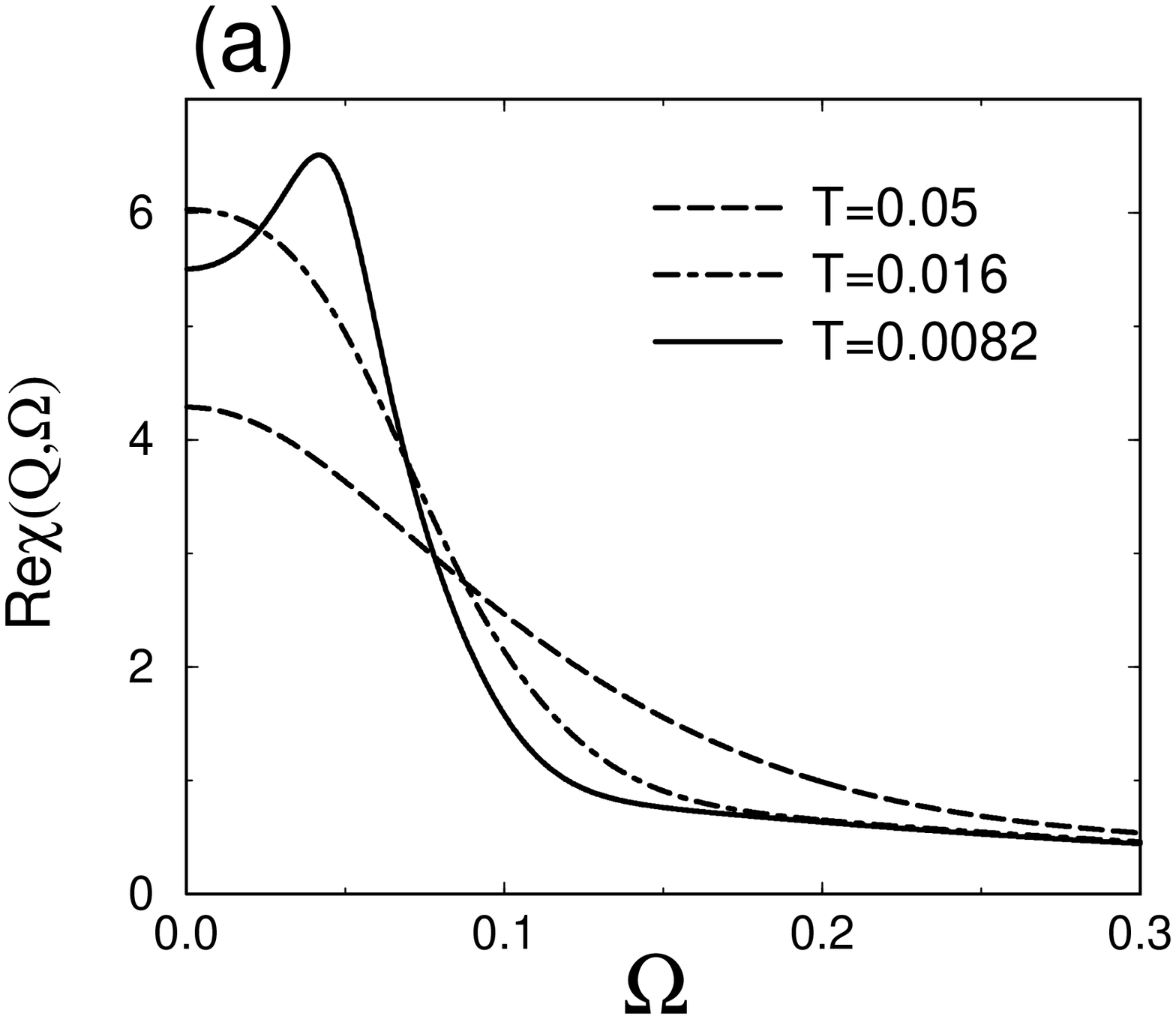}$$
   \epsfysize=6.5cm
    $$\epsffile{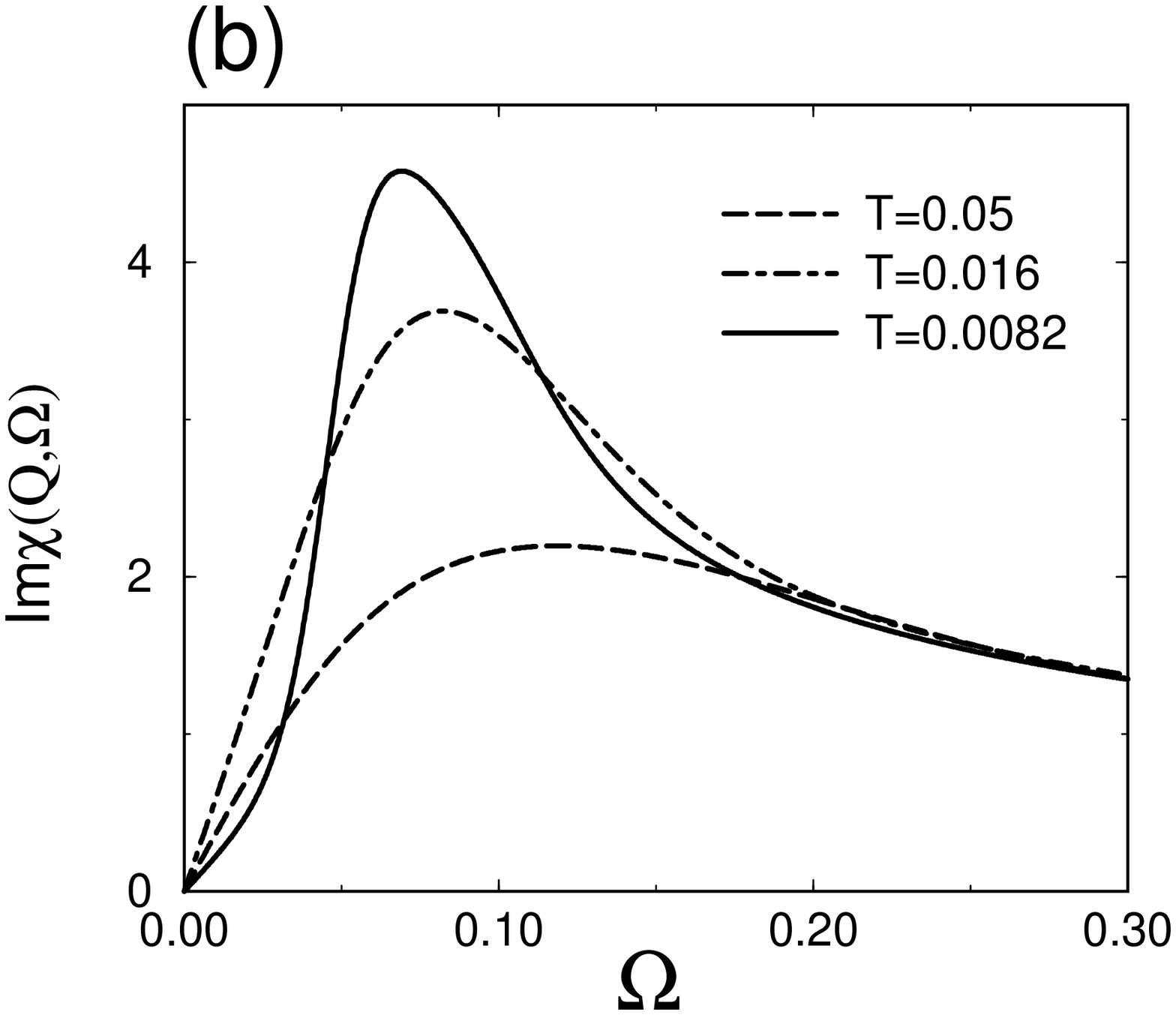}$$
    \caption{The frequency dependence of the staggard spin susceptibility 
             $\chi_{{\rm s}}^{{\rm R}} (\mbox{\boldmath$Q$}, \Omega)$ 
             calculated by the FLEX+T-matrix approximation for under-doped 
             cuprates. 
             (a) The real part. (b) The imaginary part. 
             The solid lines, the dash-dotted lines and the long-dashed lines 
             correspond to $T = 0.0082$, $T = 0.016$ and $T = 0.05$, 
             respectively. 
             }
  \end{center}
\end{figure}

 In the remaining part of this subsection, we discuss the commensurate and 
incommensurate structure of the spin fluctuations. 
 The incommensurability in the Y-based compounds 
${\rm Y}{\rm Ba}_{2}{\rm Cu}_{3}{\rm O}_{6+\delta}$ has been pointed out 
by the inelastic neutron scattering measurements,  
and has been discussed in connection with the stripe phase.~\cite{rf:stripeEX} 
 The stripe phase in the La-based compounds 
${\rm La}_{2-x}{\rm Sr}_{x}{\rm Cu}{\rm O}_{4}$ has been measured 
by the elastic neutron scattering and has been actively discussed 
lately.~\cite{rf:tranquada,rf:zaanen,rf:zachar,rf:white,rf:mizokawa,rf:yamase,
rf:machida,rf:himeda}

 It should be noticed that both commensurate and incommensurate structures 
are obtained within the FLEX approximation (Fig. 12(a) and 12(c)). 
 The incommensurability $\delta_{{\rm inc}}$ is defined by the peak position 
of the dynamical spin susceptibility 
$\mbox{\boldmath$Q$}_{{\rm p}} = (\pi \pm \delta_{{\rm inc}}, \pi)$ and 
$(\pi, \pi \pm \delta_{{\rm inc}})$.  
 Whether the spin fluctuations are commensurate or incommensurate is 
determined by the chosen parameters, $t'$, $\delta$ and $U$. 
 One of the general results in our calculation is that the incommensurability 
$\delta_{{\rm inc}}$ increases with the doping concentration $\delta$. 
 These results are qualitatively consistent with the experimental 
results.~\cite{rf:yamadaneutron} 
 The detailed agreement with the experimental results has been discussed  
by the FLEX calculation in the normal state.~\cite{rf:kuroki}

\begin{figure}[htbp]
\begin{center}
   \epsfysize=5.5cm
    $$\epsffile{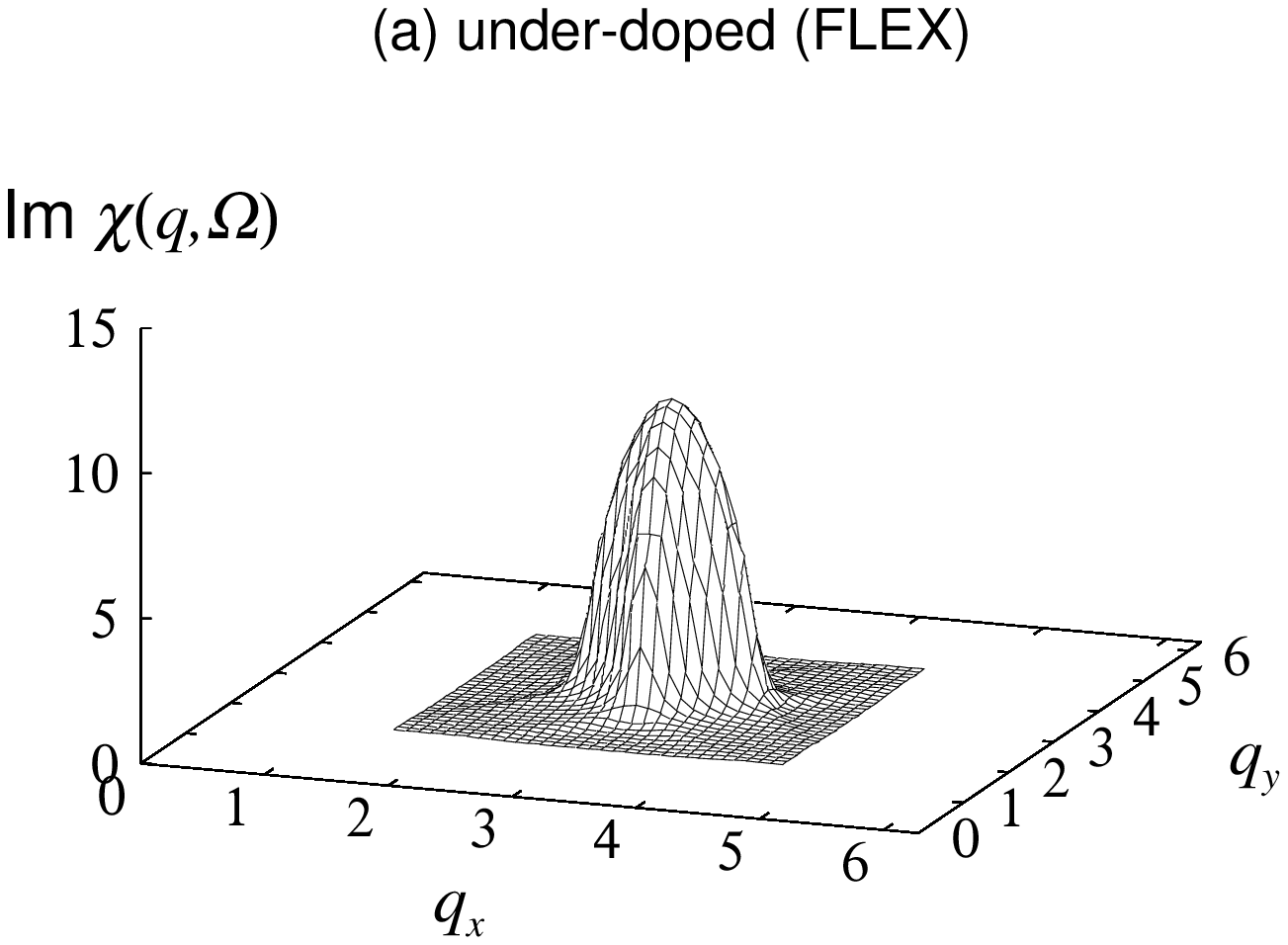}$$
   \epsfysize=5.5cm
    $$\epsffile{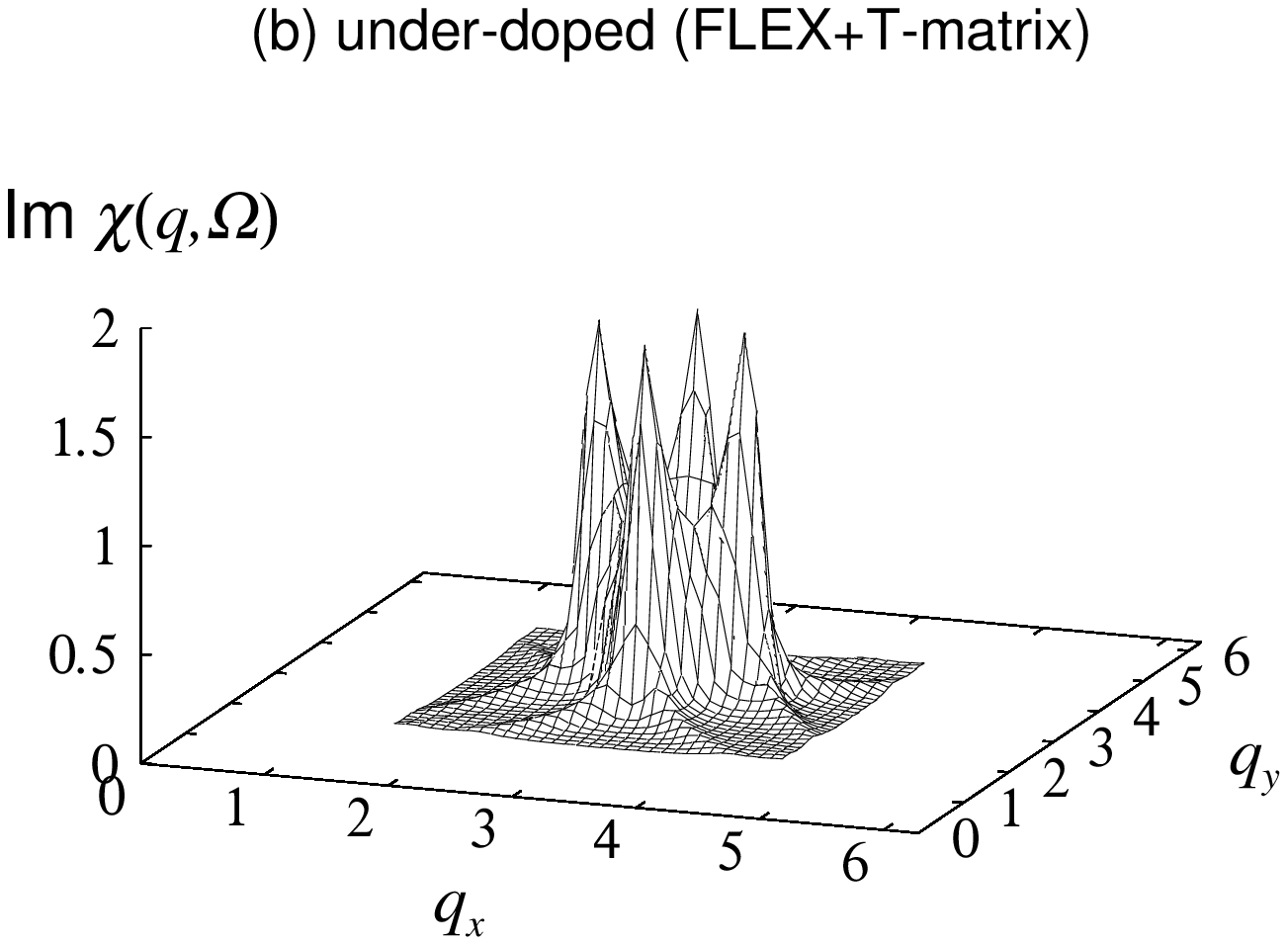}$$
   \epsfysize=5.5cm
    $$\epsffile{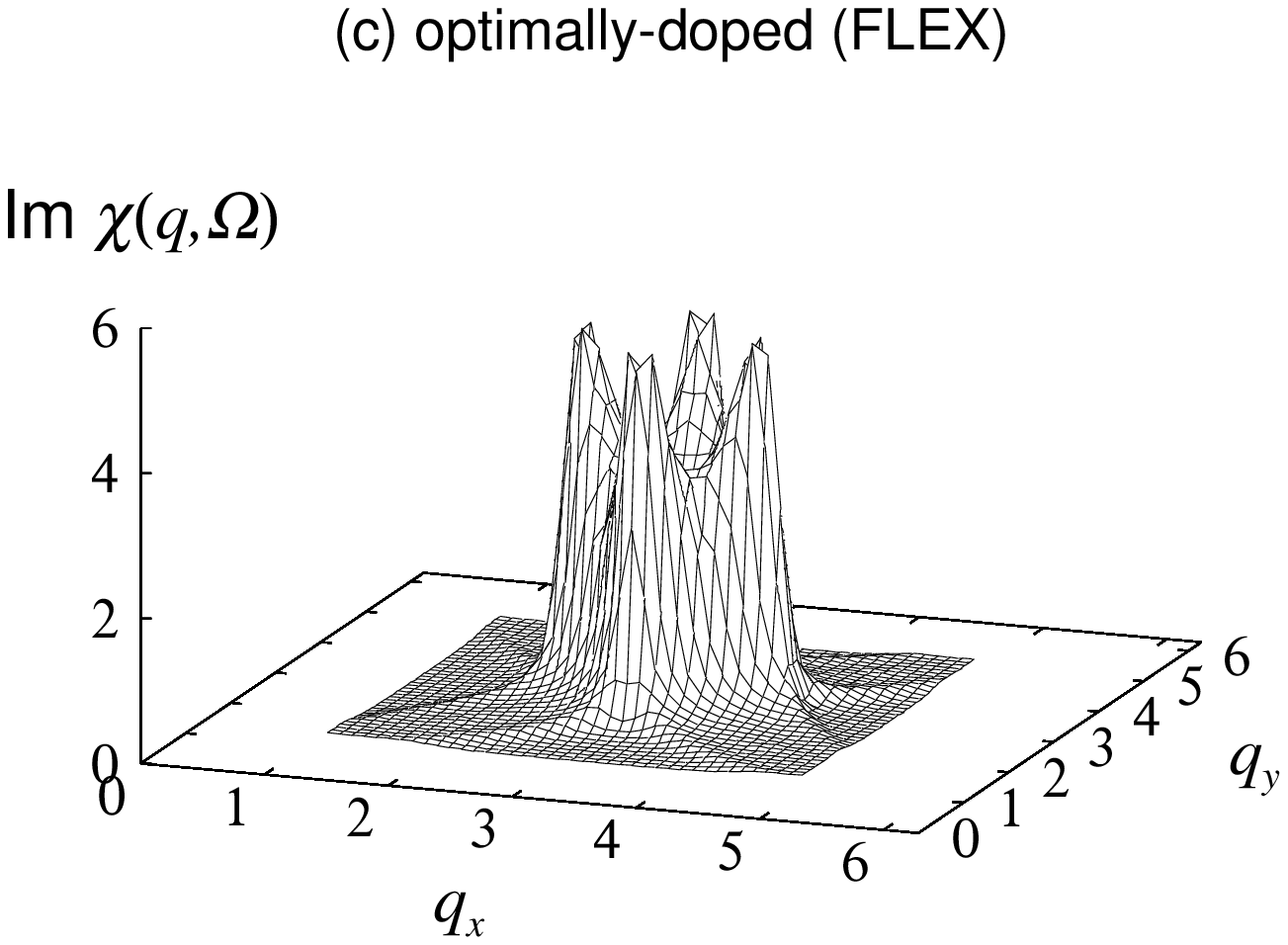}$$
   \epsfysize=5.5cm
    $$\epsffile{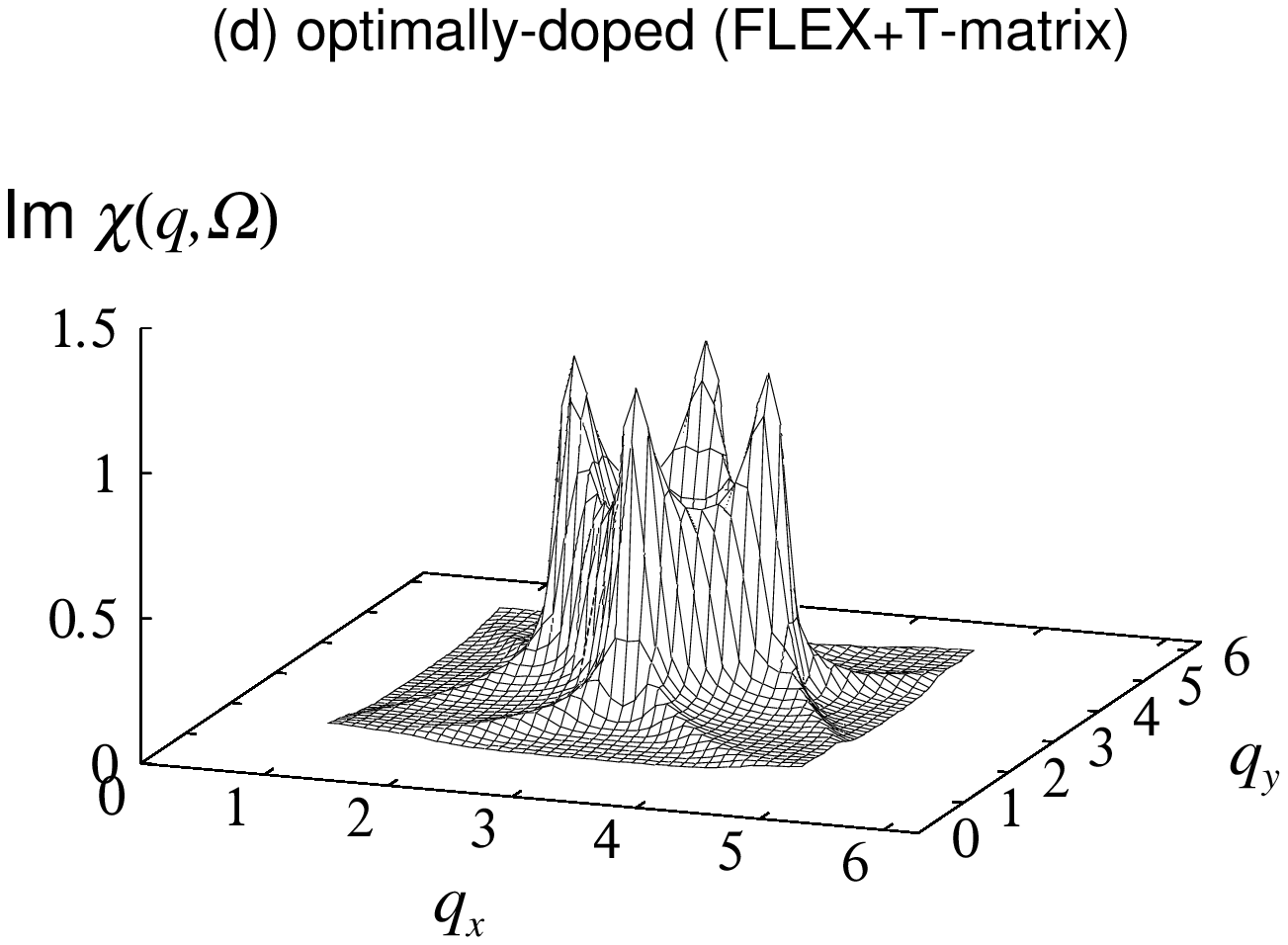}$$
    \caption{The momentum dependence of the dynamical spin susceptibility 
             ${\rm Im}\chi_{{\rm s}}^{{\rm R}} (\mbox{\boldmath$q$}, \Omega)$ 
             at $\Omega=0.01$. 
             The results for under-doped cuprates ($\delta=0.095$ and 
             $T=0.010$) (a) by the FLEX approximation and   
             (b) by the FLEX+T-matrix approximation.  
             The results for optimally-doped cuprates ($\delta=0.156$ and 
             $T=0.0078$)             
             (c) by the FLEX approximation and  
             (d) by the FLEX+T-matrix approximation.  
             }
  \end{center}
\end{figure}

 The other general result is that the superconducting fluctuations enhance the 
incommensurability. Although the commensurate peak is obtained by the FLEX 
approximation in the under-doped region (Fig. 12(a)), it becomes 
incommensurate owing to the superconducting fluctuations (Fig. 12(b)). 
 The incommensurability $\delta_{{\rm inc}}$ increases in the FLEX+T-matrix 
approximation in the optimally-doped region (Fig. 12(d)).  
 The above effect also originates in the pseudogap formation around $(\pi,0)$. 
 The results well explain the experimental results of the inelastic neutron 
scattering. The neutron scattering experiments have shown that the spin 
fluctuations change from incommensurate to commensurate with increasing 
the measured frequency.~\cite{rf:stripeEX}  
 Although the measurements have been carried out in the superconducting state 
in many cases, the qualitatively similar behaviors as those in the pseudogap 
state are expected in the superconducting state. 
 This is because the pseudogap and the superconducting gap have the same 
$d_{x^{2}-y^{2}}$-wave form. 
 The effect of the pseudogap or the superconducting gap disappears 
when the measurement is done at the higher frequency than the energy scale 
of the gap. 
 Thus, the incommensurate peak in the inelastic neutron scattering can be 
explained without any assumption of the charge and/or spin spatial modulation. 
 The calculation including the superconducting order is desired for the 
detailed agreement with the experimental results.

\subsection{Electron-doped cuprates}

 In this subsection, we apply the above calculation to the electron-doped 
cuprates. 
 It is well known that not only the hole-doped cuprates but also the 
electron-doped cuprates such as 
${\rm Nd}_{2-x}{\rm Ce}_{x}{\rm Cu}{\rm O}_{4-y}$ and 
${\rm Pr}_{2-x}{\rm Ce}_{x}{\rm Cu}{\rm O}_{4-y}$ are 
the superconductors.~\cite{rf:electrondopesuper} 
 The electron-hole symmetry is expected within the simple Hubbard model 
including only the nearest-neighbor hopping $t$. 
 However, some different properties from the hole-doped cuprates have been  
pointed out for the electron-doped cuprates. 
 The anti-ferromagnetic ordered state is robust 
against the electron doping rather than the hole doping. 
 The relatively low superconducting critical temperature $T_{{\rm c}}$ 
is observed in the narrow doping range.~\cite{rf:electrondopesuper}

 It has been suggested for a long time that the electron-doped cuprates
are an $s$-wave or the other node-less superconductor. 
 Some experiments have supported the node-less superconductivity. 
The exponential dependence of the magnetic penetration depth has been 
reported.~\cite{rf:electron-dope_S} 
 The absence of the zero bias conductance peak is also 
reported~\cite{rf:kashiwaya} while it should exist in the 
$d$-wave superconductor.~\cite{rf:tanaka} 
 The possibility of some node-less superconductivity has been proposed  
theoretically.~\cite{rf:Stheory} 
 However, some recent experimental results support the $d_{x^{2}-y^{2}}$-wave 
superconductivity also in the electron-doped cuprates. 
 The power law of the magnetic penetration depth~\cite{rf:electron-dope_D} and 
the zero bias conductance peak~\cite{rf:hayashi} are shown. 
 The $d_{x^{2}-y^{2}}$-wave form of the superconducting gap is directly 
measured by the ARPES.~\cite{rf:takahashi} 
 Moreover, the phase sensitive evidence for the $d$-wave superconductivity 
is also reported by the SQUID microscope.~\cite{rf:tsuei} 

 We consider that the hole- and electron-doped cuprates should be understood 
comprehensively by the theory including their respective characters. 
 The anomalous properties of the Hall coefficient are well explained for 
not only hole- but also electron-doped cuprates on the basis of the same 
formalism treating the spin fluctuations.~\cite{rf:kontani,rf:kanki}  
 Therefore, it is natural to expect the same mechanism for the 
superconductivity arising from the spin fluctuations. 
 Actually, the instability of the $d$-wave superconductivity mediated 
by the spin fluctuations has been reported 
theoretically.~\cite{rf:kontani,rf:kondo2,rf:manske}

 We consider that the application to the electron-doped cuprates is an 
important test of the theories on High-$T_{{\rm c}}$ cuprates.   
 Here, it is shown that our theory properly describes the important 
properties of the electron-doped cuprates. 
 The electron-hole asymmetric properties  
are introduced by the next-nearest-neighbor hopping term $t'$ 
in our model. 
The features of the electron-hole 
asymmetry are naturally explained by our calculations. 
 Moreover, some theoretical expectations are proposed below. 
 The experimental verification is desirable.

\begin{figure}[htbp]
\begin{center}
   \epsfysize=6cm
    $$\epsffile{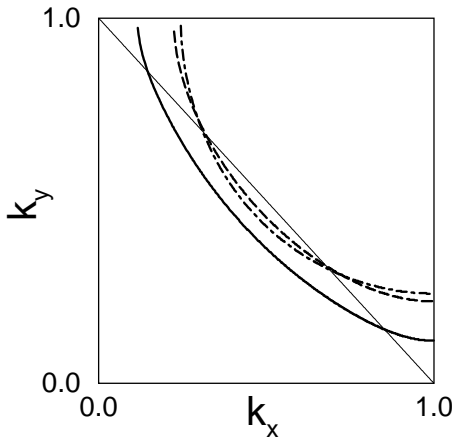}$$
    \caption{The Fermi surface of the hole- and the electron-doped cuprates. 
             The thick solid line shows the Fermi surface of the hole-doped 
             case $\delta=0.10$. 
             The long-dashed and the dash-dotted lines show the 
             Fermi surface of the electron-doped case $\delta=-0.10$. 
             The long-dashed and the dash-dotted lines correspond to case 
             $t'=-0.25t$ and $t'=-0.35t$, respectively. 
             The thin solid line shows the magnetic Brillouin zone. 
             It should be mentioned that this figure shows the non-interacting
             Fermi surface. Actually, the Fermi surface is transformed by 
             the anti-ferromagnetic spin fluctuations.~\cite{rf:yanaseTR}
             }
  \end{center}
\end{figure}

 The main difference between the electron- and hole-doped cuprates 
results from the shape of the Fermi surface. 
 The distance from the Fermi surface to the Van-Hove singularity is especially 
important. 
 The Fermi surface of the electron-doped cuprates is obtained by lifting the 
chemical potential $\mu$. The typical Fermi surface is shown in Fig. 13. 
 The Fermi level is lifted from the Van-Hove singularity 
$(\pi,0)$ where the dispersion is flat. Therefore, the DOS is 
rather small in the electron-doped cuprates. The small DOS means that the 
electron correlation is effectively weak. 
 On the other hand, the nesting around $(\pi/2,\pi/2)$ is enhanced 
and the tendency toward the anti-ferromagnetic order is robust. 
 In the light of the $d$-$p$ model, the carrier is confined in the Cu-site 
in the electron-doped cases, while it is mainly in the O-site 
in the hole-doped ones. 
 This fact probably contributes to the robustness of the 
anti-ferromagnetic order. 
 This difference may affect the parameter of the Hubbard model which is an 
effective model of the $d$-$p$ model in the metallic phase. 
 However, the essential difference is included in the properties of the band 
structure since only the vicinity of the Fermi surface is important for the 
low energy physics.

 Since we define the doping concentration as $\delta = 1-n$, it is negative 
in the electron-doped case. 
 We should note that the numerical calculation is not so correct for 
the electron-doped cases as for the hole-doped ones because of some reasons. 
 The main reason is the finite size effects which are serious owing to 
the large velocity, the weak correlation and the low temperature. 
 The finite size effects on the spin- and superconducting fluctuations are 
also serious by the same reasons. 
 Therefore, we divide the first Brillouin zone into $128 \times 128$ lattice 
points in the calculation for the electron-doped cases.

\begin{figure}[htbp]
\begin{center}
   \epsfysize=5.5cm
    $$\epsffile{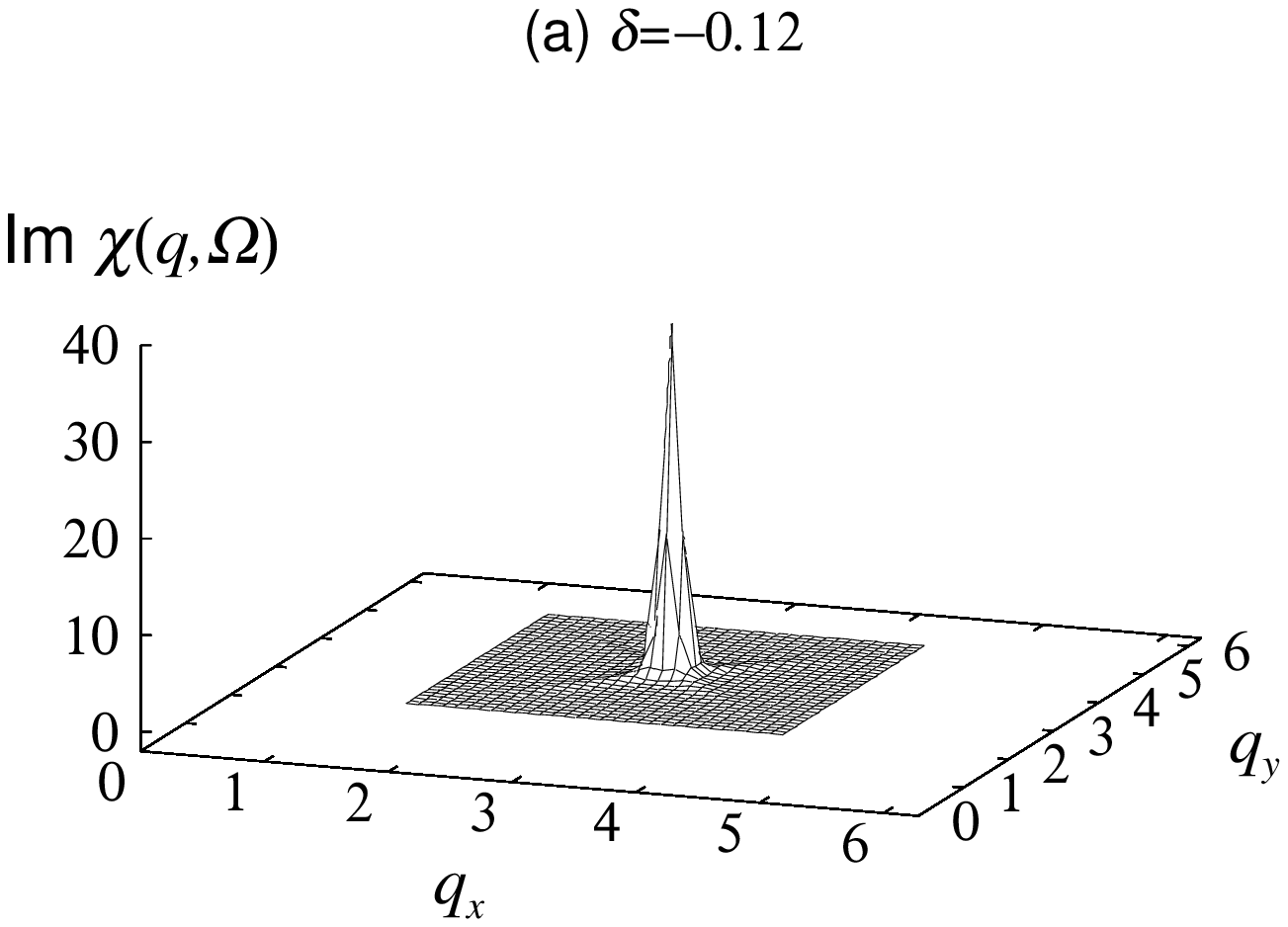}$$
   \epsfysize=5.5cm
    $$\epsffile{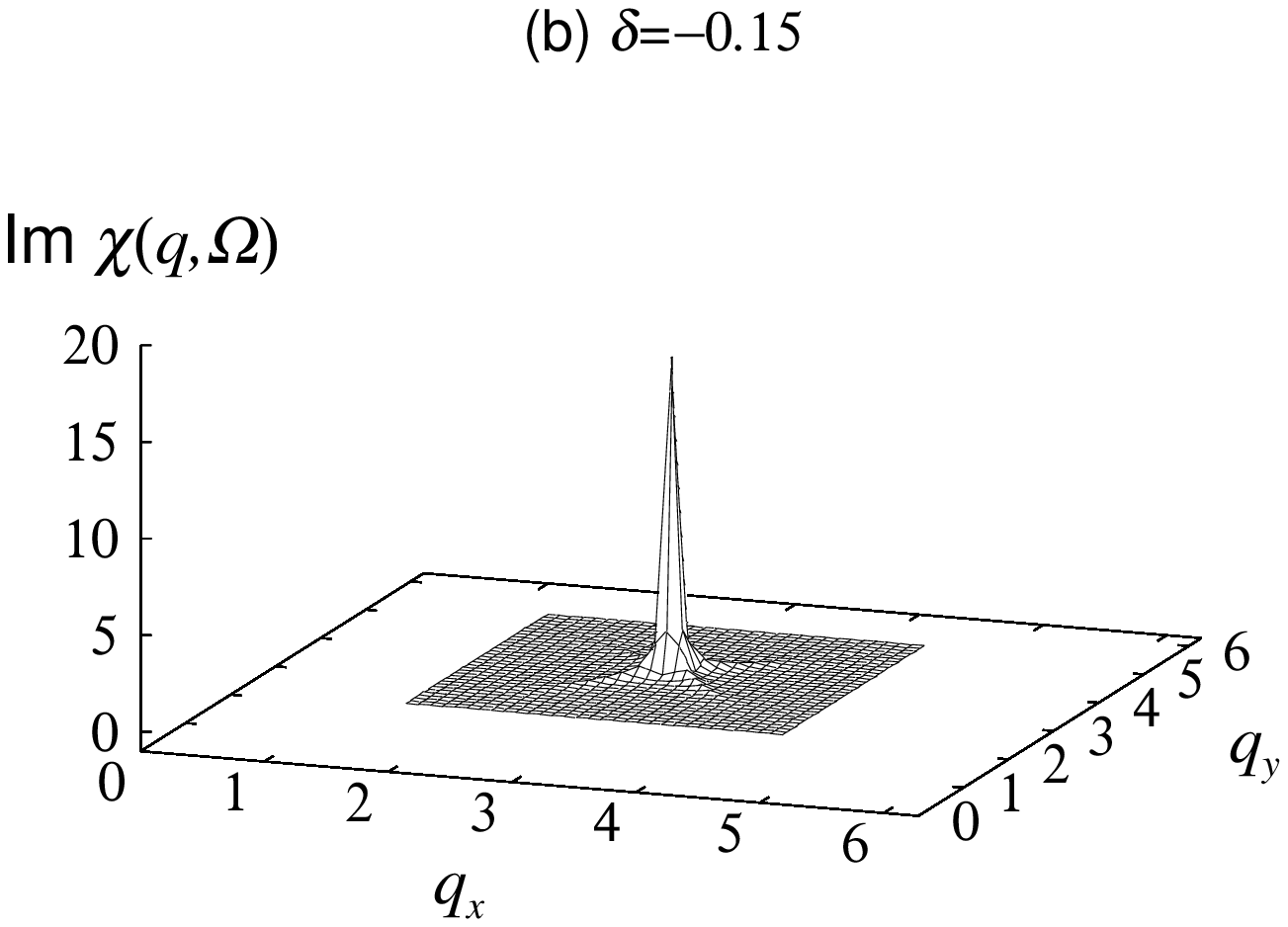}$$
   \epsfysize=5.5cm
    $$\epsffile{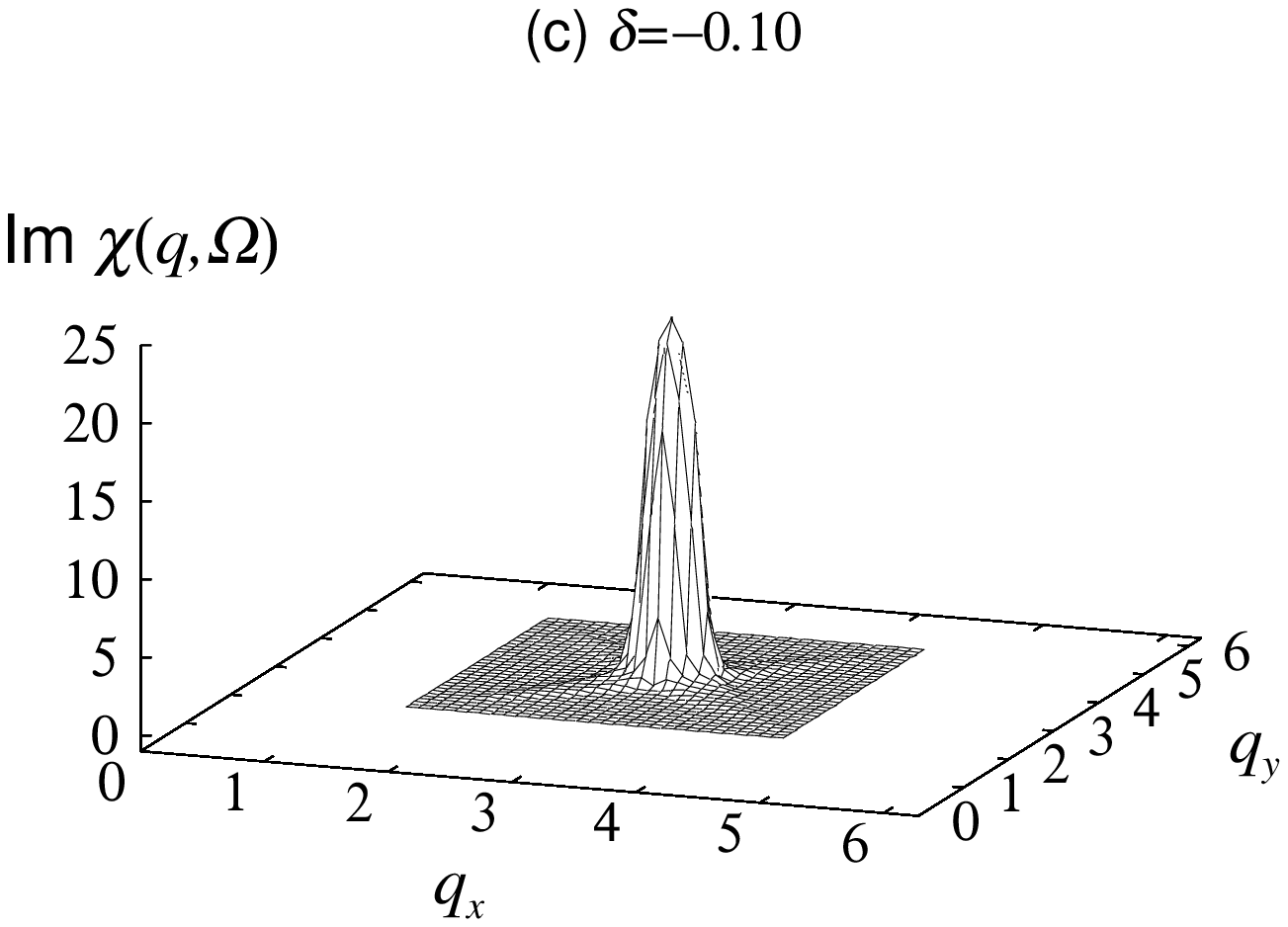}$$
   \epsfysize=5.5cm
    $$\epsffile{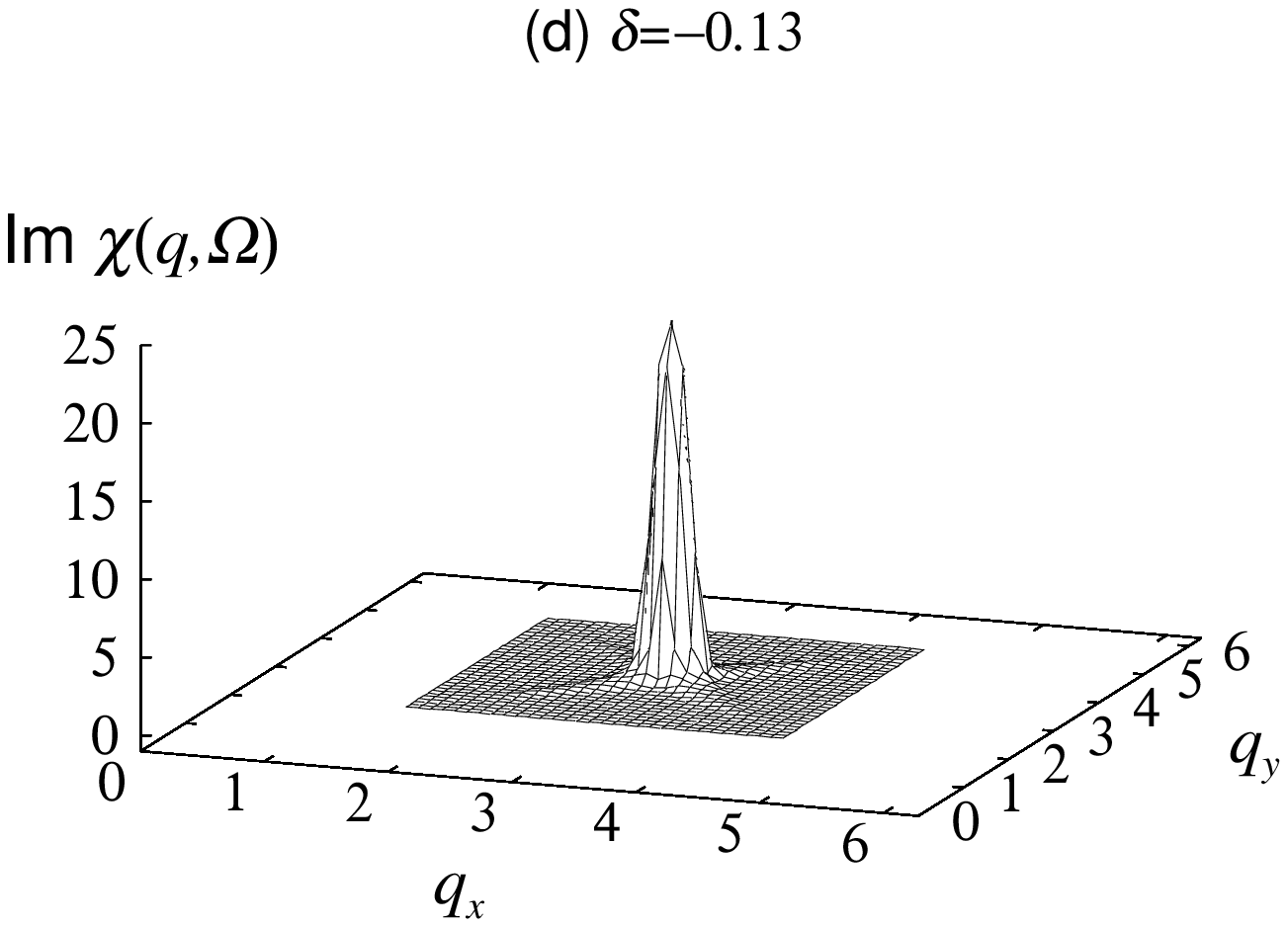}$$
    \caption{The momentum dependence of the dynamical spin susceptibility 
             ${\rm Im}\chi_{{\rm s}}^{{\rm R}} (\mbox{\boldmath$q$}, \Omega)$ 
             at $\Omega=0.01$ calculated by the FLEX approximation.   
             (a) $\delta=-0.123$ and  
             (b) $\delta=-0.150$  
             at $t'=-0.25t$ and $T=0.01$.
             (c) $\delta=-0.104$ and  
             (d) $\delta=-0.130$  
             at $t'=-0.35t$ and $T=0.005$.
             }
  \end{center}
\end{figure}

 For example, the difference from the hole-doped cuprates appears in the 
magnetic properties. We show the results of the FLEX approximation 
for the momentum dependence of the dynamical spin susceptibility 
in Figs. 14 (a) and (b). 
 In these figures, we use the same parameters $t$, $t'$ and $U$ as those of 
the hole-doped cases. 
 A remarkable feature of the spin correlation is that 
the range of the spin fluctuations is narrow in the momentum space.  
 The width of the range and the strength of the anti-ferromagnetic correlation 
are reduced by increasing the electron-doping. 

 It is notable that the FLEX calculation is difficult untill the 
electron-doping $|\delta| \sim 0.12 $ because the anti-ferromagnetic 
correlation is too strong. 
 This fact implies that the anti-ferromagnetic order is robust in the 
electron-doped cuprates than in the hole-doped ones. 
 The narrow range of the spin fluctuations means that the mode coupling 
effects are weak in the electron-doped cuprates. This favors the 
anti-ferromagnetic order. 
 The other factors probably contribute to the robustness of the 
anti-ferromagnetism. For example, the anti-ferromagnetic order is robust 
when the three dimensionality (or the interlayer coupling) is strong. 
 Moreover, the frustration due to the carrier doping is weak when the 
carrier enters the Cu-site.

 Hereafter, we discuss the superconductivity and the pseudogap in the 
electron-doped cuprates. 
 The superconductivity mediated by the spin fluctuations is also derived by 
using the FLEX approximation. 
 The most favorable superconductivity is the $d_{x^{2}-y^{2}}$-wave. 
 However, the superconducting critical temperature is very low. The 
critical temperature $T_{{\rm c}}$ higher than $0.002$ is not obtained 
in case of the parameter $t'=0.25 t$. 
 This is mainly because the pairing interaction mediated by the spin 
fluctuations is weak in the electron-doped case. The narrow range of the 
spin fluctuations in the momentum space weakens the pairing interaction. 
 It should be noticed that the pairing interaction results from the 
spin fluctuations in the wide region around $\mbox{\boldmath$Q$} = (\pi,\pi)$. 
 In addition, the small DOS reduces the critical temperature much more. 
 Moreover, the Fermi surface in the electron-doped case 
is disadvantageous to the $d_{x^{2}-y^{2}}$-wave superconductivity, 
compared with the hole-doped case.

 We can obtain a higher critical temperature $T_{{\rm c}}$ by choosing the 
hopping parameter $t'$ so as to reproduce the Fermi surface of 
${\rm Nd}_{2-x}{\rm Ce}_{x}{\rm Cu}{\rm O}_{4-y}$ with more accuracy.  
 Here, we choose $t' = 0.35 t$ and $U=2.0$. 
 The commensurate spin fluctuations over a wider range are obtained 
(Figs. 14(c) and (d)).  
 The spin fluctuations give rise to the $d_{x^{2}-y^{2}}$-wave 
superconductivity. 
 However, the critical temperature is low and the doping range in which the 
superconductivity occurs is remarkably narrow compared with the hole-doped 
ones (see the inset of Fig. 15). 
 The maximum value of the obtained critical temperature is 
$T_{{\rm c}} = 0.0045$. The doping range is $\delta = -0.102 \sim -0.107$. 
 The results for the superconductivity are qualitatively consistent with 
the experimental results.~\cite{rf:electrondopesuper}

\begin{figure}[htbp]
\begin{center}
   \epsfysize=6.5cm
    $$\epsffile{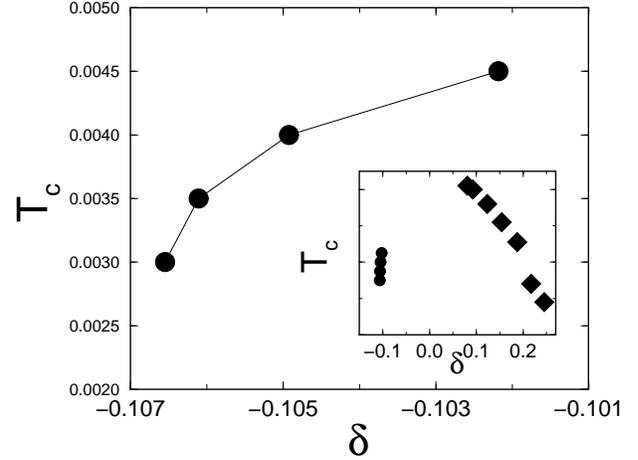}$$
    \caption{The obtained critical temperature $T_{{\rm c}}$ in the 
             electron-doped case. 
             The inset shows the phase diagram including both hole- and 
             electron-doped cases. 
             }
  \end{center}
\end{figure}

 We show the momentum dependence of the wave function of the Cooper pairs 
$\phi (\mbox{\boldmath$k$}, {\rm i} \omega_{n})$ in Fig. 16. 
 Fig. 16(a) clearly shows the $d$-wave superconductivity with the node 
in the diagonal direction. 
 It is a notable difference between the electron- and hole-doped cuprates 
that a wave function shows the sharp momentum dependence in the 
electron-doped case (Fig. 16(a)). 
 This is because the quasi-particle and the Fermi surface are 
more clearly defined in the electron-doped cases. 
 In the hole-doped cuprates, the many low energy states lie around  
$(\pi,0)$ where the quasi-particles are broad. 
 On the other hand, the Fermi surface is removed from $(\pi,0)$ and 
the electron correlation is effectively weak in the electron-doped case. 
 Therefore, the order parameter has a large value only in the vicinity of 
the Fermi surface.

 The effectively weak correlation is consistent with the $T$-square 
resistivity in the electron-doped cuprates.~\cite{rf:electrondopesuper} 
 Our calculation shows the $\omega$-square dependence of the imaginary 
self-energy 
${\rm Im} {\mit{\it \Sigma}}_{{\rm F}}^{{\rm R}}(\mbox{\boldmath$k$}, \omega)$
in the electron-doped case. 
 This is consistent with the experimental results~\cite{rf:takahashi} and 
implies the $T$-square resistivity. 
 The incommensurate spin fluctuations have been regarded to give rise to  
the above momentum dependence of the wave function 
in the electron-doped case.~\cite{rf:manske} However, 
it is actually not the main reason. Actually, the spin fluctuations are always 
commensurate in the electron-doped cuprates within our calculations.

 Thus, the comprehensive understanding for the phase diagram including the 
particle-hole asymmetry furthermore support 
the $d_{x^{2}-y^{2}}$-wave superconductivity in the electron-doped cuprates.

\begin{figure}[htbp]
\begin{center}
   \epsfysize=5.5cm
    $$\epsffile{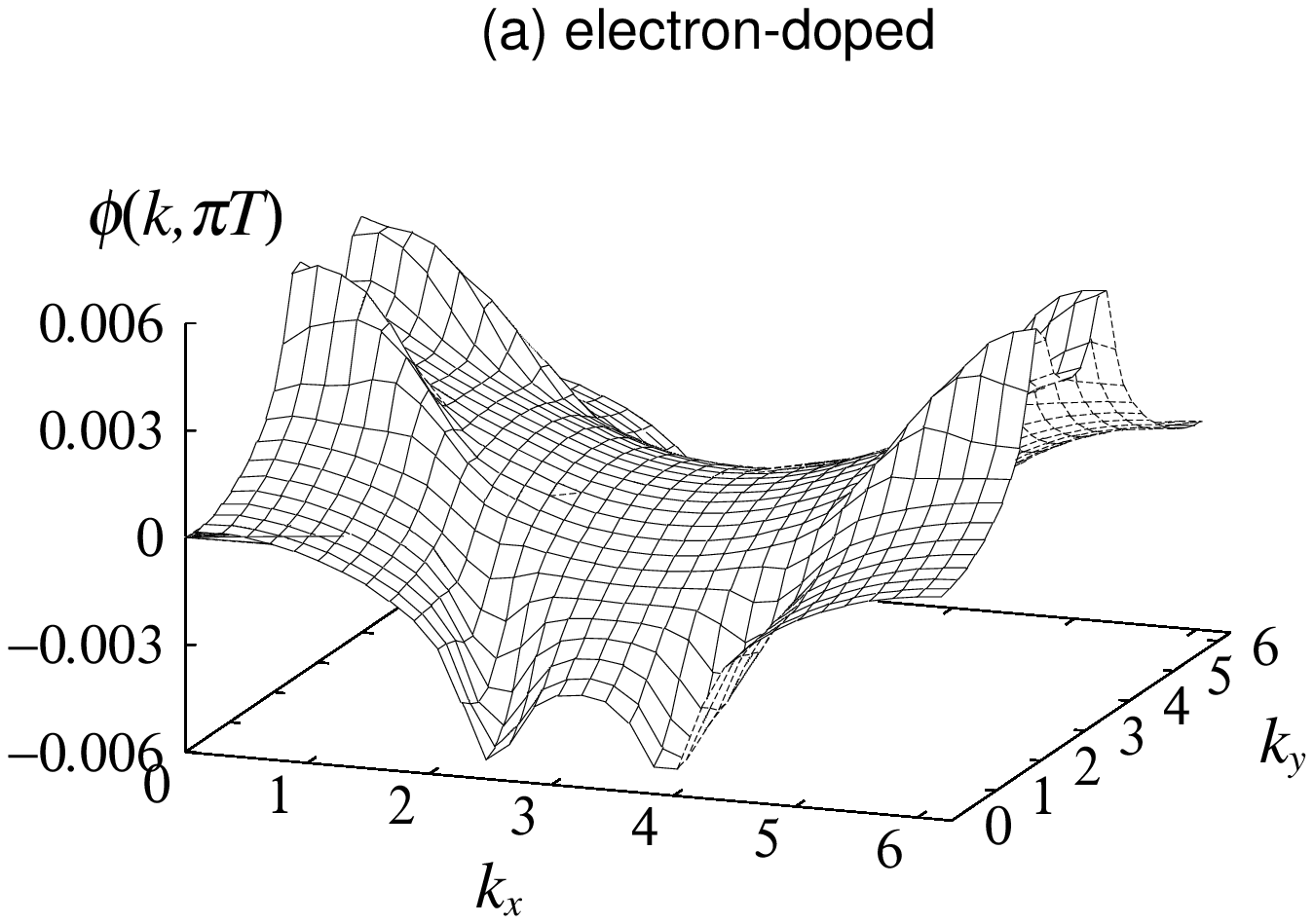}$$
   \epsfysize=5.5cm
    $$\epsffile{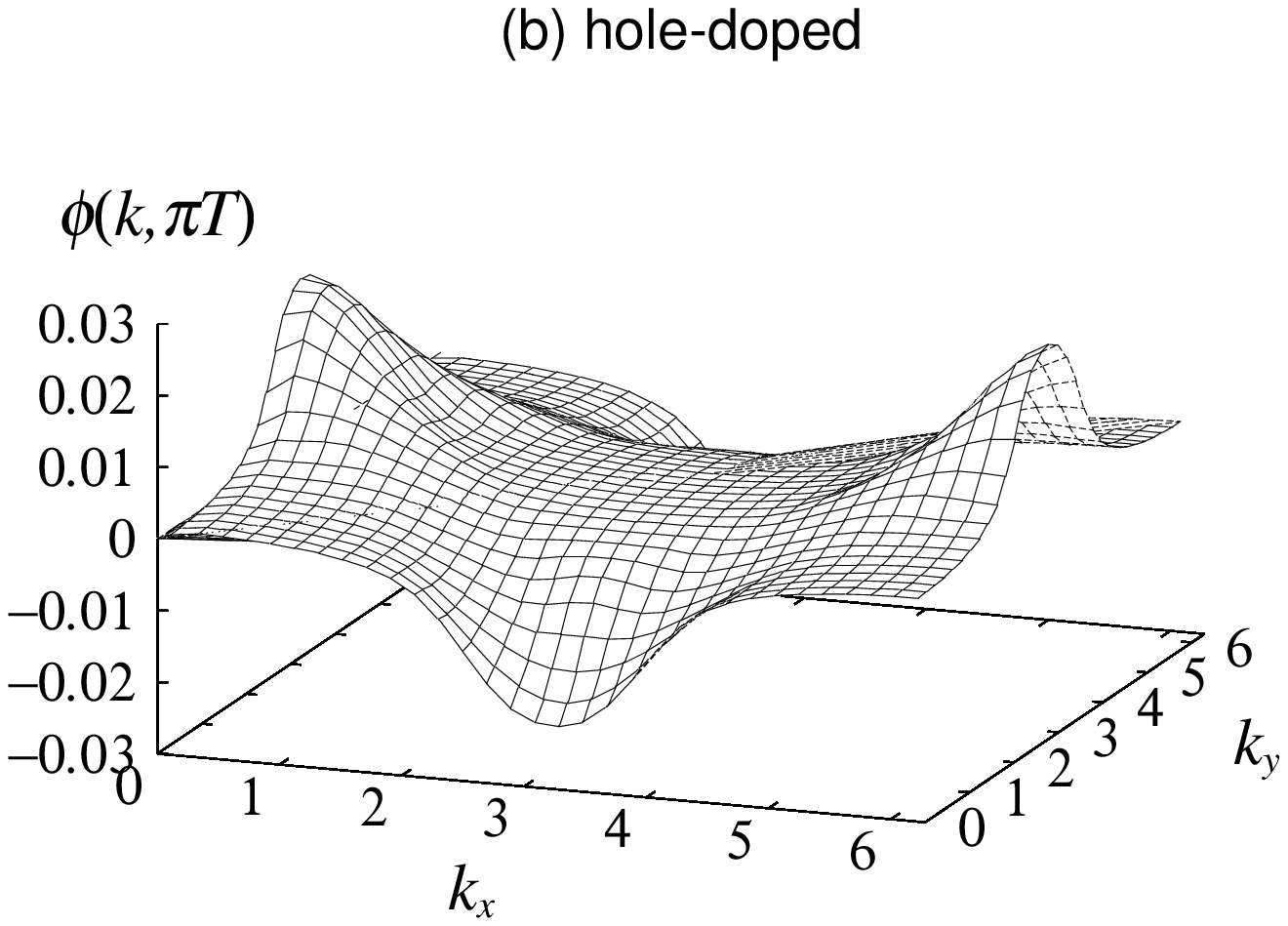}$$
    \caption{The momentum dependence of the wave function 
             $\phi (\mbox{\boldmath$k$}, {\rm i} \omega_{n})$ at 
             $\omega_{n} = \pi T$ 
             (a) in the electron-doped cuprates 
                ($\delta = -0.105$ and $T_{{\rm c}} = 0.0040$) and 
             (b) in hole-doped cuprates  
                ($\delta=0.093$ and $T_{{\rm c}}=0.0080$).
             }
  \end{center}
\end{figure}

 From the above results, we expect that the pseudogap phenomena are not so 
remarkable in the electron-doped cuprates. 
 Because of the low $T_{{\rm c}}$ and the weak electron correlation, 
the superconducting coupling $T_{{\rm c}}/\varepsilon_{{\rm F}}$ is small. 
 Actually, our calculation gives the large TDGL parameter $b = 30$ 
for the parameter set used in Fig. 16(a). 
 Therefore, the superconducting fluctuations are not so significant as well 
as in the standard BCS superconductor. 

 The calculated self-energy shows only the weak effects which are the same 
order as those in over-doped cuprates.  
 For example, the damping 
$- {\rm Im} {\mit{\it \Sigma}}^{{\rm R}}(\mbox{\boldmath$k$}, 0)$ at $T=0.005$ 
and $\delta=-0.104$ is about $1/40$ of that for the under-doped case 
in Fig. 6.  These results indicate that the pseudogap arising from the 
superconducting fluctuations is weak in the electron-doped cuprates, 
even if it is seen by some experiments. 
 The results are consistent with the experimental results of 
ARPES~\cite{rf:takahashi} and the neutron scattering~\cite{rf:kurahashi} 
which do not show the pseudogap in the electron-doped cuprates. 
 The weak effects give rise to the slight pseudogap in our calculation. 
 However, the finite size effect seriously overestimates the effects of 
the superconducting fluctuations because of the large $b$. 
 Therefore, the effects are precisely much weaker.

 Thus, our calculation explains the doping dependence of 
High-$T_{{\rm c}}$ cuprates including the particle-hole asymmetry.  
 The comprehensive understanding including both the hole- and electron-doped 
systems rather supports the pairing scenario for the pseudogap.

\section{Self-Consistent Calculation}

 In this section, we carry out the self-consistent calculation including the 
spin fluctuations, superconducting fluctuations and the single-particle 
properties. In the self-consistent calculation (We call the SC-FLEX+T-matrix 
calculation), eqs. (2.3)-(2.8) and (3.5)-(3.9) are solved self-consistently 
where the fully dressed Green function  
$G (\mbox{\boldmath$k$}, {\rm i} \omega_{n}) 
 = ( {\rm i} \omega_{n} - \varepsilon_{\mbox{{\scriptsize \boldmath$k$}}} - 
{\mit{\it \Sigma}} (\mbox{\boldmath$k$}, {\rm i} \omega_{n}))^{-1} $ is used. 
 By the self-consistent calculation, we calculate the critical temperature 
$T_{{\rm c}}$ reduced by the superconducting fluctuations. 
 As is described in the previous papers,~\cite{rf:yanasePG,rf:yanaseSC} 
there exists the singularity arising from the two-dimensionality. 
 The singularity is actually removed by the 
weak three dimensionality which surely exists in the real systems. 
 Therefore, we determine the critical temperature $T_{{\rm c}}$ as the 
temperature at which $\lambda(\mbox{\boldmath$0$}, 0) = 0.98$ in order to 
avoid the unphysical singularity. 
 This determination corresponds to that the dimensional crossover from the 
two-dimension to the three-dimension occurs around 
$\lambda(\mbox{\boldmath$0$}, 0) = 0.98$. 
 The finite critical temperature is obtained by this operation. 
The method of the determination makes no significant difference 
on the following results. 
 For example, the critical temperature is not so different 
(about $\sim 10\%$ lower) 
even if we choose $\lambda(\mbox{\boldmath$0$}, 0) = 0.99$ 
as the critical point. 
 In particular, the similar doping dependence is obtained. 
 Our choice of the parameter $\lambda(\mbox{\boldmath$0$}, 0) = 0.98$ is 
owing to the reliability of the numerical calculation. 
 Hereafter, we consider the hole-doped cases.

\subsection{Spin fluctuation and superconducting fluctuation}

 First, we clarify the relation between the spin fluctuations and the 
superconducting fluctuations. The two fluctuations complicatedly couple to 
each other through the single particle properties. 
 Here, we digest the important factors of the relations. 

 The superconducting fluctuations result from 
the spin fluctuations which act as the $d$-wave attractive interaction. 
 The spin fluctuations give rise to the renormalization of the quasi-particles 
especially around $(\pi,0)$. The renormalization reduces the effective Fermi 
energy $\varepsilon_{{\rm F}}$ for the $d$-wave superconductivity as well as 
the critical temperature $T_{{\rm c}}$. Therefore, the strong superconducting 
fluctuations are obtained at the reasonable temperature. 

 The effects of the superconducting fluctuations on the spin fluctuations have 
been investigated in \S3.4. The superconducting fluctuations suppress the low 
frequency spin fluctuations and the anti-ferromagnetic order (see Fig. 11). 
 A residual question is whether the feedback effects of the pseudogap on the 
spin fluctuations suppress the superconductivity or not. 

 In order to answer the question, we calculate the feedback effect on the 
critical temperature $T_{{\rm c}}$.  
 We carry out the Modified FLEX (M-FLEX) calculation in which the fully dressed
Green function is used only in eq. (2.6). In the other equations 
$G^{{\rm F}} (\mbox{\boldmath$k$}, {\rm i} \omega_{n})
 = ( {\rm i} \omega_{n} - \varepsilon_{\mbox{{\scriptsize \boldmath$k$}}} - 
{\mit{\it \Sigma}}_{{\rm F}} (\mbox{\boldmath$k$}, {\rm i} \omega_{n}))^{-1} $ 
is used. Thus, the effects of the superconducting fluctuations on the spin 
fluctuations are included, however those on the single particle properties are 
not included. 
 The eqs. (2.3)-(2.8) and (3.5)-(3.9) are solved self-consistently 
in the M-FLEX approximation. 
 The results of the FLEX, M-FLEX and SC-FLEX+T-matrix calculations are 
shown in Table. I. 
 Here, we determine the critical temperature by the condition 
$\lambda(\mbox{\boldmath$0$}, 0) = 0.98$ for an equity.

\begin{table}[htb]
  \begin{center}
\begin{tabular}{|c||c|c|c|} \hline
&FLEX&M-FLEX&SC-FLEX+T-matrix \\ 
\hline 
$T_{{\rm c}}$ & $0.0084$ & $0.0098$ & $0.0031$ \\ 
\hline
$g \phi_{{\rm max}}^{2}$ & $24.18$  & $14.23$ & 20.1529  \\
\hline
$\gamma_{{\rm h}}$ & $0.06970$ & $0.02747$ & 0.04277       \\
\hline 
$\gamma_{{\rm c}}$ & $0.00995$ & $0.00971$ &  0.00298      \\
\hline 
\end{tabular}
 \caption{The comparison among the FLEX, M-FLEX and SC-FLEX+T-matrix 
          approximations. The critical temperature $T_{{\rm c}}$, 
          the effective pairing interaction $g \phi_{{\rm max}}^{2}$,  
          the damping at 'Hot spot' $\gamma_{{\rm h}}$ and that at 
          'Cold spot'$\gamma_{{\rm c}}$ at $T=T_{{\rm c}}$ are shown. 
          The parameters are $U=1.6$ and $\delta = 0.083 \sim 0.096$.  
          Here, the self-energy 
          in the M-FLEX approximation is that due to the spin fluctuations 
        ${\mit{\it \Sigma}}_{{\rm F}}^{{\rm R}}(\mbox{\boldmath$k$}, \omega)$.
          } 
  \end{center}
\end{table}

 In order to understand the results, it is important that the 
spin fluctuations have not only the paring effect but also 
the de-pairing effect. The former is from the relatively wide frequency region,
and the latter is from the low frequency component. The pseudogap remarkably 
suppresses the low frequency component, however the total weight is not so 
reduced (Fig. 11). Thus, the pseudogap reduces the de-pairing effect 
rather than the pairing effect. Therefore, the higher critical temperature 
$T_{{\rm c}} = 0.0098$ is obtained by the M-FLEX calculation where 
$T_{{\rm c}} = 0.0084$ is obtained by the FLEX calculation. 
 In other words, the feedback effects are advantageous to the 
superconductivity. 
 Therefore, the feedback effects do not suppress the superconducting 
fluctuations and the pseudogap. 
 Thus, only the properties of the low frequency component 
are not sufficient in order to understand the relation between the spin and 
superconducting fluctuations.

 Needless to say, the pseudogap in the single particle properties 
reduces the critical temperature. The lower critical temperature 
$T_{{\rm c}} = 0.0031$ is obtained by the SC-FLEX+T-matrix calculation. 
 The de-paring effect from the superconducting fluctuations is rather 
drastic than that from the spin fluctuations. (Therefore, the pseudogap is 
easily caused by the superconducting fluctuations.) 

 In order to make the above understanding clear, we show the quantities 
$g \phi_{{\rm max}}^{2}$, $\gamma_{{\rm h}}$ and $\gamma_{{\rm c}}$ 
in Table. I. 
 Here, the quantity $\phi_{{\rm max}}$ is the maximum value of the order 
parameter $\phi(\mbox{\boldmath$k$}, {\rm i} \omega_{n})$ and the quantity 
$g \phi_{{\rm max}}^{2}$ represents the strength of the pairing interaction. 
 The damping at the 'Hot spot' $\gamma_{{\rm h}} = 
- {\rm Im} {\mit{\it \Sigma}}^{{\rm R}}(\mbox{\boldmath$k$}_{{\rm h}}, 0)$ 
at $T=T_{{\rm c}}$ represents the strength of the de-pairing effect. Here, 
$\mbox{\boldmath$k$}_{{\rm h}} = (0.98\pi, 0.02\pi)$. 
 In should be noticed that the obtained value 
$\gamma_{{\rm h}}$ by the M-FLEX calculation has the minimum value in 
Table. I, although the temperature is highest. This fact shows that 
the de-pairing effect of the spin fluctuations are suppressed by the 
superconducting fluctuations.

 The damping at the 'Cold spot' $\gamma_{{\rm c}} = 
- {\rm Im} {\mit{\it \Sigma}}^{{\rm R}}(\mbox{\boldmath$k$}_{{\rm c}}, 0)$ 
determines the in-plane transport.~\cite{rf:stojkovic,rf:yanaseTR,rf:ioffe} 
Here, $\mbox{\boldmath$k$}_{{\rm c}} = (0.45\pi, 0.42\pi)$. 
 The Table. I shows that the value $\gamma_{{\rm c}}$ is not so reduced in  
the M-FLEX calculation from the FLEX calculation. 
 Thus, the feedback effects through the spin fluctuations are small around 
the 'Cold spot'. 
 Since the self-energy from the superconducting fluctuations 
${\mit{\it \Sigma}}_{\rm S}^{{\rm R}}(\mbox{\boldmath$k$}_{{\rm c}}, \omega)$ 
vanishes at the 'Cold spot', we can understand that the effect of the 
pseudogap on the in-plane transport is small,~\cite{rf:yanaseSC,rf:yanaseTR} 
as is observed experimentally.~\cite{rf:transport}  
 The slope of the T-linear resistivity is slightly reduced by the feedback 
effects. 
 The anisotropy $\gamma_{{\rm h}}/\gamma_{{\rm c}}$ increases due to the 
superconducting fluctuations.  
 This is important for the incoherent {\it c}-axis transport in the pseudogap 
state because the {\it c}-axis transport is determined by the 
'Hot spot'.~\cite{rf:yanaseTR,rf:ioffe}

\subsection{Results of the SC-FLEX+T-matrix calculation}

 In this subsection, we show the results of the SC-FLEX+T-matrix 
approximation. The qualitatively similar results to the FLEX+T-matrix 
approximation are obtained, although the effects of the superconducting 
fluctuations are reduced by the self-consistency. 
 It is notable that the calculation treating a stronger electron-electron 
interaction $U$ is possible because the superconducting fluctuations suppress 
the anti-ferromagnetic order. 
 Here, we choose the interaction $U=2.4$. 

 In Fig. 17, we show the self-energy calculated for the under-doped case 
$\delta=0.073$. 
 The competition between the pseudogap and the Fermi liquid behavior is shown. 
 The small imaginary part and the positive slope of the real part are obtained
in the narrow region around the Fermi level.  
 The anomalous features due to the superconducting fluctuations are shown 
in larger energy scale. 
 These features are qualitatively the same as those obtained within the model 
with an attractive interaction.~\cite{rf:yanasePG} 
 It should be noticed that these behaviors are given 
by the self-consistency between the superconducting fluctuations and the 
single particle properties. 
 The necessity of the behaviors has been explained in ref. 19.

\begin{figure}[htbp]
\begin{center}
   \epsfysize=6.5cm
    $$\epsffile{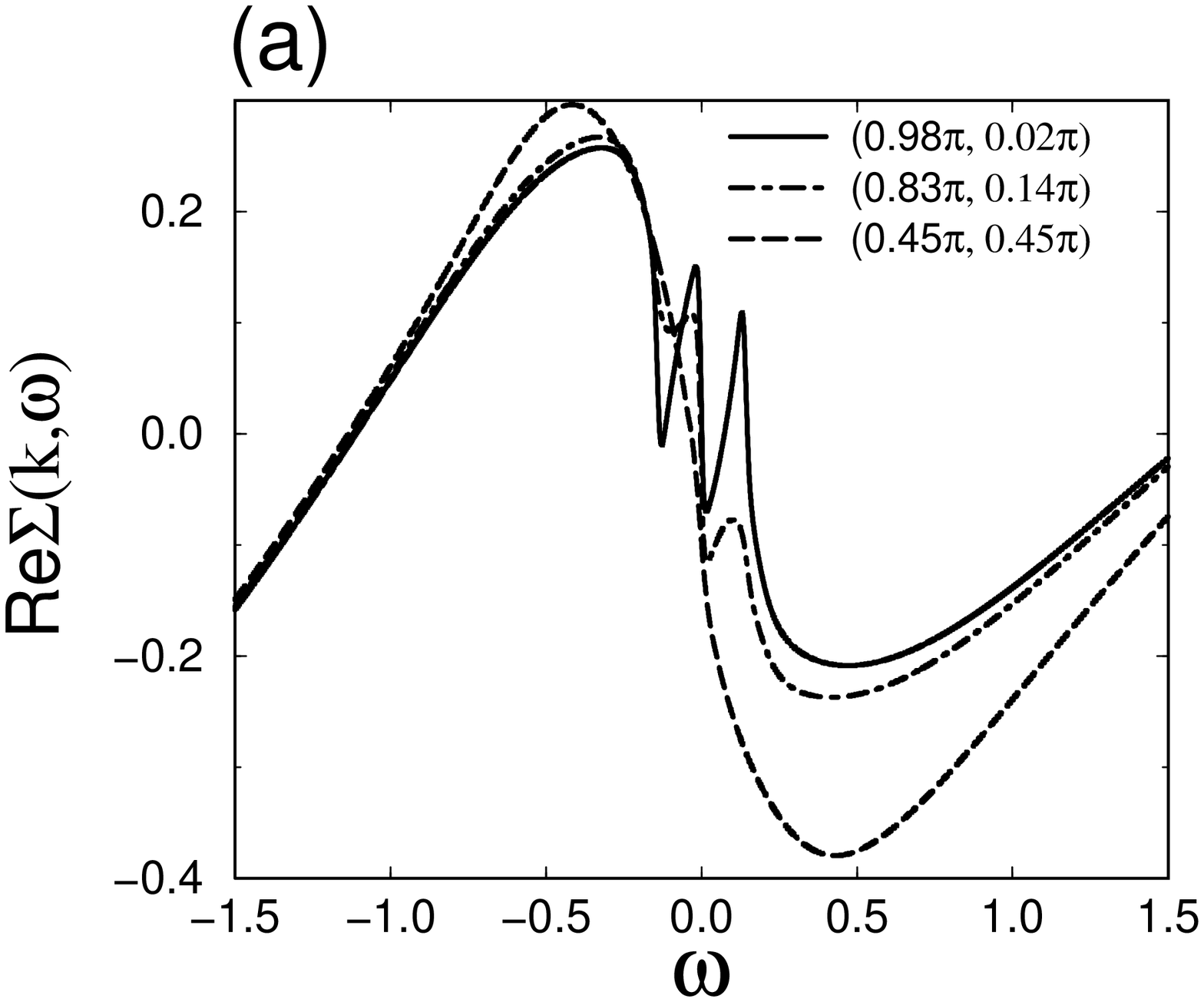}$$
   \epsfysize=6.5cm
    $$\epsffile{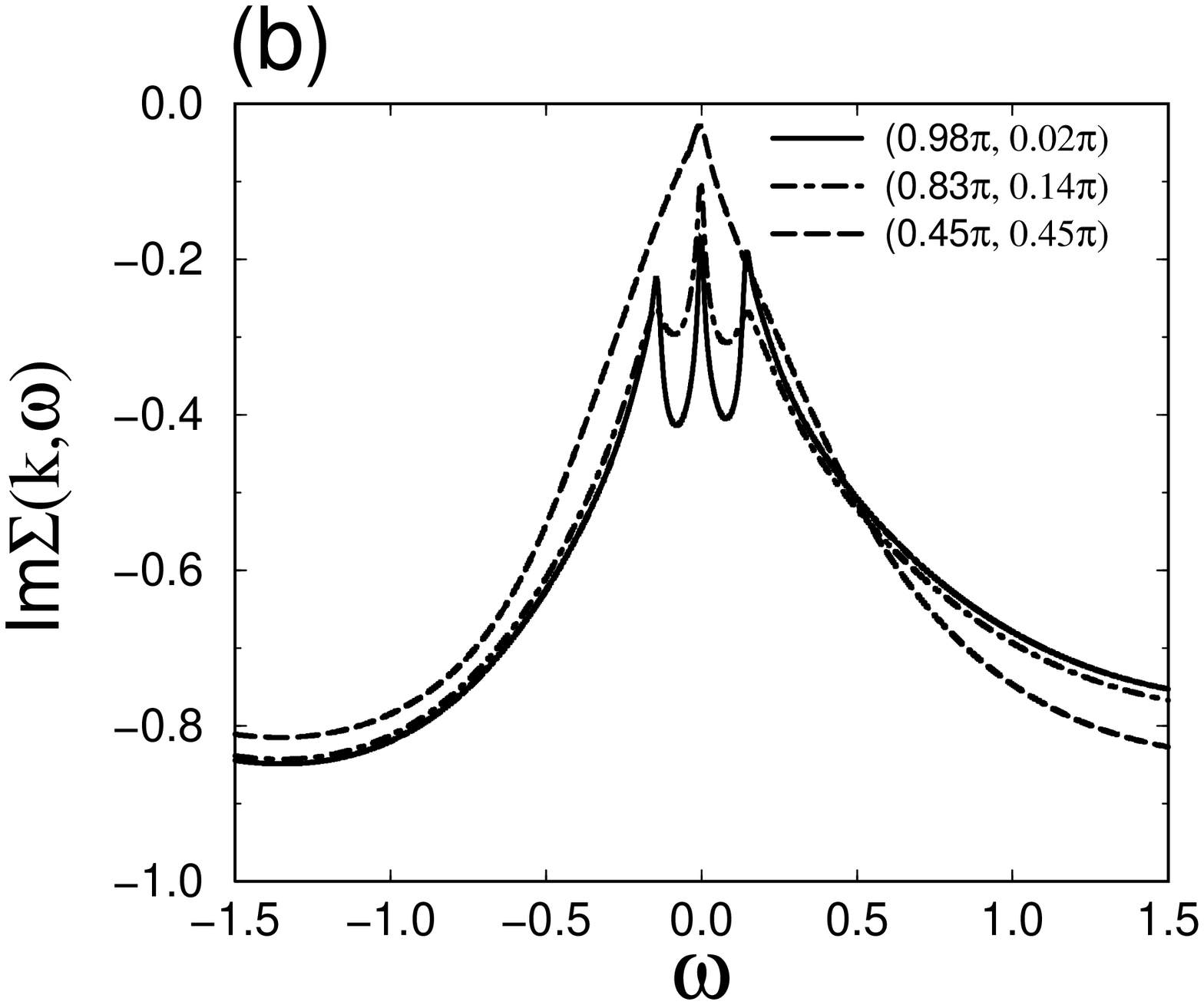}$$
    \caption{The self-energy obtained by the SC-FLEX+T-matrix approximation. 
             (a) The real part. 
             (b) The imaginary part. 
             Here, parameters are $U=2.4$, $\delta=0.073$ and $T=0.004$. 
             The solid, dash-dotted and long-dashed lines are correspond to 
             $(\frac{63}{64}\pi,\frac{1}{64}\pi)$, 
             $(\frac{53}{64}\pi,\frac{9}{64}\pi)$ and 
             $(\frac{29}{64}\pi,\frac{29}{64}\pi)$, respectively. 
             }
  \end{center}
\end{figure}

 The obtained spectral weight is shown in Fig. 18. 
 The peak is shown near the Fermi level at the momentum around $(\pi,0)$. 
 This reflects the Fermi liquid behavior of the self-energy in Fig. 17. 
 The typical three peak structure shown in the under-doped 
case $\delta = 0.073$ (Fig. 18(a)) is a common feature to the results  
based on the model with an attractive interaction.~\cite{rf:yanasePG} 
 It should be noticed that the weight around the Fermi level is remarkably 
reduced by the superconducting fluctuations 
which give the large imaginary part of the self-energy around the Fermi level. 
 The spectral weight at the low energy region shifts to the 
high frequency region. 
 The suppression of the spectral weight near the Fermi level corresponds to 
the pseudogap.

 The small weight near the Fermi level may not be observed by ARPES because 
the weight is so small. 
 The ARPES measurements have their resolving power
about the momentum and the energy. 
 The thermal broadening arising from the Fermi distribution function 
$f(\omega)$ affects the ARPES furthermore. 
 Anyway, the suppression of the spectral weight near the Fermi surface 
is a reliable result and consistent with the calculated result.

 As is shown in the inset of Fig. 18(a), the usual single peak structure is 
obtained by neglecting the self-energy due to the superconducting 
fluctuations 
${\mit{\it \Sigma}}_{{\rm S}}(\mbox{\boldmath$k$}, {\rm i} \omega_{n})$, 
namely $G (\mbox{\boldmath$k$}, {\rm i} \omega_{n}) = 
( {\rm i} \omega_{n} - \varepsilon_{\mbox{{\scriptsize \boldmath$k$}}} - 
{\mit{\it \Sigma}}_{{\rm F}} (\mbox{\boldmath$k$}, {\rm i} \omega_{n}))^{-1} $.
 The spectral weight is recovered and the three peak structure vanishes 
when the momentum leaves $(\pi,0)$ along the Fermi surface. 
 Similarly, the effects of the superconducting fluctuations are reduced by the 
hole-doping (Fig. 18(b)). The single peak structure is obtained in the 
over-doped case $\delta = 0.165$ (the inset in Fig. 18(b)). 
 In other words, the pseudogap is reduced by the hole-doping because the 
superconducting fluctuations are suppressed.

\begin{figure}[htbp]
\begin{center}
   \epsfysize=6.5cm
    $$\epsffile{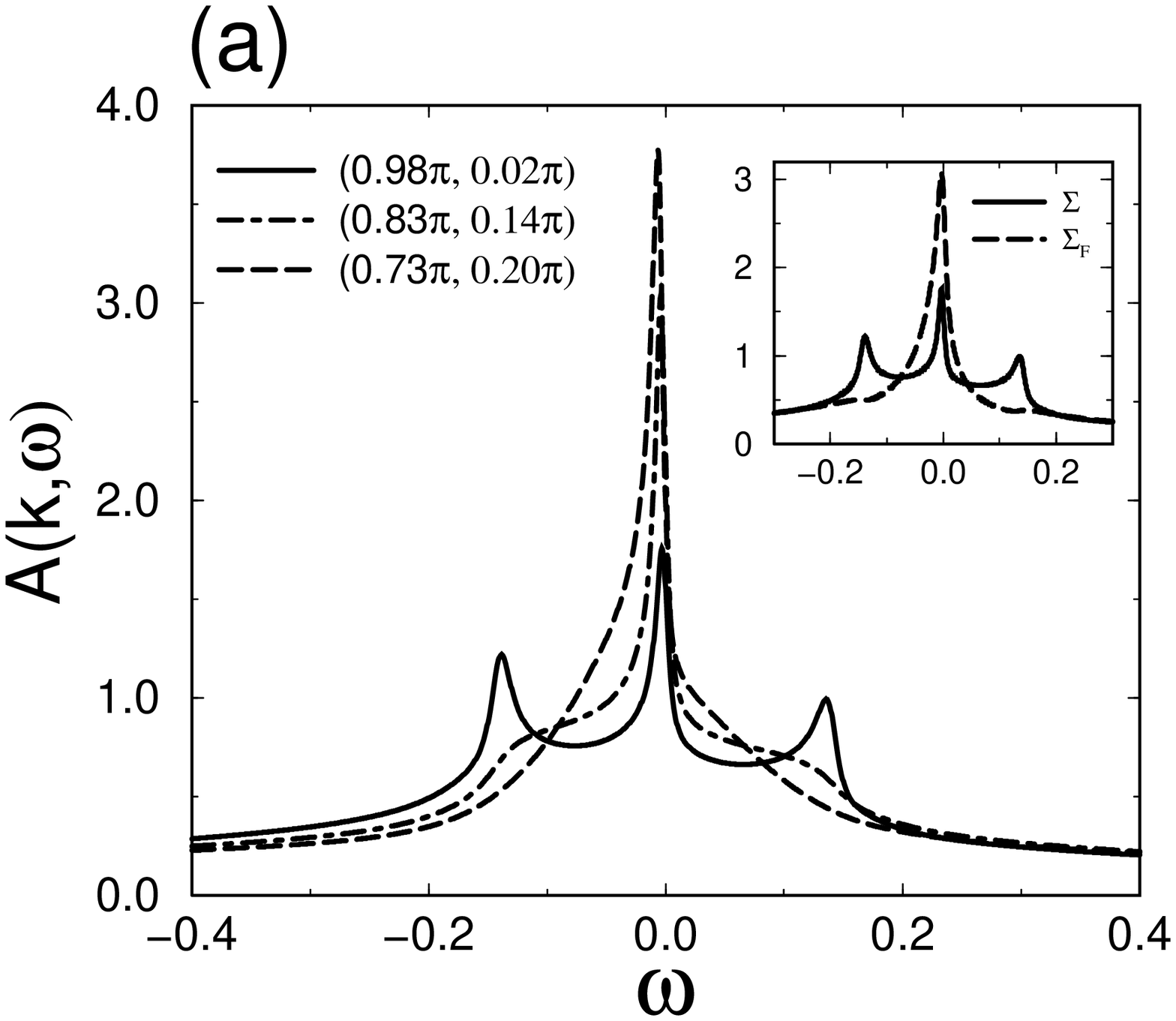}$$
   \epsfysize=6.5cm
    $$\epsffile{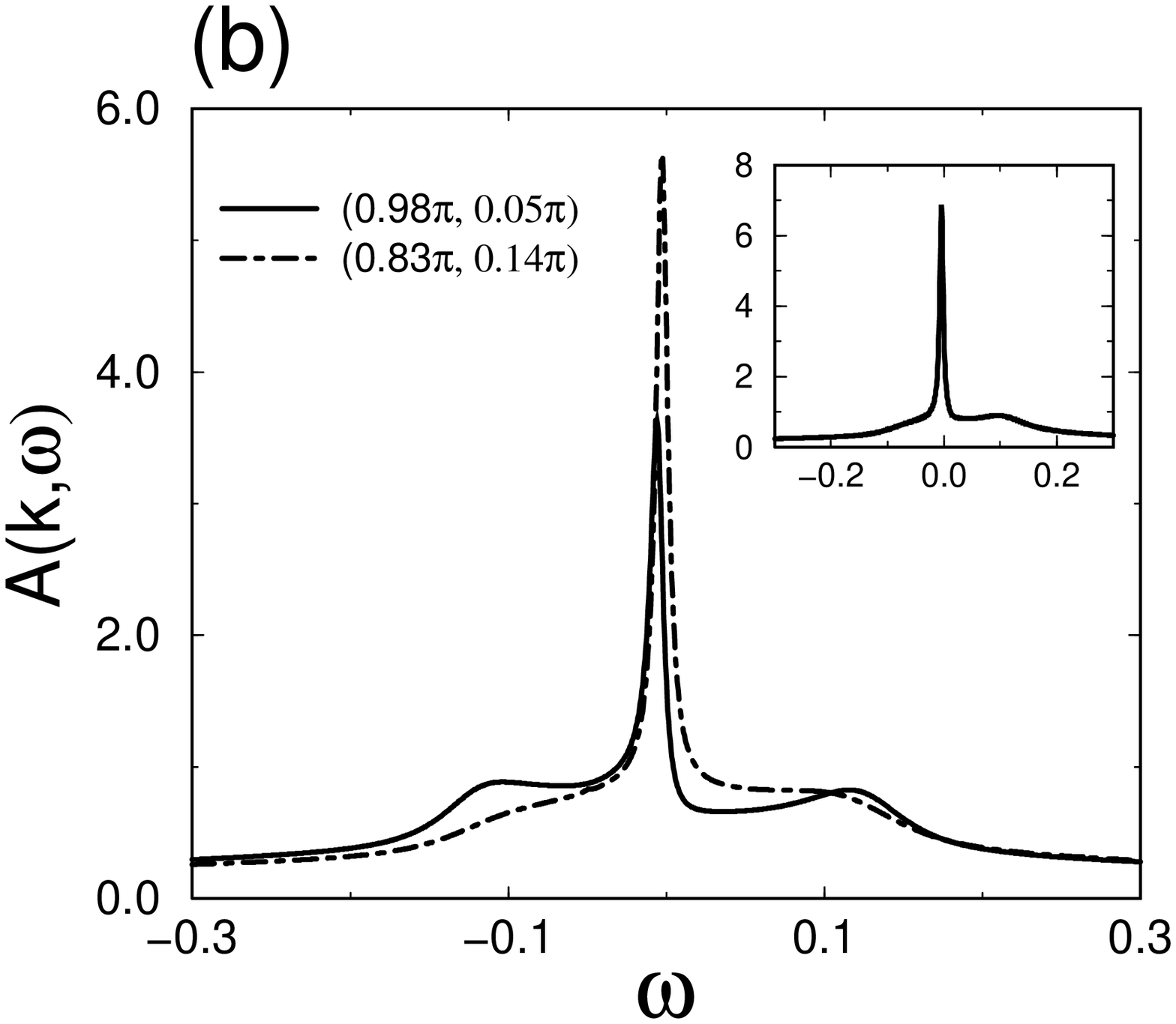}$$
    \caption{The single particle spectral weight obtained by the 
             SC-FLEX+T-matrix approximation. 
             (a) The under-doped case, $\delta=0.073$ and $T=0.004$. 
                 Here, $T_{{\rm c}}=0.0032$. 
                 The solid, dash-dotted and long-dashed lines are correspond to
                 $(\frac{63}{64}\pi,\frac{1}{64}\pi)$, 
                 $(\frac{53}{64}\pi,\frac{9}{64}\pi)$ and 
                 $(\frac{47}{64}\pi,\frac{13}{64}\pi)$, respectively. 
                 In the inset, the solid line shows the same result as in the 
                 main figure. The long-dashed lines show the result 
                 by neglecting the effects of the superconducting fluctuations 
${\mit{\it \Sigma}}_{{\rm S}}(\mbox{\boldmath$k$}, {\rm i} \omega_{n})$.  
             (b) The optimally-doped case, $\delta=0.119$ and $T=0.004$. 
                 Here, $T_{{\rm c}}=0.0035$. 
                 The solid and dash-dotted lines correspond to 
                 $(\frac{63}{64}\pi,\frac{3}{64}\pi)$ and  
                 $(\frac{53}{64}\pi,\frac{9}{64}\pi)$, respectively.  
                 The inset shows the result for the over-doped case, 
                 $\delta=0.165$, $T=0.0035$ and 
                 $\mbox{\boldmath$k$} = (\frac{63}{64}\pi,\frac{5}{64}\pi)$. 
             }
  \end{center}
\end{figure}

 The above effect of the superconducting fluctuations becomes more clear 
by showing the DOS in Fig. 19. 
 The DOS near the Fermi level is reduced by the superconducting 
fluctuations and the gap structure appears in the under-doped case 
(solid line). 
 It is confirmed by showing the result in which 
${\mit{\it \Sigma}}_{{\rm S}}(\mbox{\boldmath$k$}, {\rm i} \omega_{n})$ is 
neglected (long-dashed line) that the pseudogap in the DOS is caused 
by the superconducting fluctuations. 
 The pseudogap is suppressed by the hole-doping, similarly (dash-dotted line). 
 Thus, the pseudogap is properly obtained by the self-consistent calculation
including their doping dependence.

\begin{figure}[htbp]
\begin{center}
   \epsfysize=6.5cm
    $$\epsffile{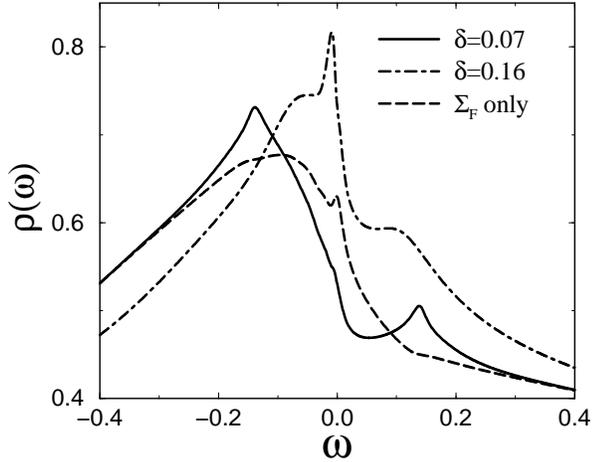}$$
    \caption{The DOS obtained by the 
             SC-FLEX+T-matrix approximation. 
                 The solid and dash-dotted lines show the under-doped case 
                 ($\delta=0.073$ and $T=0.004$) and the over-doped case 
                 ($\delta=0.165$ and $T=0.0035$), respectively. 
                 The long-dashed line shows the result for the under-doped 
                 case obtained by neglecting the self-energy 
${\mit{\it \Sigma}}_{{\rm S}}(\mbox{\boldmath$k$}, {\rm i} \omega_{n})$.  
             }
  \end{center}
\end{figure}

 At last, we show the obtained phase diagram in Fig. 20. 
 The superconducting critical temperature $T_{{\rm c}}$ suppressed 
by the superconducting fluctuations is calculated 
by the self-consistent calculation. 
 It should be noticed that the suppression of $T_{{\rm c}}$ from the mean 
field value becomes remarkable with under-doping. 
 This is a natural result because the 
superconducting fluctuations become strong with under-doping. 
 In other words, the pseudogap develops with under-doping and the reduced DOS 
gives the reduced critical temperature. 
 An important result is the following; 
 The critical temperature has the maximum value 
at $\delta \sim 0.11$ and decreases with under-doping in the SC-FLEX+T-matrix 
calculation for $U=2.4$, whereas $T_{{\rm c}}$ goes on increasing in the 
FLEX calculation. 
 In other words, the mean field critical temperature $T_{{\rm c}}^{\rm MF}$ 
develops with under-doping. 
 However, the decreasing $T_{{\rm c}}$ in the under-doped region is obtained 
by considering the superconducting fluctuations.

\begin{figure}[htbp]
\begin{center}
   \epsfysize=7cm
    $$\epsffile{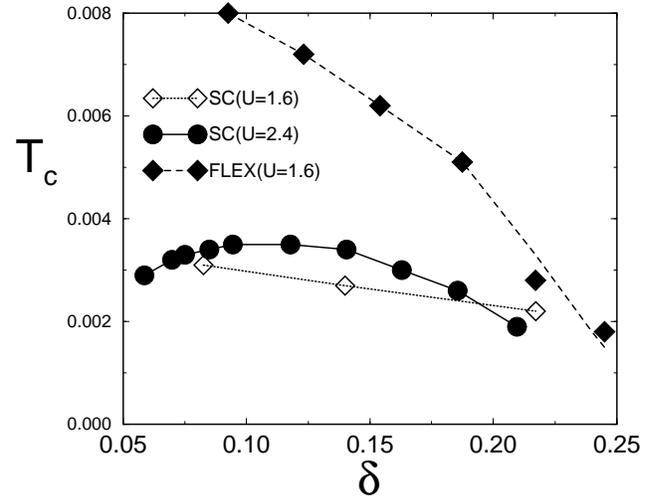}$$
    \caption{The phase diagram obtained by the FLEX and the 
             SC-FLEX+T-matrix approximations. 
             The closed circles show the results of the SC-FLEX+T-matrix 
             approximation for $U=2.4$. 
             The closed and open diamonds show the results of the FLEX and the 
             SC-FLEX+T-matrix approximations for $U=1.6$. 
             } 
  \end{center}
\end{figure}

 It is notable to write again that the strength of the superconducting 
coupling is indicated by the ratio 
$T_{{\rm c}}^{\rm MF}/\varepsilon_{{\rm F}}$, and not by $T_{{\rm c}}$. 
 Since the effective Fermi energy $\varepsilon_{{\rm F}}$ decreases, 
 the superconducting coupling becomes strong with under-doping in spite of the 
decreasing $T_{{\rm c}}$.

 We can see from Fig. 20 that the critical temperature for $U=1.6$ does not 
decrease with under-doping even in the SC-FLEX+T-matrix calculation. 
 Thus, the strong renormalization of the quasi-particles due to the 
strong electron correlation plays an important role for describing the 
under-doped cuprates. The other important thing is that we can treat the 
strong correlation by considering the superconducting fluctuations which 
suppresses the anti-ferromagnetic order. 
 If neglecting the superconducting fluctuations, the system rapidly approaches 
the magnetic order.  
 Because the magnetic order is suppressed by the superconducting 
fluctuations, the situation with the strong anti-ferromagnetic 
spin fluctuations persists in the under-doped region. 
 This fact also contributes to the phase diagram shown in Fig 20. 

 The enhancement of the electron correlation effects with under-doping 
is necessary in order to obtain the phase diagram in which the critical 
temperature $T_{{\rm c}}$ decreases with under-doping. 
 This doping dependence is probably underestimated in the FLEX approximation. 
 The limitation of our calculation mainly arises from this fact. 
 However, the doping dependence is partially included even in the 
FLEX approximation. Therefore, we can show the under-doped region in Fig. 20 
although the decrease of the critical temperature $T_{{\rm c}}$ 
is not sufficient. 
 This fact implies that the more sophisticated calculation which describes 
the stronger renormalization will gives the more decrease of the critical 
temperature $T_{{\rm c}}$. 
 Anyway, the superconducting fluctuations play an important role in describing 
the under-doped region in the phase diagram.

\section{Conclusion and Discussion}

 In this paper, we have investigated the pseudogap phenomena starting 
from the Hubbard model. 
 The superconducting fluctuations have been derived microscopically
from the electron-electron correlation. 
 We have succeeded in deriving the pseudogap phenomena by using the 
microscopic calculation. 

 The renormalized quasi-particles and the pairing interaction via the 
anti-ferromagnetic spin fluctuations are calculated by the FLEX approximation. 
 The anti-ferromagnetic spin fluctuations give rise to the strong 
renormalization around the 'Hot spot'. Therefore, the effective Fermi energy 
$\varepsilon_{{\rm F}}$ for the $d$-wave symmetry is remarkably renormalized. 
 Because of the high critical temperature $T_{{\rm c}}$ and the renormalized 
Fermi energy $\varepsilon_{{\rm F}}$, 
the superconducting coupling $T_{{\rm c}}/\varepsilon_{{\rm F}}$ 
becomes strong in the under-doped region. 
 As a result, the pseudogap phenomena occurs owing to the self-energy 
correction by the strong superconducting fluctuations. 
 The results support the scenario which is described within the model with 
a $d$-wave attractive 
interaction.~\cite{rf:yanasePG,rf:jujo,rf:jujoyanase,rf:yanaseSC,rf:yanaseMG}

 An important progress is that the pseudogap is described as a phenomenon 
near the Fermi surface. 
 Because the obtained pairing interaction affects only the quasi-particles 
near the Fermi surface, it is naturally described that the superconducting 
fluctuations and the pseudogap have the small energy scale compared with the 
electron correlation. 
 These results support the picture in which the pairing interaction becomes 
effectively strong for the renormalized quasi-particles.

 Moreover, it should be emphasized that the calculation in this paper properly 
describes the doping dependence of the pseudogap phenomena. 
 The superconducting coupling becomes weak with increasing the hole-doping, 
since the Fermi energy increases and the critical temperature decreases.  
 Therefore, the superconducting fluctuations become insignificant 
with hole-doping. 
 The obtained doping dependence of the TDGL parameter $b$ has confirmed the 
scenario and is consistent with the the magnetic field dependence of the 
pseudogap phenomena.~\cite{rf:yanaseMG}  

 The application of the calculation to the electron-doped case gives 
a consistent understanding with the experimental 
results.~\cite{rf:electrondopesuper,rf:electron-dope_D,rf:hayashi,rf:takahashi,rf:tsuei,rf:kurahashi}
 We have obtained the $d_{x^{2}-y^{2}}$-wave superconductivity 
with low critical temperatures within the narrow doping range. 
 It is shown that the superconducting fluctuations are weak and the pseudogap 
is not observed in the electron-doped case. 
 The difference from the hole-doped case is derived from the particle-hole 
asymmetry of the band structure. 
 The comprehensive understanding including the particle-hole asymmetry 
furthermore supports our scenario for High-$T_{{\rm c}}$ cuprates. 
 
 The pseudogap in the magnetic properties has been investigated and 
explained properly. We have calculated the quantities observed by the NMR 
and the neutron scattering. The obtained results have naturally explained the 
characteristics of each quantities in the pseudogap state 
which are subtly different from each other. 

 We have clarified the relation between the spin fluctuations and the 
superconducting fluctuations which are complicatedly connected with 
each other. It is shown that the feedback effects of the superconducting 
fluctuations on the spin fluctuations enhance the superconductivity. 
 Therefore, the feedback effect does not suppress 
the superconducting fluctuations. 

 Of course, the superconducting fluctuations themselves reduce the critical 
temperature. 
 We have calculated the reduced critical temperature by the self-consistent 
calculation including the single particle properties and the spin- and 
superconducting fluctuations. 
 Qualitatively the similar results for the pseudogap have been obtained by the 
self-consistent calculation. 
 The calculated critical temperature shows the maximum near 
$\delta \sim 0.11$, and decreases with under-doping, although it keeps 
increasing in the FLEX calculation. 
 Thus, we have succeeded in describing the under-doped region by considering 
the strong superconducting fluctuations in the strongly correlated 
electron systems.

 Finally, it should be stressed that the calculation in this paper starts 
from the Fermi liquid state and properly describes the High-$T_{{\rm c}}$ 
cuprates including the under-doped region. 
 The comprehensive understanding of High-$T_{{\rm c}}$ cuprates has been 
obtained including their doping dependence from the over-doped region to the 
under-doped one.

 The important and remained problem is the description of the under-doped 
limit. The calculation in this paper can not be applied to the limit 
because the FLEX approximation does not describe the Mott transition. 
 The FLEX approximation probably underestimates the electron correlation and 
overestimates the anti-ferromagnetic correlation. 
 Therefore, the anti-ferromagnetic correlation may be too strong 
in the under-doped region in this paper. 
 The momentum independent component of the electron 
correlation is probably stronger and the anti-ferromagnetic correlation is 
less strong in the actual state. 
 The calculation beyond the FLEX approximation is desirable in order to 
describe the under-doped region more precisely.

 However, it is notable that the essence of the pseudogap phenomena 
in High-$T_{{\rm c}}$ cuprates has been explained in this paper 
on the basis of the strong superconducting fluctuations. 
 It is a strong evidence for the pairing scenario that the sufficiently 
strong superconducting fluctuations and the pseudogap phenomena are derived 
from the Hubbard model.

\section*{Acknowledgements}

 The authors are grateful to Dr. S. Koikegami for providing the program 
of the FLEX calculation, and to Professor M. Ogata, Professor K. Yamada, 
Professor T. Takahashi and Mr. T. Sato for fruitful discussions. 
 Numerical computation in this work was partly carried out 
at the Yukawa Institute Computer Facility. 
 The present work was partly supported by a Grant-In-Aid for Scientific 
Research from the Ministry of Education, Science, Sports and Culture, Japan. 
 One of the authors (Y.Y) has been supported by a Research Fellowships of the 
Japan Society for the Promotion of Science for Young Scientists.

\end{document}